\documentclass[aps,prb,twocolumn,superscriptaddress]{revtex4-1}

\usepackage{physics}
\usepackage{bbold}
\usepackage{natbib}    
\usepackage{color}         
\usepackage{graphicx}      
\usepackage{epsfig}          
\usepackage{bm}            
\usepackage{amsmath}
\usepackage{thumbpdf}
\usepackage[belowskip=-14pt,aboveskip=0pt]{caption}
\usepackage{subcaption}
\usepackage{hyperref} 
\usepackage{placeins}
\usepackage{multirow}
\begin{document}

\title{Surface Magnetization in Antiferromagnets: Classification, example materials, and relation to magnetoelectric responses }

\author{Sophie F. Weber*}
\affiliation{Materials Theory, ETH Z\"{u}rich, Wolfgang-Pauli-Strasse 27, 8093 Z\"{u}rich, Switzerland}
\author{Andrea Urru*}
\affiliation{Materials Theory, ETH Z\"{u}rich, Wolfgang-Pauli-Strasse 27, 8093 Z\"{u}rich, Switzerland}
\author{Sayantika Bhowal}
\affiliation{Materials Theory, ETH Z\"{u}rich, Wolfgang-Pauli-Strasse 27, 8093 Z\"{u}rich, Switzerland}
\author{Claude Ederer}
\affiliation{Materials Theory, ETH Z\"{u}rich, Wolfgang-Pauli-Strasse 27, 8093 Z\"{u}rich, Switzerland}
\author{Nicola A. Spaldin}
\affiliation{Materials Theory, ETH Z\"{u}rich, Wolfgang-Pauli-Strasse 27, 8093 Z\"{u}rich, Switzerland}
*These authors contributed equally to the manuscript.
\date{\today}
\begin{abstract}
We use symmetry analysis and density functional theory to characterize antiferromagnetic materials which have a finite equilibrium magnetization density on particular surface terminations. A nonzero magnetic dipole moment per unit area or ``surface magnetization" can arise on particular surfaces of many antiferromagnets due to the bulk magnetic symmetries. Such surface magnetization plays an essential role in numerous device applications, from random-access magnetoelectric memory to exchange bias. However, at this point a universal description of antiferromagnetic surface magnetization is lacking. We first introduce a classification system based on whether the surface magnetization is sensitive or robust to roughness, and on whether the surface of interest is magnetically compensated or uncompensated in the bulk magnetic ground state. We show that uncompensated surface magnetization can be conveniently described in terms of magnetoelectric multipoles at the local-moment, unit cell level, and demonstrate that the symmetry of the multivalued ``multipolization lattice" distinguishes between roughness-robust and roughness-sensitive surface magnetization. We then demonstrate that magnetization on bulk-compensated surfaces arises due to magnetoelectric multipoles (in addition to higher-order magnetic terms) at the atomic site level. These can further be understood in terms of bulk magnetoelectric responses, arising from the effective electric field resulting from the surface termination. We also show with density functional calculations that nominally compensated surfaces in $\mathrm{Cr_2O_3}$ and $\mathrm{FeF_2}$ develop a finite magnetization density at the surface, in agreement with our predictions based on both group theory and the linear and higher-order magnetoelectric response tensors. Our analysis provides a comprehensive basis for understanding the surface magnetic properties in antiferromagnets, and has important implications for phenomena such as exchange bias coupling.  

\end{abstract}

\maketitle
\section*{Introduction}
\label{sec:intro}
\begin{figure}
\includegraphics{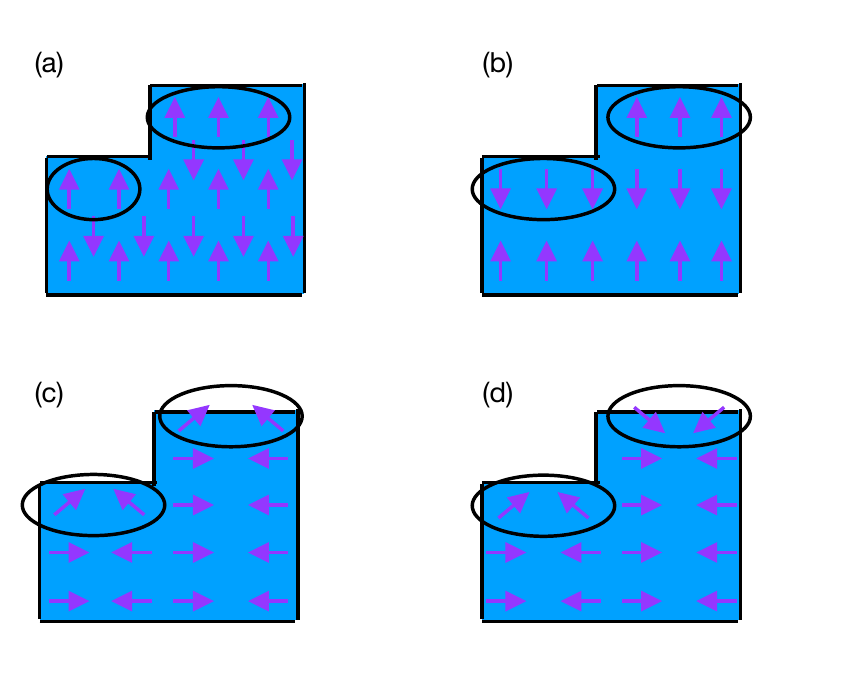}
\caption{\label{fig:surfM_types} Categories of surface magnetization on a surface with roughness, or steps, as described in the text. Circled arrows indicate the magnetic dipoles closest to the vacuum-terminated surface. (a) Roughness-robust, uncompensated. (b) Roughness-sensitive, uncompensated (c) Roughness-robust, induced. (d) Roughness-sensitive, induced.}
\end{figure}

By definition, antiferromagnets (AFMs) have zero net magnetization since the magnetic dipole moments of the oppositely oriented equivalent sublattices sum to zero over the bulk magnetic unit cell. However, theoretical arguments\cite{Andreev1996,Belashchenko2010} and subsequent experimental measurements\cite{Appel2019,Wornle2021} have definitively demonstrated that AFMs with particular bulk symmetries posses a finite magnetic dipole moment per unit area (which we refer to as ``surface magnetization" throughout this work) on certain surfaces. Such surface magnetization enhances the utility of spintronics-based devices that use bulk AFM domains as logical bits. Specifically, since the direction of surface magnetization is connected to the underlying bulk AFM domain, it is a directly detectable indicator of the domain state\cite{Hedrich2021}. Additionally, surface magnetization is believed to play a major role in exchange bias, whereby the magnetization direction of a ferromagnet (FM) is pinned by exchange coupling to an adjacent AFM\cite{Nogues1999}. Exchange bias is exploited in a wide variety of applications, from magnetic random access memory cells to giant magnetoresistive read heads\cite{Borisov2005}.\\
\indent Linear magnetoelectric (ME) AFMs are one experimentally attractive class of AFMs identified as possessing an symmetry-required uncompensated magnetization on particular surface planes\cite{Andreev1996,Belashchenko2010}. The combination of broken inversion and time-reversal symmetries in linear ME AFMs implies that for certain crystallographic directions, an applied electric field $\mathbf{E}$ will induce a finite magnetization $\mathbf{M}$ in the bulk AFM via the equation $\mathbf{M}_i=\sum_{j}\alpha_{ji}\mathbf{E}_j$, where $\alpha$ is the linear ME response tensor\cite{Astrov1960}. Since a surface normal $\hat{\mathbf{n}}$ reduces the bulk symmetry in the same way as an electric field $\mathbf{E}$\cite{Belashchenko2010}, a vacuum-terminated surface in an ME AFM should develop the same magnetization components as those induced by an electric field applied to a bulk sample in the direction parallel to the surface normal\cite{Belashchenko2010,He2010}.\\
\indent The most intuitive, well-known category of surface magnetization arises in linear ME AFMs when the surface planes contain in-plane ferromagnetically ordered atomic magnetic moments, which thus result in a non-vanishing magnetic dipole per unit area. In this case, the magnetic dipoles at the surface maintain the same ordering as they would in the bulk, and the nonzero surface magnetization arises simply due to the vacuum termination of the periodic AFM order. Moreover, the bulk magnetic symmetries of linear ME AFMs imply that the sign and direction of this surface magnetization is robust in the presence of surface roughness or atomic steps\cite{Belashchenko2010}. We refer to this category, depicted with a cartoon in Fig. \ref{fig:surfM_types}(a) as ``roughness-robust, uncompensated" surface magnetization throughout this manuscript.\\
\indent The prototypical linear ME AFM is $\mathrm{Cr_2O_3}$ (chromia)\cite{Dzyaloshinskii:1960,Astrov1960}, and its uncompensated, roughness-robust surface magnetization  has been studied experimentally and theoretically on the $(001)$ surface perpendicular to the N\'{e}el vector direction. The $(001)$ surface magnetization of $\mathrm{Cr_2O_3}$ has been imaged using photoemission electron microscopy\cite{Wu2011}, and the magnitude has been estimated by measuring the stray magnetic field from $(001)$ domains using nitrogen vacancy scanning magnetometry\cite{Appel2019,Wornle2021}. The uncompensated surface magnetization on $(001)$ $\mathrm{Cr_2O_3}$ is highly promising for spintronics applications, because the direction of the magnetic dipoles at the surface can be readily switched (along with switching of the underlying bulk AFM domain) by simultaneous application of electric and magnetic fields (``magnetoelectric annealing")\cite{Borisov2005,He2010}.\\
\indent However, surface magnetization arising from uncompensated magnetic dipole moments in linear ME AFMs is just one category. Indeed, Andreev pointed out nearly three decades ago that for a general surface of \emph{any} AFM (not just linear MEs), a vanishing surface magnetization is thermodynamically unstable\cite{Andreev1996}. In fact, to \emph{not} have a finite surface magnetization, the vacuum-terminated plane must have an orientation which retains particular bulk symmetry operations of the AFM, for example, with its surface normal parallel to a rotation axis or a mirror plane.\\
\indent One variety of surface magnetization outside of the uncompensated, roughness-robust case of $(001)$ $\mathrm{Cr_2O_3}$ consists of uncompensated magnetic dipoles that are ferromagnetically ordered only on a perfectly flat surface. In these cases, the direction of surface magnetization switches between atomic steps, thus averaging to zero in the presence of roughness. The symmetry requirements for this ``roughness-sensitive" uncompensated surface magnetization, depicted in Fig. \ref{fig:surfM_types}(b), are significantly less strict than those for roughness-robust, uncompensated surface magnetization, and while this category was pointed out decades ago by Andreev\cite{Andreev1996}, to our knowledge it has not been examined further.\\
\indent Counter-intuitively, surface magnetization can even arise on AFM surface planes showing an in-plane AFM order, which formally results in a vanishing magnetic dipole per unit area, but symmetry reduction due to the surface termination allows for induced components of the magnetic moments, producing a net ferromagnetic (or ferrimagnetic) order and, thus, a non-zero surface magnetization. We refer to this as ``induced" surface magnetization, which in principle can be either roughness-robust or roughness-sensitive (Figs. \ref{fig:surfM_types}(c) and (d) respectively). Crucially, induced surface magnetization, even the roughness-robust category, can occur in both linear ME AFMs and in centrosymmetric AFMs for which the linear ME response is forbidden.  
We note that a few specific material examples of induced surface magnetization, such as the surfaces perpendicular to $(001)$ in $\mathrm{Cr_2O_3}$\cite{Belashchenko2010} or surface magnetization in non-linear ME, centrosymmetric $\mathrm{FeF_2}$\cite{Lapa2020}, both of which we analyze in detail in this manuscript, have been mentioned in the literature. However, a unifying description of different surface magnetization categories, their distinguishing properties, and \emph{ab initio} calculations demonstrating the existence of such surface magnetization, is lacking.\\
\indent Our first goal in Part I (Sec. \ref{sec:partI}) of this work is to introduce a classification scheme of AFM surface magnetization in terms of group-theory symmetry requirements. In Secs. \ref{subsec:all_SurfaceM} and \ref{subsec:rough}, we review and extend an elegant point-group symmetry formalism, developed by Belashchenko in 2010\cite{Belashchenko2010,Lapa2020}, to identify surfaces in bulk AFMs with equilibrium magnetization. We apply this formalism to identify and distinguish surface magnetization that is robust to surface roughness, treated in Refs. \citenum{Belashchenko2010} and \citenum{Lapa2020}, and surface magnetization that is symmetry allowed but which averages to zero in the presence of roughness or atomic steps (the roughness-sensitive case). Moreover, we show that in both roughness-robust and roughness-sensitive families, a net surface magnetization can further be classified as either uncompensated or induced, as described above.\\
\indent We then move on to describe how AFM surface magnetization can be understood and distinguished by a multipole-based analysis complementary to the group theory formalism. First, in Part II (Sec. \ref{sec:partII}) we focus on magnetically uncompensated surface magnetization, analyzing both the roughness-robust and roughness-sensitive cases in terms  of magnetoelectric multipoles. In Sec.\ref{sec:multpol} we remind the reader of the decomposition of the ME multipole tensor into a unit cell level ``local-moment" contribution and a sum of ``atomic-site" contributions. We then discuss how uncompensated surface magnetization can be quantitatively described by the local-moment contribution to the ``magnetoelectric multipolization"\cite{Spaldin2021}. Furthermore, we show how the symmetry of the bulk local-moment multipolization lattice provides a method of distinguishing roughness-robust and roughness-sensitive cases which is complementary to the group theory description. In Sec. \ref{subsec:partII_egs} we work through the explicit examples of roughness-robust, uncompensated surface magnetization in $(001)$ $\mathrm{Cr_2O_3}$ and roughness-sensitive uncompensated surface magnetization in $(111)$ $\mathrm{NiO}$ and $(001)$ $\mathrm{Fe_2O_3}$\\
\indent In Part III (Sec. \ref{sec:partIII}), we move to surface magnetization that is induced by symmetry-lowering at a surface. In Sec. \ref{ME_response} we show that induced surface magnetization can be described by multipoles of the magnetization density at the \emph{atomic site level} as opposed to the unit-cell level, local-moment origin of the uncompensated surface magnetization in Part II \ref{sec:partII}. We extend the discussion of multipoles to second and third order beyond the linear ME multipoles, and show how these are linked to first-, second-, and third-order ME responses. In Sec. \ref{subsec:induced_robus} we analyze roughness-robust, induced surface magnetization. We show the connection between linear, second-order (bilinear), and third-order (trilinear) ME responses, the corresponding ferroically ordered magnetic multipoles, and the induced roughness-robust surface magnetization. We take Cr$_2$O$_3$ $(\bar{1}20)$, and $(100)$ and FeF$_2$'s $(110)$ surfaces as examples for which we confirm our symmetry-based predictions with \textit{ab initio} calculations. In Sec. \ref{subsec:induced_sense} we discuss how roughness-sensitive, induced surface magnetization arises due to each surface plane having ferroically ordered atomic-site multipoles which change sign on adjacent planes. Finally, in Sec. \ref{subsec:trivial} we show with symmetry arguments and density functional theory (DFT) calculations that the $(1\bar{1}0)$ surface of $\mathrm{NiO}$ has zero compensated or induced surface magnetization due to the in-plane antiferroic ordering of both its dipoles and all of its atomic-site multipoles. Sec. \ref{conclusions} contains our conclusions.\\
\indent Our work introduces a unifying symmetry formalism that synthesizes multiple distinct literature reports on AFM surface magnetization\cite{Marchenko1981, Dzyaloshinskii1992, Andreev1996, Belashchenko2010,Lapa2020}. We demonstrate that surface magnetization can also be described in terms of bulk magnetic multipoles, implying that multipoles at various orders may serve as bulk order parameters for surface magnetization, and conversely surface magnetization may indicate the existence of hidden bulk multipoles. Finally, we provide quantitative DFT calculations indicating that symmetry-allowed, induced surface magnetization can stabilize nominally compensated surfaces, which to our knowledge are the first calculations of their kind. Overall, our findings show that the phase space of bulk AFMs and Miller planes with surface magnetization that can be exploited in applications should be vastly expanded compared to the current cases under consideration. \\  
\section{\label{sec:partI}Part I: Group-theory description and classification of surface magnetization in antiferromagnets}
To start, we introduce the group-theoretical procedure we will use to determine firstly whether a certain Miller plane for a particular AFM will have a finite equilibrium surface magnetization, and secondly whether the surface magnetization is roughness-robust or roughness-sensitive. We emphasize that the group-theory procedure to identify roughness-robust surface magnetization has already been developed in Refs. \citenum{Belashchenko2010} and \citenum{Lapa2020}. Here, we extend the procedure to include, and distinguish, the roughness-sensitive categories of surface magnetism\cite{Andreev1996}.\\
\subsection{\label{subsec:all_SurfaceM}Identification of surfaces compatible with magnetization in the atomically flat limit}
\begin{figure*}
\includegraphics{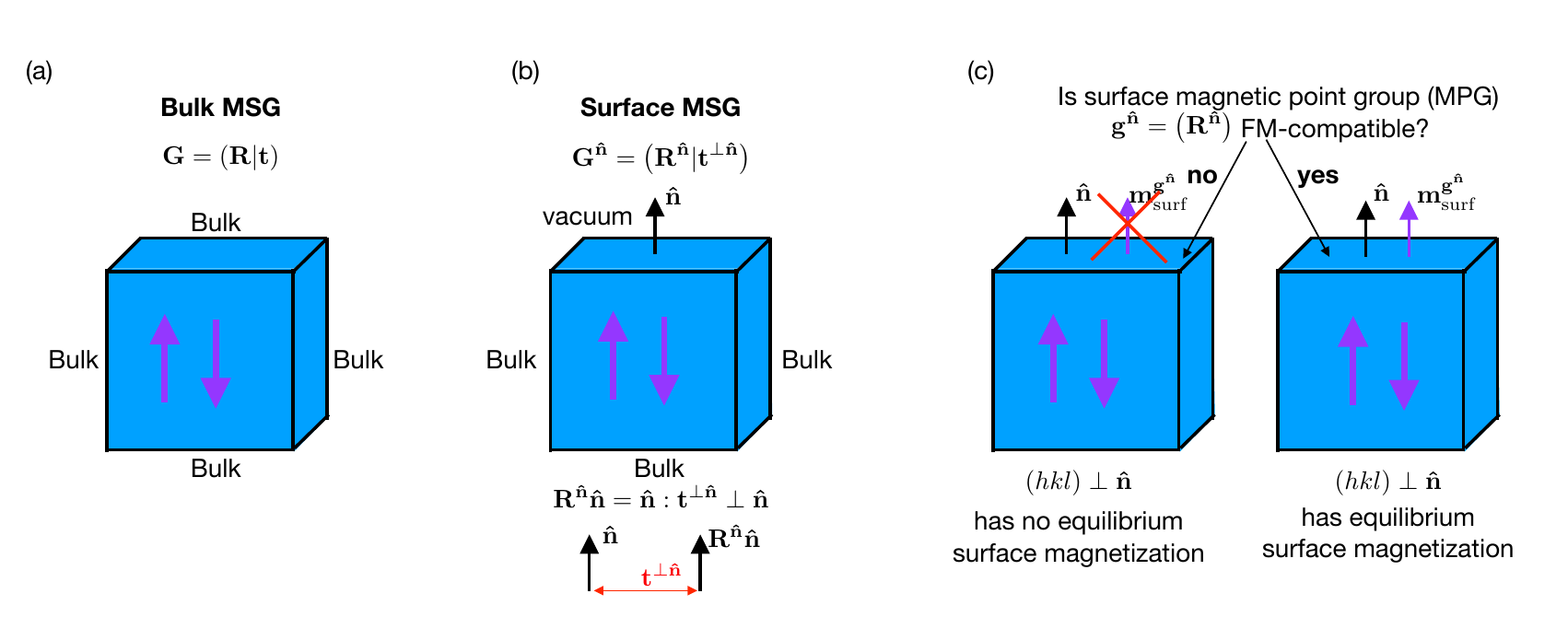}
\caption{\label{fig:surfM_cartoon} Group theory procedure to identify whether a given Miller plane $(hkl)\perp\mathbf{\hat{n}}$ has a symmetry-allowed surface magnetization, assuming that the surface has no atomic steps. (a) The bulk magnetic space group (MSG) $\mathbf{G}$ is identified for the bulk AFM which is periodic in all directions. (b) The surface MSG $\mathbf{G^{\hat{n}}}$ is identified as the subgroup of $\mathbf{G}$ with operations which leave the surface normal $\mathbf{\hat{n}}$ invariant, modulo translations parallel to the surface. (c) If the magnetic point group (MPG) $\mathbf{g^{\hat{n}}}$ corresponding to $\mathbf{G^{\hat{n}}}$ is compatible with ferromagnetism, then $(hkl)\perp\mathbf{\hat{n}}$ has an equilibrium surface magnetization $\mathbf{m}_{\mathrm{surf}}^{\mathbf{g^{\hat{n}}}}$.} 
\end{figure*}
\indent Let us consider a generic bulk AFM, periodic in all spatial directions, with magnetic space group (MSG) $\mathbf{G}=\left(\mathbf{R}|\mathbf{t}\right)$. Here, $\mathbf{R}$ is the set of point group operations in $\mathbf{G}$ (some of which may contain time-reversal $\Theta$), and $\mathbf{t}$ are the fractional space group translations. Note that in this work, for simplicity we focus exclusively on collinear AFMs; the extension to noncollinear AFMs is straightforward. Now, we take a particular Miller plane $(hkl)$, and create a surface by terminating the solid and interfacing it with vacuum. Assuming a macroscopically flat termination with no steps (we will later relax this requirement), the unit vector surface normal $\mathbf{\hat{n}}\perp (hkl)$, which is perpendicular to the vacuum-terminated surface, is all that is needed to characterize the subgroup of the bulk MSG that defines the two-dimensional surface in question. This ``surface" MSG, which we will denote $\mathbf{G^{\hat{n}}}$, is merely the set of space group operations $\left(\mathbf{R^{\hat{n}}}|\mathbf{t}^{\perp\mathbf{\hat{n}}}\right)$ in $\mathbf{G}$ that leave the polar vector $\mathbf{\hat{n}}$ invariant, modulo translations parallel to the surface, or equivalently, perpendicular to $\mathbf{\hat{n}}$:
\begin{equation}
(\mathbf{R}_i|\mathbf{t}_i)\in \mathbf{G^{\hat{n}}}\iff \mathbf{R}_i\mathbf{\hat{n}}=\mathbf{\hat{n}}\land \mathbf{t}_i\perp\mathbf{\hat{n}},
\label{eq:surf_MSG}
\end{equation}
where $(\mathbf{R}_i|\mathbf{t}_i)$ is the $i^{\mathrm{th}}$ element of the bulk MSG $\mathbf{G}$. Note that the surface MSG must have broken inversion symmetry $\mathcal{I}$, since $\mathcal{I}$ will always reverse the direction of $\mathbf{\hat{n}}$. Other operations in $\mathbf{G}$ which are preserved or broken in $\mathbf{G^{\hat{n}}}$ depend on the relative orientations of $\mathbf{\hat{n}}$ and the principle axes of the bulk MSG operations.\\
\indent Once we have identified the surface MSG $\mathbf{G^{\hat{n}}}$ for the Miller plane of interest, we next consider the corresponding magnetic point group (MPG) $\mathbf{g^{\hat{n}}}=(\mathbf{R^{\hat{n}}})$ which contains the point group elements of $\mathbf{G^{\hat{n}}}$ without the accompanying translations. We then check whether $\mathbf{g^{\hat{n}}}$ allows for ferromagnetism. $31$ out of the $58$ MPGs in solids are FM-compatible,  meaning that at least one component of magnetization is left invariant under all operations of the point group\cite{Schmid1973}. In three dimensions, an FM-compatible MPG implies that a finite magnetic dipole moment \emph{per unit volume} is allowed. In the present formalism, $\mathbf{g^{\hat{n}}}$ describes the two-dimensional surface perpendicular to $\mathbf{\hat{n}}$: in this case, an FM-compatible surface MPG allows for a nonzero magnetic dipole moment \emph{per unit area}, i.e. a surface magnetization. Conversely, if $\mathbf{g^{\hat{n}}}$ is not FM-compatible, an equilibrium surface magnetization on $(hkl)\perp\mathbf{\hat{n}}$ is symmetry-forbidden. This simple procedure to identify atomically smooth surfaces of AFMs which are expected to have nonzero surface magnetization is illustrated with cartoons in Fig. \ref{fig:surfM_cartoon}(a)-(c).\\
\indent Note that in contrast to our procedure in Fig. \ref{fig:surfM_cartoon}, Refs. \citenum{Belashchenko2010} and \citenum{Lapa2020} define the surface MPG by considering all point group operations in $\mathbf{G}$ which leave the direction of $\mathbf{\hat{n}}$ unchanged, irrespective of the directions of the corresponding space group translations with respect to the surface. In some cases this definition of the surface MPG yields a higher-symmetry point group than that shown in Fig. \ref{fig:surfM_cartoon}(c), and it excludes surface magnetization that exists for atomically smooth surfaces only. Our formalism in Fig. \ref{fig:surfM_cartoon} on the other hand allows for this category. In what follows, we explain how to determine whether the surface magnetization identified in Fig. \ref{fig:surfM_cartoon} is robust in the presence of atomic steps, or whether it averages to zero if roughness is introduced.
\subsection{\label{subsec:rough}``Roughness-robust" versus ``roughness-sensitive" surface magnetization}
In the previous section, by restricting the surface MSG to operations which only involve translations parallel to the surface $\perp\mathbf{\hat{n}}$, we showed how to identify AFM surfaces which have a macroscopic magnetization in the limit of an ideally smooth plane with no roughness or steps. Next, we impose an additional criterion to distinguish between FM-compatible surfaces  whose macroscopic magnetization is robust to roughness (corresponding to the subset which Refs. \citenum{Belashchenko2010} and \citenum{Lapa2020} identify), and those whose magnetization is roughness-sensitive.\\
\begin{figure}
\includegraphics{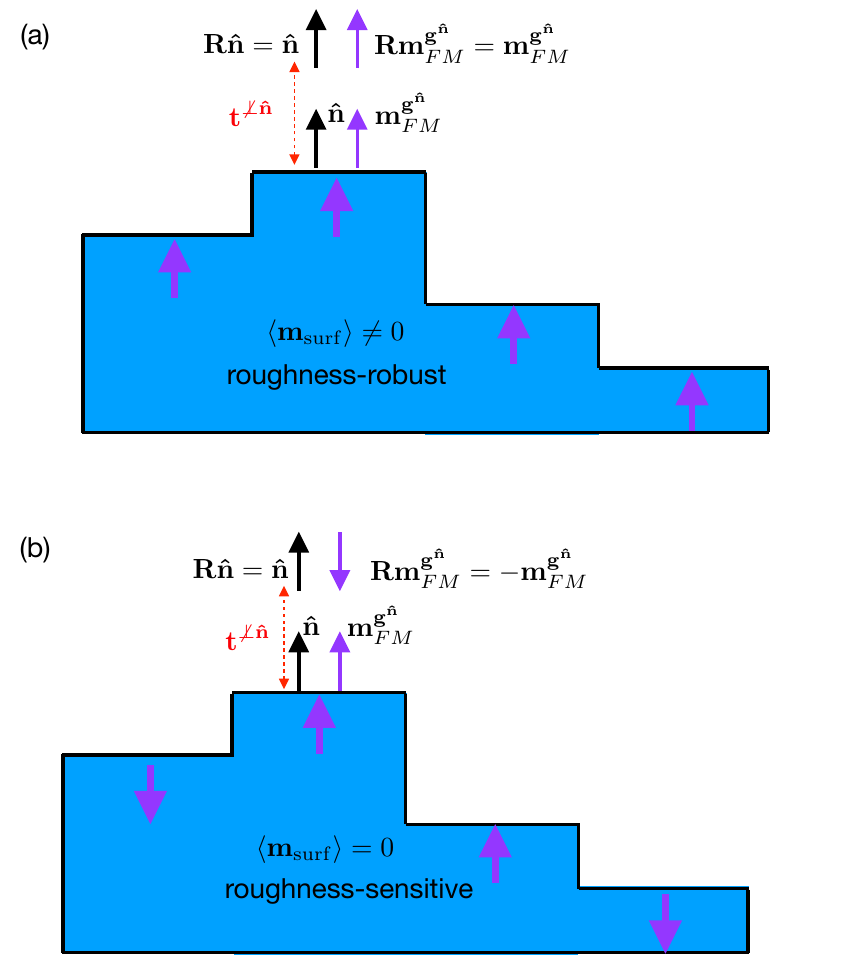}
\caption{\label{fig:roughness} ``Roughness-robust" versus ``roughness-sensitive" surface magnetization. (a) Roughness-robust: any operation in the bulk MSG preserving the direction of $\mathbf{\hat{n}}$ and containing a translation component $\parallel\mathbf{\hat{n}}$ leaves $\mathbf{m}_{\text{FM}}^{\mathbf{g^{\hat{n}}}}$ invariant. Then, the sign and direction of surface magnetization is constant for all atomic steps. (b) Roughness-sensitive: at least one operation in $\mathbf{G}$ with $\mathbf{t}\not\perp\mathbf{\hat{n}}$ reverses the direction of $\mathbf{m}_{\text{FM}}^{\mathbf{g^{\hat{n}}}}$. Then the surface magnetization will average to zero over a macroscopic area in the presence of roughness.} 
\end{figure}
\indent Assume that for some bulk AFM we have already identified a particular Miller plane characterized by surface normal $\mathbf{\hat{n}}$ which has equilibrium surface magnetization according to the procedure in Fig. \ref{fig:surfM_cartoon}. Now, let us go back and again consider the higher symmetry bulk MSG $\mathbf{G}$. We want to check whether there exists in $\mathbf{G}$ at least one operation $(\mathbf{R}_i|\mathbf{t}_i)\notin \mathbf{G^{\hat{n}}}$ for which
\begin{equation}
\mathbf{R}_i\mathbf{\hat{n}}=\mathbf{\hat{n}}\label{eq:roughness_nreq},
\end{equation}
\begin{equation}
\mathbf{t}_i\cdot\mathbf{\hat{n}}\neq 0\label{eq:roughness_treq},
\end{equation}
and
\begin{equation}
\mathbf{R}_i\mathbf{m}_{\mathrm{FM}}^{\mathbf{g^{\hat{n}}}}=-\mathbf{m}_{\mathrm{FM}}^{\mathbf{g^{\hat{n}}}}\label{eq:roughness_mreq}.
\end{equation}
Here, $\mathbf{m}_{\mathrm{FM}}^{\mathbf{g^{\hat{n}}}}$ is the direction of magnetization which remains invariant under the operations of the surface MPG $\mathbf{g^{\hat{n}}}$, i.e., the direction along which ferromagnetism is allowed. If Eqs. \eqref{eq:roughness_nreq}-\eqref{eq:roughness_mreq} all hold, then there exist atomic steps on surface $(hkl)\perp\mathbf{\hat{n}}$, connected by $\mathbf{t}_i$, which are energetically and symmetrically equivalent, but which have opposite signs of surface magnetization\cite{Andreev1996}. Thus, for a macroscopic region with surface roughness in thermodynamic equilibrium, steps with opposite directions and equal magnitudes of magnetization will occur with equal probability such that the total magnetization averaged over the entire surface area is zero. Conversely, if all bulk MSG operations which involve $\mathbf{t}\not\perp\mathbf{\hat{n}}$ leave $\mathbf{m}_{\mathrm{FM}}^{\mathbf{g^{\hat{n}}}}$ unchanged, then all atomic steps in the presence of roughness will have the same direction of surface magnetization, hence the surface magnetization is ``roughness-robust." This is illustrated in Fig. \ref{fig:roughness}.\\
\indent Surfaces with the roughness-robust category of surface magnetization in Fig. \ref{fig:roughness}(a) have some clear practical advantages. To start, the experimental challenge of synthesizing atomically smooth surfaces either in thin film form or in bulk is nontrivial. Nevertheless, synthesis methods are constantly improving, particularly in the field of thin film deposition, and it is often possible to achieve atomically smooth surfaces even for problematic compounds such as metal oxides\cite{Nunn2021,Sun2020}.\\
\indent Moreover, as we will see more explicitly later, roughness-sensitive surface magnetization can occur in AFMs forbidding the linear ME effect, and in these cases it cannot be controlled by ME annealing. But for very thin AFM films in which the ratio of uncompensated surface magnetic dipoles to compensated dipoles in the bulk is fairly large, it can be possible to control the surface magnetization with a reasonably sized magnetic field alone. Thus, identification of roughness-sensitive surface magnetization and further studies of its properties are of practical interest given the considerable increase in the number of materials with usable surface magnetization this could afford, \\
\begin{figure*}
\includegraphics{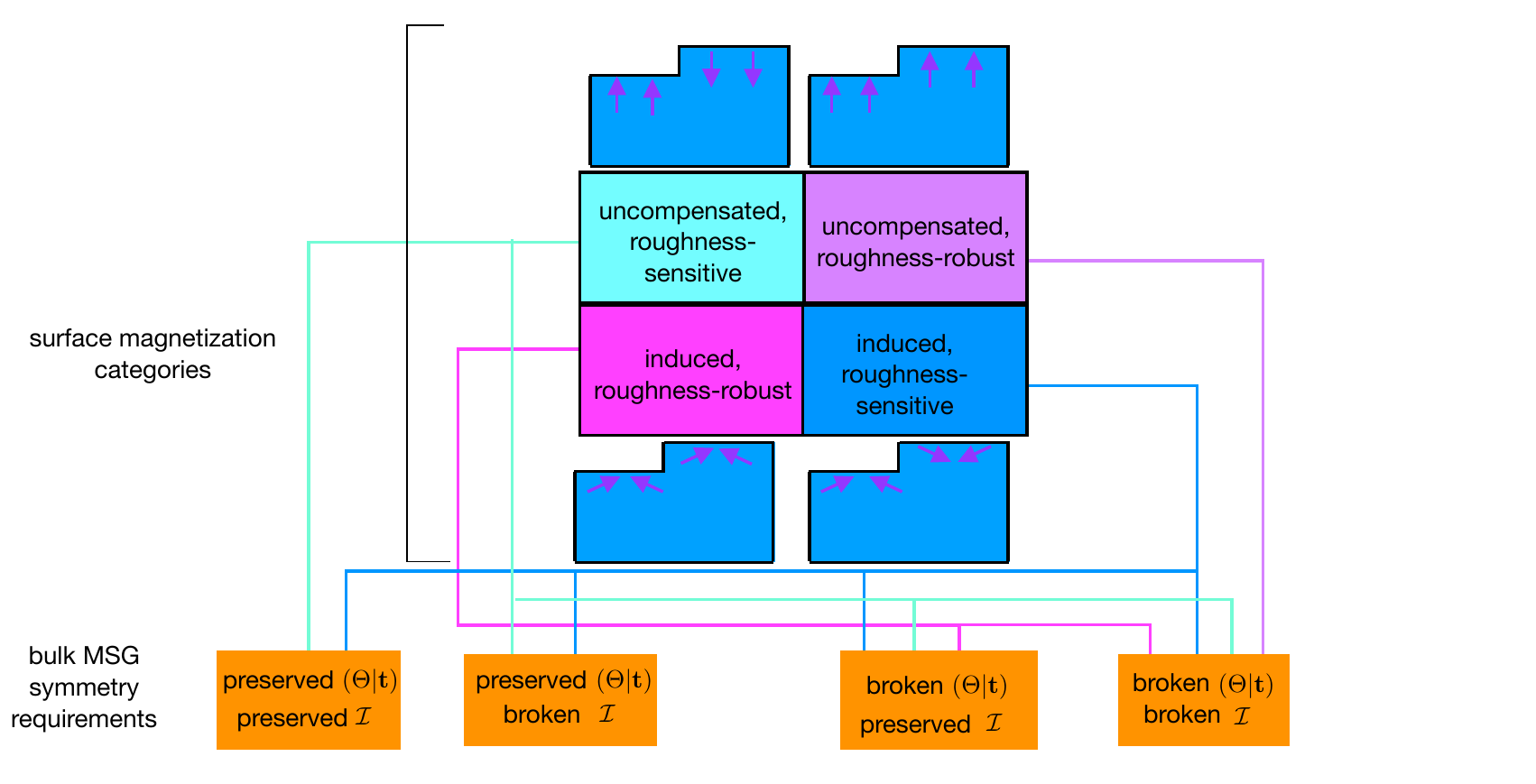}
\caption{\label{fig:categories} Flowchart summarizing the categories of surface magnetization that can exist for certain Miller planes in a given bulk AFM. We distinguish the bulk AFM types based on the absence or presence of $\Theta$ and $\mathcal{I}$ in the bulk MSG. We also provide cartoons depicting the behavior of the magnetization in each of the categories on a surface with atomic steps. $\left(\Theta|\mathbf{t}\right)$ represents any operations in the bulk MSG containing time-reversal, that is, it includes pure time-reversal $\left(\Theta|0\right)$ as well as $\Theta$ combined with a translation $\mathbf{t}$.}
\end{figure*}
\indent The group-theory formalism which we have described thus far is general, identifying all AFM Miller planes having a symmetry-allowed finite equilibrium surface magnetization which may either exist despite roughness or may require an atomically flat surface to be realized. Fig. \ref{fig:categories} gives a flowchart categorizing these four types of surface magnetization, and indicating which are allowed, given the absence or presence of inversion and time-reversal symmetries, (as well as combined time-reversal plus translation space group symmetries), in the bulk MSG. The importance of the inversion symmetry requirements will become more clear when we introduce the multipole-based description of surface magnetization in Parts II and III. We can see that the roughness-robust, uncompensated surface magnetization has the strictest symmetry requirements for the bulk AFM, namely having a bulk MSG with broken time-reversal and inversion symmetries. Note that these are the same symmetries which must be broken for a linear ME response. On the other hand, the roughness-sensitive categories, both induced and uncompensated, can occur for bulk AFMs with all possible combinations of inversion and time-reversal symmetries.\\
\indent While Eqs. \eqref{eq:roughness_nreq}-\eqref{eq:roughness_mreq} distinguish between roughness-robust and roughness-sensitive cases, they do not distinguish between uncompensated and induced surface magnetization categories. In the following parts of the manuscript, we will introduce a complementary description of AFM surface magnetization, combined with concrete material examples, based on the multipoles of the bulk magnetization density, which will help us to classify and understand each of the four categories. We will see that this multipole-based formalism is intimately connected to the group-theory description we have reviewed and modified in Part I.\\
\section{\label{sec:partII}Part II: Uncompensated surface magnetization in terms of local-moment multipolization }
In Part II of this manuscript, we will develop a semi-quantitative description of uncompensated AFM surface magnetization (both roughness-robust and roughness-sensitive) in terms of a ``local-moment" multipole tensor which involves summing over the positions and dipole moments of magnetic atoms in the bulk unit cell which tiles the slab containing the surface in question. This ``multipolization" formalism was already proposed in Ref. \citenum{Spaldin2021} to predict the value of uncompensated, roughness-robust surface magnetization in linear MEs. Here, we show that this local-moment multipolization can also describe and distinguish roughness-sensitive uncompensated surface magnetization, even when the bulk MSG forbids a ME response. We first review the description of bulk magnetization density in terms of a multipole expansion, and then investigate the implications of the form of the local-moment multipolization on uncompensated surface magnetization .\\
\subsection{\label{sec:multpol}Magnetoelectric multipoles as bulk indicators of surface magnetization}
While in many cases one can approximate the magnetic order of a solid as a series of point magnetic dipole moments localized on the magnetic ions, the full magnetization density is more complex. It is convenient to describe the asymmetry beyond the usual magnetic dipole of a generic, inhomogenous magnetization density in terms of a multipole expansion. Usually, one constructs the form of the magnetic multipoles at each order via an expansion of a magnetization density $\boldsymbol{\mu}(\mathbf{r})$ interacting with a spatially varying magnetic field $\mathbf{H}(\mathbf{r})$\cite{Spaldin2021,Bhowal2022}:
\begin{align}
E_{\text{int}} &= -\int \boldsymbol{\mu}(\mathbf{r})\cdot\mathbf{H}(\mathbf{r})d^3\mathbf{r} \nonumber \\
&= -\int \boldsymbol{\mu}(\mathbf{r})\cdot \mathbf{H}(0)d^3\mathbf{r}-\int r_i\mu_j(\mathbf{r})\partial_i H_j(0) d^3\mathbf{r} \nonumber \\
& \, \, \, \, \, \,   -\int r_i r_j \mu_k(\mathbf{r})\partial_i \partial_j H_k(0) d^3\mathbf{r} - \dots,
\label{eq:multpol_Eint}
\end{align}
where $(i,j)=1,2,3$ refer to Cartesian components. The first term of the expansion in Eq. \eqref{eq:multpol_Eint} is the magnetic dipole whereas the the coefficient of the second term, 
\begin{equation}
\mathcal{M}_{ij}=\int r_i\mu_j(\mathbf{r})d^3\mathbf{r},
\label{eq:Memultpol}
\end{equation}
is known as the ME multipole tensor and describes the first order of asymmetry in magnetization density\cite{Ederer2007,Suzuki2018,Spaldin2008,Spaldin2013},. We briefly mention here that the rank-3 tensor which gives the coefficient of the third term in Eq. \eqref{eq:multpol_Eint},
\begin{equation}
\mathcal{M}_{ijk}=\int r_i r_j\mu_k(\mathbf{r})d^3\mathbf{r},
\label{eq:octupole}
\end{equation}
is called the magnetic octupole\cite{Urru2022,Bhowal2022}. Note that Eq. \eqref{eq:octupole} is even under inversion symmetry $\mathcal{I}$, and in Part III (Sec. \ref{sec:partIII}) we will show that ferroically ordered octupoles explain the existence of roughness-robust, \emph{induced} surface magnetization in centrosymmetric AFMs. For uncompensated surface magnetization however, the lower-order ME multipoles are sufficient to describe both roughness-sensitive and roughness-robust cases, so we will concentrate exclusively on $\mathcal{M}_{ij}$ in this section.\\
\indent It is evident that the ME multipole tensor Eq. \eqref{eq:Memultpol} simultaneously breaks inversion symmetry and time-reversal due to its dependence on both position and magnetization. In fact, one can show rigorously that nonzero values of the $\mathcal{M}_{ij}$ tensor imply a corresponding nonzero linear ME response $\alpha_{ij}$ in the presence of an applied electric field\cite{Spaldin2008}, where as a reminder the linear ME response tensor $\alpha$ is defined by 
\begin{equation}
\mathbf{M}_i=\sum_j\alpha_{ji}\mathbf{E}_j,
\label{eq:MEtensor}
\end{equation}
with $\mathbf{E}$ the applied electric field and $\mathbf{M}$ the induced magnetization in the bulk. Thus, the linear ME effect, which is a thermodynamic response, can be conveniently understood in terms of the ME multipole tensor, which characterizes the bulk magnetization density in the absence of applied fields.\\
\begin{figure}
\includegraphics{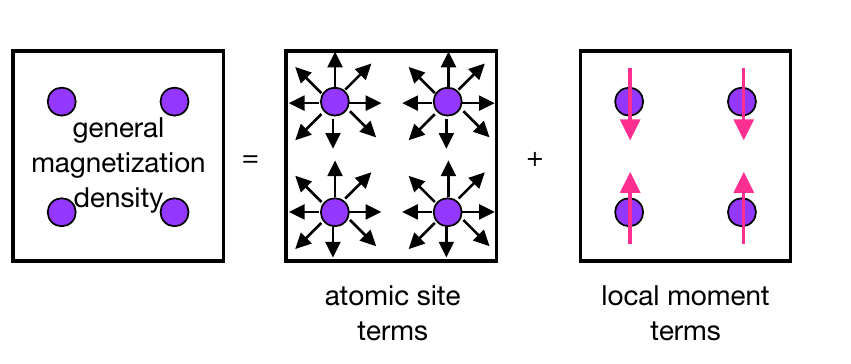}
\caption{\label{fig:lm_vs_as} local-moment and atomic-site contributions to a magnetization density. Any inhomogenous magnetization may be decomposed into a sum of atomic-site terms, which capture magnetization asymmetry around the atomic nuclei (purple circles), and a local-moment contribution due to the arrangement of magnetic dipoles centered on the ions in the unit cell.}
\end{figure}
\indent Crucially for our discussion of surface magnetization categories, $\mathcal{M}$ can be decomposed into a sum of ``local-moment" (lm) and ``atomic-site" terms in the following way \cite{Bhowal2021}:
\begin{align}
\mathcal{M}_{ij} &\approx \sum_{\alpha} \int_{\Omega_{\alpha}}(r_i-R_{i, \alpha}) \mu_j(\mathbf{r}) d^3\mathbf{r} \nonumber \\
& \, \, \, \, \, \, +\sum_{\alpha} \int_{\Omega_{\alpha}} R_{i,\alpha}\mu_j(\mathbf{r})d^3\mathbf{r} \nonumber \\
&=\sum_{\alpha} \int_{\Omega_{\alpha}}(r_i-R_{i,\alpha}) \mu_j(\mathbf{r})d^3\mathbf{r} + \sum_{\alpha} R_{i,\alpha} m_{j,\alpha} \nonumber \\
&=\mathcal{M}_{ij}^{\text{as}}+\mathcal{M}_{ij}^{\text{lm}}.
\label{eq:loc_as}
\end{align}
Here, $\mathbf{R}_{\alpha}$ is the position of atom $\alpha$, the sum is over magnetic atoms in the unit cell, and $\Omega_{\alpha}$ is a spherical region around atom $\alpha$. Note that Eq. \eqref{eq:loc_as} neglects contributions to $\mathcal{M}_{ij}$ due to magnetization density in interstitial regions, which we assume to be negligible. The first term in the expression, the ``atomic-site" (as) term, originates from asymmetry in the magnetic density around an atomic site, and the second term, called the ``local-moment" term, describes contributions to $\mathcal{M}$ due to the asymmetric arrangement of local dipole moments across all magnetic sites at the unit cell level. When considering $\mathcal{M}_{ij}^{\text{lm}}$ in an extended solid, we normalize the tensor by the volume $V$ of the magnetic unit cell:
\begin{equation}
\widetilde{\mathcal{M}}_{ij}^{\text{lm}}=\frac{1}{V}\sum_{\alpha}R_{i,\alpha}m_{j,\alpha},
\label{eq:M_lm}
\end{equation}
In Part III (Sec. \ref{sec:partIII}), we will show in detail that the atomic-site contribution to the multipolization, as well as higher-order terms in the multipole expansion\cite{Urru2022,Bhowal2022} of Eq. \eqref{eq:multpol_Eint}, are bulk indicators for the existence of \emph{induced} surface magnetization. In this part, however, we discuss only the lm contribution given by Eq. \eqref{eq:M_lm}, which as we will see, is necessarily nonzero for surface orientations corresponding to uncompensated surface magnetization in AFMs.\\
\indent We motivate the following discussion by a brief comparison with the case of charge on the surfaces of materials with non-zero electric polarization. Recall that within the modern theory of polarization, there is a correspondence between the bulk electric polarization vector, $\mathbf{P}_{\mathrm{bulk}}$, and the bound surface charge density $\sigma_{\mathrm{surf}}$ for a perpendicular surface, that is, $\mathbf{P}_{\mathrm{bulk}}\cdot \hat{\mathbf{n}}=\sigma_{\mathrm{surf}}$\cite{Vanderbilt1993}. However, the periodicity of a bulk crystal implies that $\mathbf{P}_{\mathrm{bulk}}$ is only defined modulo a ``polarization quantum" corresponding to translating one electron by a lattice vector\cite{KingSmith1993}. Therefore, rather than a single value of $\mathbf{P}_{\mathrm{bulk}}$, a given bulk crystal has a polarization ``lattice" with values separated by the polarization quantum $\mathbf{P}_q=e\mathbf{R}/V$, with $\mathbf{R}$ a Bravais lattice vector. Therefore, at first glance the correspondence to surface charge for a given termination $\mathbf{P}_{\mathrm{bulk}}\cdot \hat{\mathbf{n}}=\sigma_{\mathrm{surf}}$ seems ambiguous. This apparent issue is solved by the fact that selecting a specific surface termination dictates a particular basis choice for the bulk unit cell; that which periodically tiles the semi-infinite solid containing the surface of interest\cite{Stengel2011}. Thus, the component of $\mathbf{P}_{\mathrm{bulk}}$ along the surface normal $\mathbf{P}_{\mathrm{bulk}}\cdot \hat{\mathbf{n}}$ calculated for this particular unit cell choice is single valued, and the surface charge is determined unambiguously. Moreover, in general, for a given Miller plane $(hkl)$, the atomic termination which is electrostatically stable or nonpolar is defined by the bulk unit cell for which $\mathbf{P}_{\mathrm{bulk}}\cdot \hat{\mathbf{n}}=\sigma_{\mathrm{surf}}=0$\cite{Nakagawa2006}.\\
\indent While $\mathbf{P}_{\mathrm{bulk}}$ has units of electric charge per unit area, the components of the volume-normalized $\widetilde{\mathcal{M}}^{\text{lm}}$ have units of magnetic dipole moment per unit area. Thus, just as the $1\mathrm{st}$-order term (polarization) in a multipole expansion of a bulk charge density in an electric field gives rise to a $0\mathrm{th}$-order term (charge) associated with its surface, in Ref. \citenum{Spaldin2021}, we argued that the $2\mathrm{nd}$-order term (the multipolization) in the multipole expansion in Eq. \eqref{eq:multpol_Eint} should lead to a $1\mathrm{st}$-order term (magnetic dipole per unit area) associated with a surface. Specifically, the unit cell-normalized bulk multipolization component $\mathcal{M}_{ij}$ should correspond to the magnetic dipole density per unit area along Cartesian direction $\hat{j}$ with surface unit normal along $\hat{i}$. This surface magnetization can have contributions from both the as and lm contributions of the $\mathcal{M}_{ij}$ tensor in general.\\
\indent The unit cell-level lm term in Eq. \eqref{eq:M_lm}, like the unit cell-level electric polarization $\mathbf{P}_{\mathrm{bulk}}$, is multivalued in bulk due to the lattice periodicity\cite{Ederer2007,Spaldin2008,Spaldin2013,Spaldin2021} Here, the multivaluedness corresponds to translating a magnetic ion by one lattice vector $\mathbf{R}$. In fact, there can be up to $9l$ linearly independent ``multipolization increments" (assuming that that the magnetic moments do not rotate); one for each component of the $\mathcal{M}$ tensor, and one for each of $l$ magnetic atoms in the basis. But again, the choice of a specific Miller plane and atomic termination fixes the origin of the bulk unit cell used to compute Eq. \eqref{eq:M_lm} along the direction of the surface normal. Thus, the three components $j=1,2,3$ of surface magnetization associated with $\mathcal{M}_{ij}$ on surface $(hkl)\perp\hat{i}$ are all single-valued. As a technicality, note that for a lattice of monoclinic or triclinic symmetry, the out-of-plane lattice vector corresponding to different multipolization increments with different surface terminations is not exactly parallel to the surface normal. In this case, we can merely project the components of Eq. \eqref{eq:M_lm} corresponding to the out-of-plane direction onto the surface normal in order to describe the surface magnetization in an orthogonal basis.\\
\indent From the construction of Eq. \eqref{eq:M_lm} it is clear that $\widetilde{\mathcal{M}}_{ij}^{\text{lm}}$ calculated for a bulk unit cell defining any surface which is terminated by an uncompensated, FM layer of magnetic dipole moments (i.e. with a surface magnetization) must be nonzero. Moreover, the lm contribution will capture the quantitative surface magnetization expected by weighting the contribution of each magnetic moment in the unit cell by its distance from the surface. Thus, $\widetilde{\mathcal{M}}_{ij}^{\text{lm}}$ is sufficient to quantitatively describe the uncompensated contribution to the surface magnetization for a  given Miller plane.\\
\begin{figure}
\includegraphics{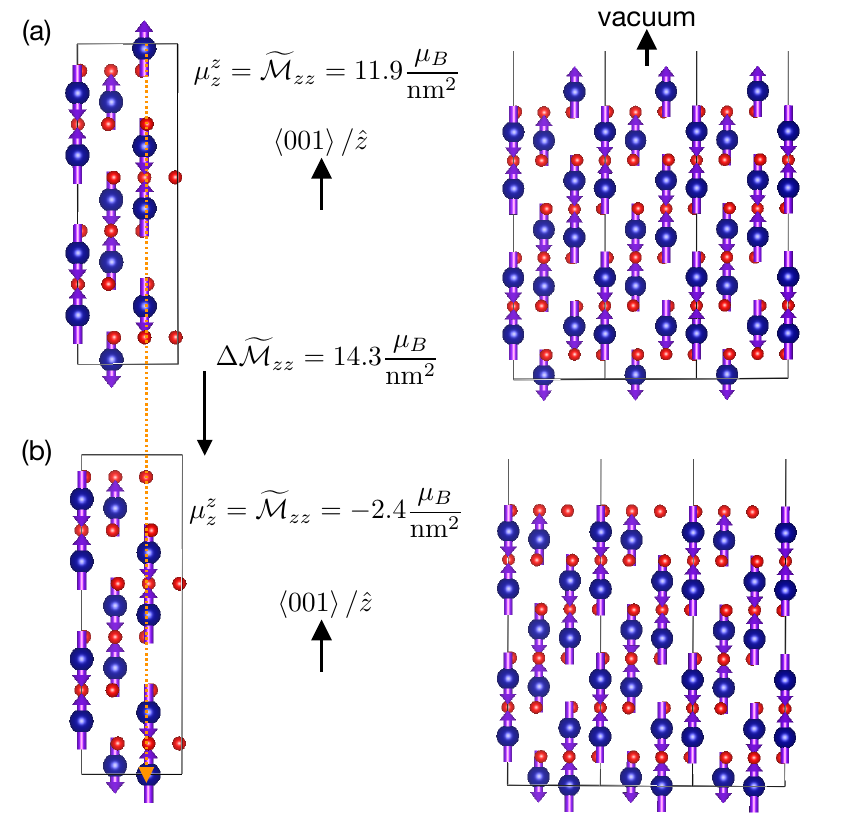}
\caption{\label{fig:surfM_mult} Uncompensated surface magnetization for two different terminations of $(001)$ $\mathrm{Cr_2O_3}$ calculated from the local-moment multipolization. (a) left: bulk unit cell defining the experimental, nonpolar termination of $(001)$ chromia (right). The $\left<001\right>$ oriented surface magnetization for this termination is $11.9$ $\mu_{\text{B}}/\mathrm{nm^2}$ from the $(\mathrm{3,3})$ component of $\widetilde{\mathcal{M}}$. (b) left: bulk unit cell related to the cell in (a) by one multipolization increment $\Delta\widetilde{\mathcal{M}}_{zz}$ (orange arrow). This basis defines the polar $(001)$ surface (right) with surface magnetization $-2.4$ $\mu_{\text{B}}/\mathrm{nm^2}$ when the surface moments are fully polarized along $\left<001\right>$ in their bulk configuration.} 
\end{figure}
\indent We illustrate the surface magnetization-multipolization correspondence concretely in Fig. \ref{fig:surfM_mult}(a) for $(001)$ $\mathrm{Cr_2O_3}$. The top left of the figure shows the 30-atom bulk unit cell in the hexagonal setting, with an ``up down up down" AFM order of the $\mathrm{Cr}$ moments along $\left<001\right>$. This surface defines the electrostatically stable surface termination (top right of Fig. \ref{fig:surfM_mult}(a)) of $(001)$ chromia, with a single $\mathrm{Cr}$ ion above the final oxygen layer. Using the $\mathrm{Cr}$ positions in the bulk unit cell and assuming the idealized $0$ $\mathrm{K}$ limit where the $\mathrm{Cr}$ moments have their formal $3$ $\mu_{\text{B}}$ value and are polarized fully along $\left<001\right>$, Eq. \eqref{eq:M_lm} yields a $z/\left<001\right>$ oriented magnetization of $11.9$ $\mu_{\text{B}}/\mathrm{nm^2}$ on the $(001)$ surface corresponding to $\mathcal{M}_{zz}$\cite{Spaldin2021,Weber2022}. On the other hand, we can instead choose the bulk unit cell on the left in Fig. \ref{fig:surfM_mult}(b), which can be obtained from the  unit cell in \ref{fig:surfM_mult}(a) by translating the topmost $\mathrm{Cr}$ ion downwards along $\left<00\bar{1}\right>$ by one lattice vector. This bulk unit cell defines the \emph{polar}, oxygen-terminated $(001)$ surface on the right in Fig. \ref{fig:surfM_mult}. Recalculating $\widetilde{\mathcal{M}}_{zz}$ using the $\mathrm{Cr}$ ion positions of this new unit cell yields a magnetization of $-2.4$ $\mu_{\text{B}}/\mathrm{nm^2}$. The difference between these two values, $11.9-(-2.4)=14.3$ $\mu_{\text{B}}/\mathrm{nm^2}$, is precisely equal to the multipolization increment for the $(3,3)$ component of $\mathcal{M}$, 
\begin{equation}
\Delta\mathcal{M}_{zz}=\frac{m_{\mathrm{Cr}}^z|\mathbf{c}|}{V},
\label{eq:multinc}
\end{equation}
where $|\mathbf{c}|$ is the value of the hexagonal lattice vector along $\left<001\right>$.\\
\indent We mention here that, because the polar surface on the right of Fig. \ref{fig:surfM_mult}(b) is electrostatically unstable, this $(001)$ atomic termination would not occur naturally during synthesis. Nevertheless, the $0$ $\mathrm{K}$-limit $(001)$ surface magnetization calculated for this termination is actually a better estimate of the value for the experimental, nonpolar termination in \ref{fig:surfM_mult}(a) \emph{at room temperature} than the $11.9$ $\mu_{\text{B}}/\mathrm{nm^2}$ calculated using the positions for the nonpolar unit cell. This is because the topmost layer of $\mathrm{Cr}$ moments on the nonpolar surface is essentially paramagnetic near the bulk N\'{e}el temperature of chromia (around $300$ $\mathrm{K}$\cite{Astrov1960}) as a result of weak exchange coupling to the bulk, and only becomes magnetically ordered at much lower temperatures. Thus, the top $\mathrm{Cr}$ ions on the nonpolar surface contribute negligibly to the surface magnetization at elevated temperatures, and so the $0$ $\mathrm{K}$ surface magnetization calculated for the effective surface in Fig. \ref{fig:surfM_mult}(b) more closely matches the room-temperature value as estimated with nitrogen vacancy magnetometry measurements\cite{Appel2019,Wornle2021}. Hence, values in the multipolization lattice other than those corresponding to nonpolar surfaces are still physically relevant. We refer the reader to Ref. \citenum{Weber2022} for a more complete discussion.\\
\indent Now, we point out a salient feature of the so-called ``multipolization increments" which differentiates between roughness robust and roughness-sensitive magnetization on uncompensated surfaces, discussed further in Secs \ref{subsec:uncomp_robus} and \ref{subsec:uncomp_rough_sens}. Like the lattice of electrical polarization values, the complete set of lm multipolization values corresponding to a certain surface with normal $\mathbf{\hat{n}}$ must be invariant under all symmetry transformations in the MSG of the bulk AFM\cite{Ederer2007,Spaldin2021b}. Therefore, if the bulk AFM has inversion symmetry, $\mathcal{I}$, the allowed values of a given component $\widetilde{\mathcal{M}}_{ij}$ from Eq. \eqref{eq:M_lm} must take one of the following two forms:
\begin{equation}
\widetilde{\mathcal{M}}_{ij}=0+n\Delta\widetilde{\mathcal{M}}_{ij},
\label{eq:centlatt_0contain}
\end{equation}
or
\begin{equation}
\widetilde{\mathcal{M}}_{ij}=\frac{\Delta\widetilde{\mathcal{M}}_{ij}}{2}+n\Delta\widetilde{\mathcal{M}}_{ij},
\label{eq:centlatt_halfquant}
\end{equation}
where n is an integer and the multipolization increment $\Delta\widetilde{\mathcal{M}}_{ij}=m^j|r^i|/V$ is the generalized form of Eq. \eqref{eq:multinc}. We refer to Eqs. \eqref{eq:centlatt_0contain} and \eqref{eq:centlatt_halfquant} as ``zero-containing" and ``half-increment-containing" multipolization lattices respectively. In fact, because the lm multipolization in Eq. \eqref{eq:M_lm} involves a summation over only the magnetic atoms in the unit cell, multipolization increments for a particular material, and the corresponding allowed surface magnetization values, must take the form of Eq. \eqref{eq:centlatt_0contain} or \eqref{eq:centlatt_halfquant} as long as the lattice occupied by the magnetic atoms is centrosymmetric, even if the bulk AFM including all magnetic and nonmagnetic atoms breaks $\mathcal{I}$.\\
\indent While it is not essential to our discussion, we should point out that of course, if a bulk AFM has $\mathcal{I}$ symmetry, a linear ME response via Eq. \eqref{eq:MEtensor} is forbidden. Thus the fact that local-moment multipolization in $\mathcal{I}$-symmetric AFMs can be nonzero from Eqs. \eqref{eq:centlatt_0contain} and \eqref{eq:centlatt_halfquant} might seem contradictory. This can be understood however by considering that the response tensor $\alpha$ is a single-valued object that must respect the bulk MSG symmetries, whereas the lm contribution to the multipolization is a multivalued object. Thus, while the presence of $\mathcal{I}$ in the bulk forces the bulk ME response to be zero, $\widetilde{\mathcal{M}}_{ij}^{\text{lm}}$ is a centrosymmetric lattice that can have nonzero values, and once $\mathcal{I}$ is broken by creating a vacuum-terminated surface, the particular increment of $\widetilde{\mathcal{M}}_{ij}^{\text{lm}}$ chosen by this atomic termination may in turn be nonzero.\\
\indent As an aside, we point out here that if a bulk magnetic lattice has $\mathcal{I}$ symmetry, then \emph{all} crystallographic directions having uncompensated magnetization must have multipolization lattices in the form of either Eq. \eqref{eq:centlatt_0contain} or \eqref{eq:centlatt_halfquant}. In principle a magnetic lattice that breaks $\mathcal{I}$ symmetry can also have such a symmetric multipolization lattice, corresponding to roughness-sensitive surface magnetization, for specific directions. This can occur, for example, if a mirror plane exists perpendicular to the direction of the surface, and the magnetization is oriented parallel to the mirror plane. Put succinctly, uncompensated magnetization in AFMs with a centrosymmetric magnetic lattice must be roughness-sensitive, but roughness-sensitive surface magnetization along a specific direction can occur in AFMs with $\mathcal{I}$-broken magnetic lattices.\\
\indent The centrosymmetry of Eqs. \eqref{eq:centlatt_0contain} and \eqref{eq:centlatt_halfquant} implies that a particular Miller plane $(hkl)\perp{\hat{i}}$ and specific atomic termination corresponding to a surface magnetization $\mu_i^{j*}=\mathcal{M}_{ij}^*$ is symmetrically equivalent to the same plane with an atomic termination having surface magnetization $-\mu_i^{j*}=-\mathcal{M}_{ij}^*$. This multipolization-based argument is in fact just a complementary way of viewing Eqs. \eqref{eq:roughness_nreq}-\eqref{eq:roughness_mreq} of Sec. \ref{subsec:rough}, which characterize roughness-sensitive surface magnetization for uncompensated surfaces within our group-theory formalism. We can conclude then that nominally uncompensated surface magnetization in bulk AFMs that are centrosymmetric \emph{must} be roughness-sensitive. Moreoever, uncompensated surface magnetization in noncentrosymmetric AFMs which have a centrosymmetric magnetic lattice will also be roughness-sensitive, except in particular cases where the nonmagnetic atoms are arranged in such a way that surfaces corresponding to equal magnitudes but opposite signs of surface magnetization   are not energetically degenerate. We emphasize again however that roughness-robust \emph{induced} surface magnetization is still possible in centrosymmetric AFMs such as $\mathrm{FeF_2}$; we will return to this case in Sec. \ref{subsec:induced_robus}). \\
\indent In contrast, AFMs with broken $\mathcal{I}$ in the bulk have noncentrosymmetric multipolization lattices
\begin{equation}
\widetilde{\mathcal{M}}_{ij}=\widetilde{\mathcal{M}}_{ij}^0+n\Delta\widetilde{\mathcal{M}}_{ij},
\label{eq:Ibroken_mult}
\end{equation}
where $\widetilde{\mathcal{M}}_{ij}^0$ is the spontaneous multipolization discussed above, and is neither zero nor half a multipolization increment. For the corresponding surfaces, a termination with surface magnetization $\mu_i^{j*}$ does not have a symmetry-connected, energetically degenerate termination with surface magnetization $-\mu_i^{j*}$. Therefore, unlike inversion-symmetric AFMs, uncompensated surface magnetization in inversion-asymmetric AFMs \emph{can} be roughness-robust.\\
\subsection{\label{subsec:partII_egs}Examples of uncompensated surface magnetization}
\subsubsection{\label{subsec:uncomp_robus}Uncompensated, roughness-robust: $(001)$ $\mathrm{Cr_2O_3}$}
\begin{figure}[t]
\includegraphics[width=0.48\textwidth]{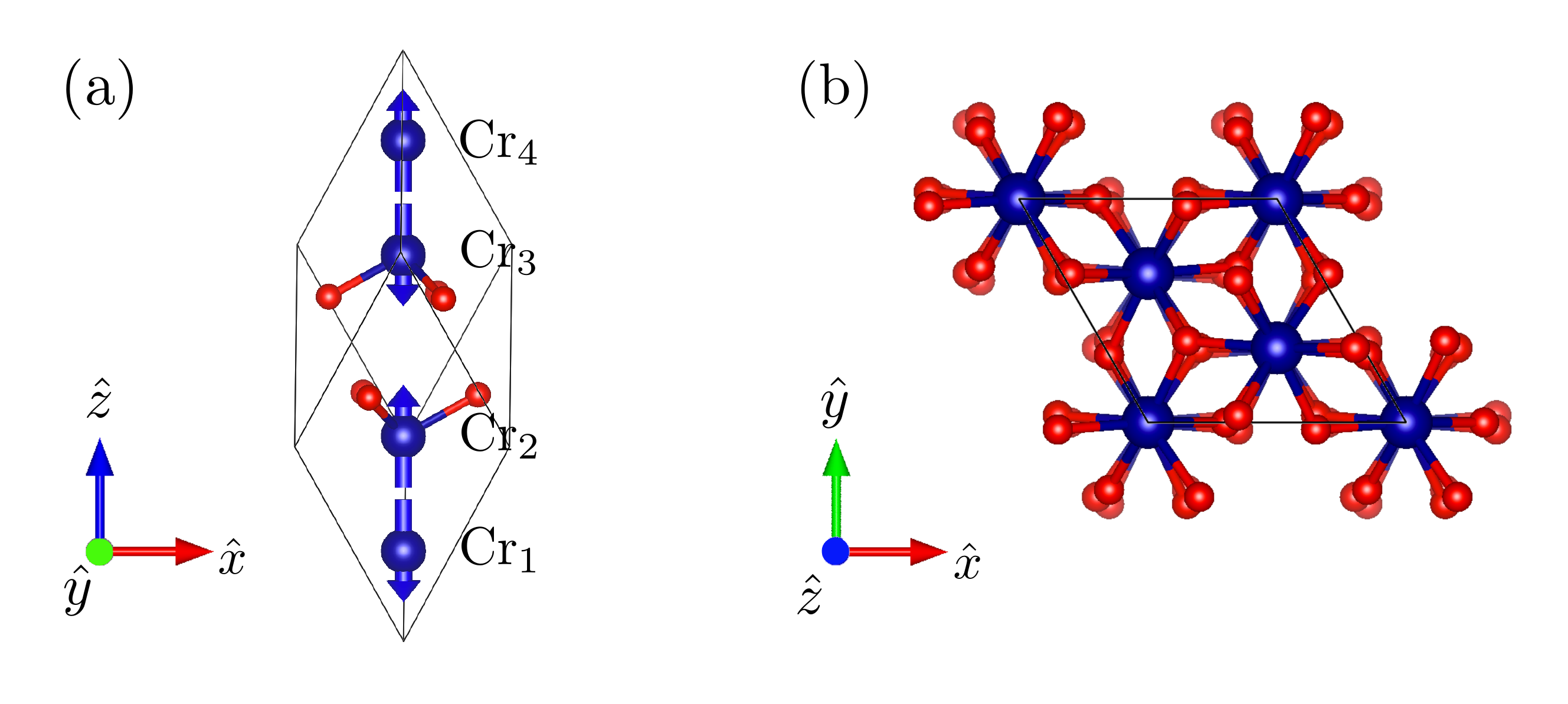}
\caption{(a) Primitive rhombohedral cell and magnetic ordering of Cr$_2$O$_3$. Cr and O atoms are identified by blue and red spheres, respectively. (b) Top view of the conventional hexagonal cell of Cr$_2$O$_3$.}
\label{fig7}
\end{figure}

We begin with the category of surface magnetization which has generated the most attention to date: uncompensated, roughness-robust surface magnetization. Surfaces which are magnetically uncompensated considering the bulk AFM order can have large dipole moments per unit area\cite{Belashchenko2010}, on the order of several $\mu_{\text{B}}/\mathrm{nm^2}$, and insensitivity to surface roughness implies that painstaking synthesis efforts are not required to realize uniform magnetization over a large surface area.\\
\indent Let us first justify the minimal symmetry requirements which we gave in Fig. \ref{fig:categories} for the bulk AFM MSG in order to have at least one Miller plane with this type of surface magnetization. First, as we already discussed in Sec. \ref{sec:multpol}, uncompensated, roughness-robust surface magnetization can only occur in AFMs with broken inversion in the bulk, since the presence of $\mathcal{I}$ implies that any surface with a termination corresponding to one surface magnetization value is energetically degenerate with a symmetry equivalent termination having the same magnitude but opposite sign.\\
\indent Next, we consider the effect of time-reversal $\Theta$. For any direction of surface normal $\mathbf{\hat{n}}$ and any direction of magnetization pseudovector $\mathbf{m}$, $(\mathbf{\hat{n}},\mathbf{m})\rightarrow(\mathbf{\hat{n}},-\mathbf{m})$ under the action of $\Theta$. Therefore, it is not possible to have a surface with roughness-robust surface magnetization if $\Theta$ or $\left(\Theta|\mathbf{t}\right)$ (with $\mathbf{t}$ being any fractional translation) is a bulk symmetry.\\
\indent The combined requirement of broken $\mathcal{I}$ and $\Theta$ implies that this type of surface magnetization can only occur in AFMs with a bulk MSG that allows for the linear ME effect. Note that the AFM is not required to actually exhibit the ME effect; for example, a metallic AFM with broken $\mathcal{I}$ and $\Theta$ still allows for uncompensated, roughness-robust surface magnetization, even though electron screening suppresses magnetoelectricity in the bulk\cite{Belashchenko2010}. Nevertheless, the intimate symmetry relation between this surface magnetization and the linear ME effect indicates that in general uncompensated, roughness-robust surface magnetization in insulators can be readily controlled with ME annealing.\\
\indent We will now work through the prototypical case of $(001)$ $\mathrm{Cr_2O_3}$ using the symmetry analysis method from Section \ref{sec:partI}. In Fig. \ref{fig7} we show the structure of Cr$_2$O$_3$, both the primitive rhombohedral cell and the top view of the conventional hexagonal cell. The bulk MSG of linear ME $\mathrm{Cr_2O_3}$ is $R\bar{3}'c'$ [167.106]\cite{Yuan2021}. Although the crystal lattice of $\mathrm{Cr_2O_3}$ is centrosymmetric, the AFM ordering breaks inversion symmetry in the MSG. Therefore, $R\bar{3}'c'$ has neither space inversion nor time-reversal as a symmetry, so we know immediately that it can host uncompensated, roughness-insensitive surface magnetization for at least some Miller planes. It contains clockwise (cw) and counterclockwise (ccw) three-fold rotations and rotoinversions about the $\left<001\right>$ axis, and two-fold rotations with corresponding mirror planes perpendicular to the rotation axes along $\left<100\right>$, $\left<110\right>$ and $\left<010\right>$. Note that we use angled brackets ``$\left<\right>$" to denote real-space directions, as opposed to square brackets ``$\left[\right]$" which refer to directions in reciprocal space. In the hexagonal basis of $\mathrm{Cr_2O_3}$, $[001]\parallel\left<001\right>$. Because $R\bar{3}'c'$ is a ``type-III" MSG\cite{Yuan2020,Yuan2021}, half of the $36$ space group operations contain time-reversal.  Indeed, although $\mathcal{I}$ and $\Theta$ are individually broken, the product $\mathcal{I}\Theta$ is a symmetry of the bulk MSG.\\
\indent Now consider the unit vector $\mathbf{\hat{n}}\parallel\left<001\right>$ perpendicular to the $(001)$ surface. The operations in $R\bar{3}'c'$ that either leave $\mathbf{\hat{n}}$ invariant modulo translations $\perp\left<001\right>$ are $\left(\mathrm{1}|0\right)$, $\left(\mathrm{3^+}_{\left<001\right>}|0\right)$, and $\left(\mathrm{3^-}_{\left<001\right>}|0\right)$. Here, left-hand terms in the parentheses are point group operations and right-hand terms are translations. ``$+$" and ``$-$" refer to cw and ccw rotations respectively. Now considering only the point group operations, the MPG for the $(001)$ surface is $\mathbf{3}$, which is indeed FM-compatible. Specifically, a generically oriented magnetization pseudovector $\mathbf{m}=\left(m_x,m_y,m_z\right)$ transforms in the following way under the point group operations in $\mathbf{3}$:
\begin{gather}
\mathrm{1}:\left(m_x,m_y,m_z\right)\rightarrow\left(m_x,m_y,m_z\right);\label{eq:eye}\\
\mathrm{3^+}_{\left<001\right>}:\left(m_x,m_y,m_z\right)\rightarrow\nonumber\\
\left(-\frac{1}{2}m_x-\frac{\sqrt{3}}{2}m_y,\frac{\sqrt{3}}{2}m_x-\frac{1}{2}m_y,m_z\right);\label{eq:3plus}\\
\mathrm{3^-}_{\left<001\right>}:\left(m_x,m_y,m_z\right)\rightarrow\nonumber\\
\left(-\frac{1}{2}m_x+\frac{\sqrt{3}}{2}m_y,-\frac{\sqrt{3}}{2}m_x-\frac{1}{2}m_y,m_z\right);\label{eq:3minus}.
\end{gather}
Note that we have chosen our Cartesian coordinate system such that $x\parallel\left<110\right>$ and $z\parallel\left<001\right>$. From Eqs. \eqref{eq:eye}-\eqref{eq:3minus} it is clear that the only invariant magnetization component is $m_z$ which is parallel to the surface normal $\left<001\right>$. Therefore, the magnetic dipole per unit area on the $(001)$ surface must be oriented along $\left<001\right>$. $\left<001\right>$ is known to be the ground state polarization direction of the N\'{e}el vector\cite{Foner1963,Stone1971,Fechner2018}, and from inspection of the nonpolar and polar surfaces in Fig. \ref{fig:surfM_mult}(a)-(b) we see that the $(001)$ surface always has an uncompensated ferromagnetic layer of $\mathrm{Cr}$ moments simply by terminating the ground state magnetic order with vacuum.\\
\indent We can also check that this surface magnetization is roughness-robust by going back to the $R\bar{3}'c'$ bulk MSG and looking for operations that leave the direction of $\mathbf{\hat{n}}$ unchanged but have a component of translation along the $\left<001\right>$ surface normal. This is indeed the case for the three combined mirror plane-time-reversal operations for which $\mathbf{\hat{n}}$ lies in the mirror plane: $\left(\sigma'_{(100)}|0,0,\frac{1}{2}\right)$, $\left(\sigma'_{(110)}|0,0,\frac{1}{2}\right)$ and $\left(\sigma'_{(001)}|0,0,\frac{1}{2}\right)$, where the prime denotes the additional action of $\Theta$. Now we have to see how magnetization behaves under these operations:
\begin{gather}
\sigma'_{(100)}:\left(m_x,m_y,m_z\right)\rightarrow \nonumber\\
\left(\frac{1}{2}m_x+\frac{\sqrt{3}}{2}m_y,\frac{\sqrt{3}}{2}m_x-\frac{1}{2}m_y,m_z\right);\label{eq:m100}\\
\sigma'_{(110)}:\left(m_x,m_y,m_z\right)\rightarrow\left(-m_x,m_y,m_z\right);\label{eq:m110}\\
\sigma'_{(010)}:\left(m_x,m_y,m_z\right)\rightarrow \nonumber\\
\left(\frac{1}{2}m_x-\frac{\sqrt{3}}{2}m_y,-\frac{\sqrt{3}}{2}m_x-\frac{1}{2}m_y,m_z\right)\label{eq:m010}.
\end{gather}
Therefore, although operations \eqref{eq:m100}-\eqref{eq:m010} take one $(001)$ surface through a half-lattice-vector step to a symmetrically equivalent surface, all three operations leave $m_z$, the FM-compatible direction of magnetization in the surface MPG, unchanged. Then, we also know that this magnetization is roughness-robust.\\
\indent Note that this group theory formalism makes no assumptions about the specific atomic termination, and the corresponding \emph{value} of $(001)$ surface magnetization. The identification of this surface as uncompensated and roughness-robust is termination independent, implying that surface magnetization cannot vanish for \emph{any} $(001)$ termination. Quantitative information about the value of surface magnetization for an experimentally realistic surface requires calculation of the multipolization lattice for chromia.\\
\indent It turns out that, due to the alternating up-down-up-down order of the $\mathrm{Cr}$ moments, as the outermost $\mathrm{Cr}$ ions (and accompanying oxygens) are iteratively translated downwards along $\left<00\bar{1}\right>$ to create new surface terminations, the multipolization along this direction simply oscillates between the two values in Fig. \ref{fig:surfM_mult}(a) and (b) ($11.9$ $\mu_{\text{B}}/\mathrm{nm^2}$ and $-2.4$ $\mu_{\text{B}}/\mathrm{nm^2}$ respectively; for the opposite bulk AFM domain, the signs of both values are reversed). Thus, $(001)$ $\mathrm{Cr_2O_3}$ has two symmetrically distinct terminations with unequal but finite magnitudes of surface magnetization, as required by the roughness-robust identification from group theory. In practice, the termination corresponding to a nonpolar surface will be energetically favorable, so experimentally all atomic steps will be symmetrically equivalent to Fig. \ref{fig:surfM_mult}(a) with the corresponding surface magnetization.\\
\indent Before closing this section, we mention one caveat regarding uncompensated, roughness-robust surface magnetization in ME AFMs. In spite of its vast technological potential, and that this category should exist for at least some surfaces of any ME AFM, to our knowledge the only such surface magnetization which has been directly measured to date is our example case, $(001)$ $\mathrm{Cr_2O_3}$. The $(001)$ surface of ME $\mathrm{Fe_2TeO_6}$ for instance was singled out theoretically as an example\cite{Belashchenko2010} but never experimentally confirmed by nitrogen vacancy magnetometry or any other methods. 
The reason for the paucity of experimentally confirmed cases of uncompensated surface magnetization in ME AFMs remains unclear. A pragmatic possibility is that $(001)$ $\mathrm{Cr_2O_3}$ has dominated interest and attention due to its relatively high N\'{e}el temperature compared to other MEs, making it and its doped variants promising candidates in applications\cite{Schlitz2018,Muduli2021,Wang2019}. Nevertheless, given the myriad uses of such surface magnetization, and the fact that the surface $\mathrm{Cr}$ ions on the $(001)$ surface are quite weakly coupled to bulk near the N\'{e}el temperature\cite{Weber2022}, pursuing other ME AFMs as well as different surfaces in chromia could reveal new candidates with more robust uncompensated surface magnetization at room temperature.

\subsubsection{\label{subsec:uncomp_rough_sens}Uncompensated, roughness-sensitive: $(111)$ $\mathrm{NiO}$ and $(001)$ $\mathrm{Fe_2O_3}$}
\begin{figure}
\includegraphics{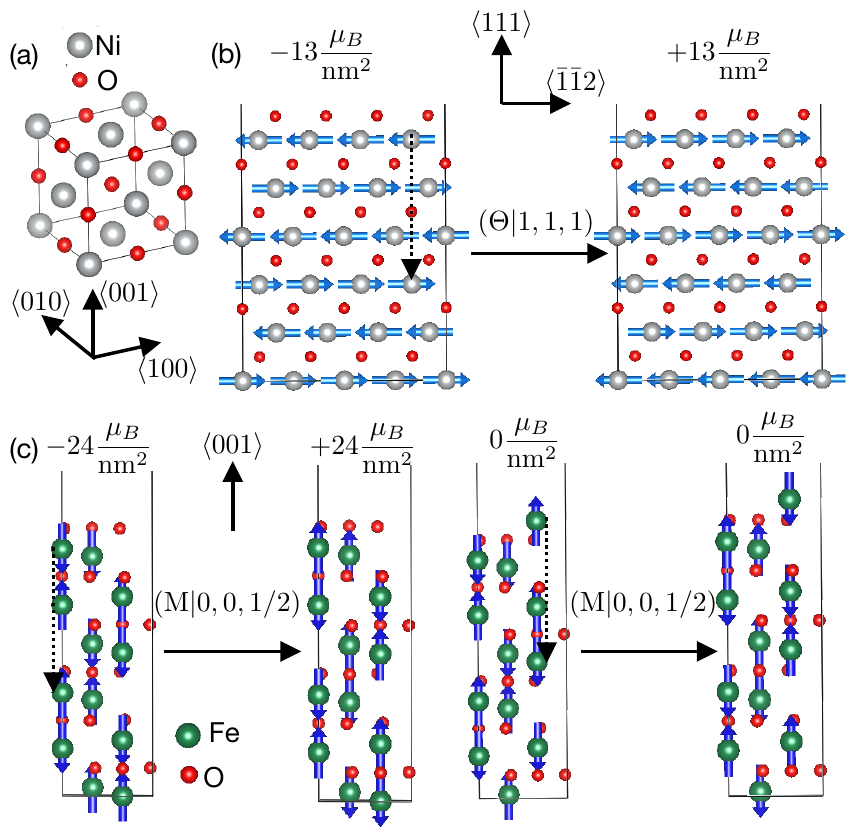}
\caption{\label{fig:NiO_Fe2O3} Uncompensated, roughness-sensitive surface magnetization in $(111)$ $\mathrm{NiO}$ and $(001)$ $\mathrm{Fe_2O_3}$. (a) Nonmagnetic unit cell of rock-salt $\mathrm{NiO}$. (b) Unreconstructed $(111)$ surface of $\mathrm{NiO}$. Left and right surfaces, with equal magnitudes and opposite signs of surface magnetization based on the local-moment multipolization, are connected by time-reversal plus a translation along the $\left<111\right>$ surface normal (dashed black arrow). $(111)$ $\mathrm{NiO}$ has a ``half-increment-containing"  multipolization lattice. (c) Two inequivalent atomic terminations for $(001)$ $\mathrm{Fe_2O_3}$. Left two figures correspond to a polar $\mathrm{O}$ termination, and symmetrically equivalent surfaces with equal magnitudes and opposite signs of surface magnetization are related by a mirror plane bisected by $[001]$ plus a translation along $[001]$. Right: Two symmetrically equivalent surfaces with the nonpolar termination, which has zero surface magnetization. $(001)$ $\mathrm{Fe_2O_3}$ has a ``zero-containing" multipolization lattice.} 
\end{figure}

We turn next to surface magnetization that is magnetically uncompensated, but which averages to zero in the presence of surface roughness (roughness-sensitive). Let us again consider the minimal symmetry requirements of the bulk MSG in order for such surface magnetization to exist on any Miller plane. In contrast to uncompensated, roughness-robust surface magnetization described in Sec. \ref{subsec:uncomp_robus}, uncompensated, roughness-sensitive surface  magnetization corresponds to a centrosymmetric local-moment multipolization lattice. Without such a centrosymmetric multipolization lattice, there cannot be symmetry-equivalent surface steps with opposite directions and equal magnitudes of magnetization.\\
\indent As we mentioned in Sec. \ref{sec:multpol} however, the multipolization lattice can take the centrosymmetric form of Eq. \eqref{eq:centlatt_0contain} or \eqref{eq:centlatt_halfquant} for specific directions even if the magnetic lattice breaks $\mathcal{I}$ symmetry, provided there is another operation in the MSG that constrains the lattice in the direction of interest. Thus, in principle, uncompensated roughness-sensitive surface magnetization can occur in MSGs with magnetic lattices having both broken and preserved $\mathcal{I}$. We point out that for all examples we investigated, the magnetic lattice has $\mathcal{I}$ symmetry. Indeed, the examples we work through explicitly ($\mathrm{NiO}$ and $\mathrm{Fe_2O_3}$) both have inversion symmetry in the bulk MSG. $\mathrm{PbNiO_3}$, a rhombohedral AFM insulator which has attracted interest due to possible ferroelectric properties\cite{Inaguma2011,Hao2012}, is an example of an AFM having a broken inversion symmetry in the bulk MSG but a centrosymmetric lattice of $\mathrm{Ni}$ ions which allows for roughness-sensitive, uncompensated surface magnetization along certain directions. The fact that we have not been able to identify any examples of roughness-sensitive surface  magnetization in AFMs with $\mathcal{I}$-broken magnetic lattices implies that the situation is likely somewhat contrived.\\
\indent We now move on to the requirements for $\Theta$ in order to have uncompensated, roughness-sensitive surface magnetization. As discussed previously, $\Theta$ reverses the direction of any magnetization pseudovector $\mathbf{m}$, while leaving the polar vector corresponding to a surface normal $\mathbf{\hat{n}}$ unchanged (that is, Eqs. \eqref{eq:roughness_nreq} and \eqref{eq:roughness_mreq} are fulfilled). Obviously then, if pure time-reversal or $\Theta$ combined with any translation parallel to the surface is a symmetry of the bulk MSG, it will be preserved in the surface MSG obtained in Fig. \ref{fig:surfM_cartoon}(b). This means the surface MPG will not be FM-compatible, and hence the Miller plane in question cannot host surface magnetization.\\
\indent Recall however that in identifying the surface MSG and corresponding MPG in Fig. \ref{fig:surfM_cartoon}, we exclude symmetries in the bulk MSG which include a component of translation perpendicular to the surface, parallel to $\mathbf{\hat{n}}$. Therefore, if a symmetry $\left(\Theta|\mathbf{t}\right)$ in the bulk MSG has $\mathbf{t}\cdot\mathbf{\hat{n}}\neq0$ (the equivalent statement to Eq. \eqref{eq:roughness_treq}), and there is no additional symmetry $\left(\Theta|\mathbf{t}'\right)$ where $\mathbf{t}'\perp \mathbf{\hat{n}}$, it is possible for the surface of interest $(hkl)\perp\mathbf{\hat{n}}$ to have a FM-compatible MPG in the atomically smooth limit. Thus, AFMs which have $\Theta$ coupled with one or more translations as a symmetry of the bulk MSG can have uncompensated, roughness-sensitive surface magnetization for surfaces which are perpendicular to the translations. This implies that type-IV AFMs, which by definition contain $\left(\Theta|\mathbf{t}\right)$ as a symmetry\cite{Yuan2021}, can still have roughness-sensitive surface magnetization. Note that type-IV AFMs were implicitly excluded in the formalism of Ref. \citenum{Belashchenko2010}.\\
\indent We now give examples of particular surfaces in two bulk AFMs which have uncompensated, roughness-sensitive surface magnetization. We first focus on $\mathrm{NiO}$. $\mathrm{NiO}$ is an insulating AFM with the rock-salt structure. The nonmagnetic, face-centered cubic unit cell is shown in Fig. \ref{fig:NiO_Fe2O3}(a). The ground state AFM order consists of FM planes stacked antiferromagnetically along the $\left<111\right>$ direction, with the $\mathrm{Ni}$ magnetic moments polarized in the planes along the $\left<11\bar{2}\right>$ and $\left<\bar{1}\bar{1}2\right>$ directions\cite{Yuan2021}. $\mathrm{NiO}$ has the bulk MSG $C_c2/c$ [15.90]. $C_c2/c$ is a type IV MSG, meaning that it contains a set of unitary symmetries plus a copy of these symmetries combined with time-reversal plus a translation. When discussing the symmetries of $\mathrm{NiO}$, it is useful to adopt the basis of a magnetic supercell in which the $\mathrm{Ni}$ moments are perpendicular to two of the lattice vectors and parallel to the third. To this end, we use a magnetic cell with lattice vectors $\mathbf{a}=\left<11\bar{2}\right>$, $\mathbf{b}=\left<1\bar{1}0\right>$, and $\mathbf{c}=\left<222\right>$ in terms of the conventional lattice vectors in Fig. \ref{fig:NiO_Fe2O3}(a). The resulting, reoriented supercell is shown in Fig. \ref{fig:NiO_Fe2O3}(b) for two opposite AFM domains.\\
\indent Now let us consider the $(111)$ surface of $\mathrm{NiO}$, which is perpendicular to the direction of AFM stacking. The rotated unit cells shown in Fig. \ref{fig:NiO_Fe2O3}(b) show an atomically smooth termination of the $(111)$ surface, with the two AFM domains having opposite orientations of the $\mathrm{Ni}$ moments in the topmost layer. Note that due to the alternating, equally spaced layers of $\mathrm{Ni^{2+}}$ and $\mathrm{O^{2-}}$ ions along $\left<111\right>$, it is not possible to create an atomically smooth nonpolar $(111)$ surface. Both $\mathrm{Ni}$ and $\mathrm{O}$ terminations are intrinsically polar, and thus the true $(111)$ surface of $\mathrm{NiO}$ almost certainly undergoes a reconstruction\cite{Barbier2000,Zhang2008}. We nevertheless consider the unreconstructed, polar $(111)$ surface here as a straightforward example of a $\Theta$-symmetric AFM surface with uncompensated roughness-sensitive magnetization. Inspection of either structure in Fig. \ref{fig:NiO_Fe2O3}(b) shows that such a surface has an uncompensated FM layer of $\mathrm{Ni}$ moments either along $\left<11\bar{2}\right>$ or $\left<\bar{1}\bar{1}2\right>$.\\
\indent We go through the same procedure as for $(001)$ $\mathrm{Cr_2O_3}$, where we first identify the space group symmetries in the bulk MSG which leave $\mathbf{\hat{n}}\parallel\left<111\right>$ unchanged except for possible translations parallel to the surface. $C_c2/c$ has a monoclinic axis along $\left<1\bar{1}0\right>$, and thus contains two-fold rotations about $\left<1\bar{1}0\right>$ and perpendicular mirror planes $\sigma_{(1\bar{1}0)}$ combined with several linearly independent translations. $\mathcal{I}$ and $\Theta$ combined with translations are also symmetries. The operations which leave $\mathbf{\hat{n}}$ invariant modulo $\mathbf{t}\perp\mathbf{\hat{n}}$ are, in addition to the identity $E$, $\left(\sigma'_{(1\bar{1}0)}|0\right)$ and $\left(\sigma'_{(1\bar{1}0)}|0,\bar{1},1\right)$. Recall that we use a magnetic supercell to describe the $C_c2/c$ symmetries, and so translations such as $\left<0\bar{1}1\right>$ in terms of the primitive lattice coordinates are within a single supercell. Now considering just the point group operation $\sigma'_{(1\bar{1}0)}$, we can look at how magnetization transforms under this symmetry, taking $x\parallel\left<\bar{1}\bar{1}2\right>$, $y\parallel\left<1\bar{1}0\right>$, and $z\parallel\left<111\right>$:
\begin{equation}
\sigma'_{(1\bar{1}0)}:\left(m_x,m_y,m_z\right)\rightarrow\left(m_x,-m_y,m_z\right).
\label{eq:Mp_1m10}   
\end{equation}
Because $x\parallel\left<\bar{1}\bar{1}2\right>$, magnetization on the $(111)$ surface is allowed along the direction of the bulk N\'{e}el vector, consistent with Fig. \ref{fig:NiO_Fe2O3}(b). Note that from Eq. \eqref{eq:Mp_1m10} we see that an additional out-of-plane component of surface magnetization can develop along $z\parallel\left<111\right>$ via canting. Indeed, the surface MPG which corresponds to the identity plus a single $\sigma'$ operation is $m'$, which is a FM-compatible MPG as expected.\\
\indent We now go back to the bulk MSG and check for symmetries which reverse the surface magnetization while translating the polar vector $\mathbf{\hat{n}}$ perpendicular to the $(111)$ surface. This is the case for the symmetry $\left(\Theta|1,1,1\right)$. The action of this symmetry is shown by the dashed black arrow in the structure on the left of Fig. \ref{fig:NiO_Fe2O3}(b), and it connects the left-hand surface with surface magnetization along $\left<11\bar{2}\right>$ to the symmetrically equivalent $(111)$ surface on the right-hand side with surface magnetization along $\left<\bar{1}\bar{1}2\right>$. Evaluating the $(1,3)$ component of the multipolization tensor using Eq. \eqref{eq:M_lm} in a coordinate system with $x\parallel\left<\bar{1}\bar{1}2\right>$ and $z\parallel\left<111\right>$ and assuming a magnetic moment of $2$ $\mu_{\text{B}}$ for $\mathrm{Ni^{2+}}$ yields an uncompensated $(111)$ surface magnetization of $-13$ $\frac{\mu_{\text{B}}}{\mathrm{nm^2}}$ ($+13$ $\frac{\mu_{\text{B}}}{\mathrm{nm^2}}$) along $\left<\bar{1}\bar{1}2\right>$ for the left (right) surface in Fig. \ref{fig:NiO_Fe2O3}(b). These two terminations are in fact related by a multipolization increment along the $\left<111\right>$ direction, and it is not possible to create an atomically smooth $(111)$ termination with a different magnitude of surface magnetization. This is the result of $(111)$ $\mathrm{NiO}$ having a centrosymmetric half-increment-containing multipolization lattice. In the presence of surface roughness, (again, assuming the idealized nonreconstructed surface), the two energetically degenerate surfaces in Fig. \ref{fig:NiO_Fe2O3}(b) will occur with equal probability and the surface magnetization will average to zero.\\
\indent Roughness-sensitive, uncompensated surface magnetization is also possible for AFM planes corresponding to zero-containing multipolization lattices. We illustrate this for the case of $(001)$ $\mathrm{Fe_2O_3}$. Below the N\'{e}el temperature $\sim955$ $\mathrm{K}$, the $\mathrm{Fe}$ moments lie in the $x$-$y$ plane with a weak FM canting, but we focus on the low-temperature phase of $\mathrm{Fe_2O_3}$ below $260$ $\mathrm{K}$ where the moments are oriented along $\left<001\right>$\cite{Spaldin2021}. $\mathrm{Fe_2O_3}$ is isostructural with $\mathrm{Cr_2O_3}$, but its ``up, up, down, down" magnetic order along the $\left<001\right>$ direction, shown in Fig. \ref{fig:NiO_Fe2O3}(c) does not break inversion symmetry, and so the linear ME effect is not allowed (although, as discussed in Ref. \citenum{Verbeek2023}, the symmetry of $\mathrm{Fe_2O_3}$ does allow for an antiferroically ordered local linear ME response). Moreover, the full bulk MSG of low-temperature $\mathrm{Fe_2O_3}$, $R\bar{3}c$ [167.103], has $\mathcal{I}$ symmetry. Therefore, any uncompensated surface magnetization must correspond to a centrosymmetric multipolization lattice. Searching for the operations in $R\bar{3}c$ which leave $\mathbf{\hat{n}}\parallel\left<001\right>$ invariant modulo translations parallel to the surface, we get the same set of cw and ccw three-fold rotations (Eqs. \eqref{eq:3plus} and \eqref{eq:3minus}) about $\left<001\right>$ as for the $R\bar{3}'c'$ MSG of $\mathrm{Cr_2O_3}$, so the surface MPG of $(001)$ $\mathrm{Fe_2O_3}$, $\mathbf{3}$, is FM-compatible and can host surface magnetization along $\left<001\right>$.\\
\indent However, if we now search for symmetries in $R\bar{3}c$ which leave the direction of $\mathbf{\hat{n}}\parallel\left<001\right>$ invariant but translate $\mathbf{\hat{n}}$ perpendicular to the $(001)$ surface, we get mirror planes $\left(\sigma_{(100)}|0,0,\frac{1}{2}\right)$, $\left(\sigma_{(110)}|0,0,\frac{1}{2}\right)$ and $\left(\sigma_{(001)}|0,0,\frac{1}{2}\right)$. Note that these symmetries are analogous to those for $(001)$ $\mathrm{Cr_2O_3}$ in Eqs. \eqref{eq:m100}-\eqref{eq:m010}, except that in this case they do not contain time-reversal. Therefore, for $(001)$ $\mathrm{Fe_2O_3}$ these operations \emph{do} switch the sign of the the surface magnetization along $z\parallel\left<001\right>$ meaning that $(001)$ $\mathrm{Fe_2O_3}$ surface magnetization is roughness-sensitive, in contrast to roughness-robust surface magnetization in $(001)$ ME $\mathrm{Cr_2O_3}$. The mirror plane operations connecting surfaces with equal magnitude and opposite signs of magnetization are shown by the orange arrow for the two pairs of inequivalent terminations in Fig. \ref{fig:NiO_Fe2O3}(c).\\
\indent If we consider the $(001)$ $\mathrm{Fe_2O_3}$ surface magnetization from the perspective of its multipolization lattice, evaluation of the $(3,3)$ term of Eq. \eqref{eq:M_lm} assuming a local moment of $\mathrm{5}$ $\mu_{\text{B}}$ for $\mathrm{Fe^{3+}}$ yields three distinct values each differing by a  multipolization increment: $+24$ $\mu_{\text{B}}/\mathrm{nm^2}$, $0$ $\mu_{\text{B}}/\mathrm{nm^2}$, and $-24$ $\mu_{\text{B}}/\mathrm{nm^2}$. Further $\left<001\right>$ translations of the topmost $\mathrm{Fe}$ moment on the $(001)$ surface will simply yield a repeat of these three values. Therefore, the multipolization lattice for $(001)$ $\mathrm{Fe_2O_3}$ is zero-containing. From the left-hand pair of structures in Fig. \ref{fig:NiO_Fe2O3}(c), it is apparent that the finite $\pm{24}$ $\mu_{\text{B}}/\mathrm{nm^2}$ surface magnetization corresponds to the polar, oxygen-terminated $(001)$ surface. The surfaces with positive and negative magnetization are symmetrically equivalent due to the mirror plane combined with a $\left<001\right>$ translation. On the right-hand side of Fig. \ref{fig:NiO_Fe2O3}(c), we see that the multipolization value of zero in fact corresponds to the nonpolar $(001)$ termination. There are again two such surfaces which are symmetrically equivalent and connected by a mirror plane, but in this case the surface magnetization for both is zero. Therefore, in the case of a zero-containing centrosymmetric multipolization  lattice, every inequivalent termination contains a pair of degenerate surfaces with equal magnitude but opposite sign of surface magnetization. Since degenerate surfaces occur with equal probability, any finite contributions would cancel. Experimentally of course, only the electrostatically stable surface on the right-hand side is likely to occur, corresponding to the zero value of the multipolization lattice in this case.\\
\indent In Part II, we have built on our work in Ref. \citenum{Spaldin2021} which proposed the local-moment multipolization as a bulk descriptor of uncompensated surface magnetization. We have further shown that the symmetry of the multipolization lattice (centrosymmetric or non-centrosymmetric) immediately identifies whether the surface magnetization is roughness-robust or roughness-sensitive and additionally yields a prediction of the quantitative, experimental values of surface magnetization in contrast to the purely qualitative group-theory procedure in Part I. Next, we move on in Part III to characterize the induced forms of surface magnetization using symmetry arguments combined with \emph{ab initio} calculations. 
 \section{\label{sec:partIII}Part III: Induced surface magnetization in terms of atomic-site multipolization} 

As anticipated in the Introduction, and following the group-theory-based guidelines discussed in Sec. \ref{sec:partI}, we remind the reader that the symmetry reduction with respect to the bulk symmetry caused by the surface termination, can allow FM order and give rise to surface magnetization. Such FM components are equivalent by symmetry to induced FM orders arising in the bulk as a consequence of an external electric field parallel to $\hat{\mathbf{n}}$, where they are described by the ME response of the material. In this Part, we will explore the connection between the surface magnetization and the bulk ME response in detail for nominally compensated surfaces. In these cases, the induced surface magnetization is the dominant effect since there are no uncompensated magnetic dipoles in the surface plane. We start by reviewing the bulk ME responses at first, second, and third order in the electric field, and outlining how these can be conveniently described using the multipoles of the magnetization density introduced in Sec. \ref{sec:multpol}. Next, we discuss as concrete examples $(\bar{1}20)$ and $(100)$ Cr$_2$O$_3$, $(110)$ FeF$_2$, and $(120)$ NiO, showing how to link the allowed bulk ME and magnetic multipoles to the induced surface magnetism.   

\subsection{Prelude: local ME responses and atomic-site multipoles} \label{ME_response}

As introduced in Secs. \ref{sec:intro} and \ref{sec:multpol},  by definition ME materials show a net change in magnetization, $\boldsymbol{\delta} \mathbf{M}$, when an external electric field $\mathbf{E}$ is applied or, viceversa, exhibit a net change in polarization, $\boldsymbol{\delta} \mathbf{P}$, in presence of an external magnetic field $\mathbf{H}$. The lowest-order, most well-known ME response is the linear ME effect \cite{Dzyaloshinskii:1960}, whereby the net change in magnetization is linear in the electric field's strength, see Eq. \eqref{eq:MEtensor}. 

The linear ME effect can be conveniently recast in terms of microscopic order parameters, called \textit{ME multipoles} \cite{Spaldin2008}. These are the ME monopole 
\begin{equation}
   a = \frac{1}{3} \int_{\Omega} \mathbf{r} \cdot  \boldsymbol{\mu}(\mathbf{r}) \, d^3 \mathbf{r} ,
\end{equation}
the ME toroidal moment vector 
\begin{equation}
    \mathbf{t} = \frac{1}{2}  \int_{\Omega} \mathbf{r} \times \boldsymbol{\mu}(\mathbf{r}) \, d^3 \mathbf{r},
\end{equation}
and the ME quadrupole tensor 
\begin{equation}
q_{ij} = \frac{1}{2} \int_{\Omega} \left[r_i \mu_j (\mathbf{r}) + r_j \mu_i (\mathbf{r}) - \frac{2}{3} \delta_{ij} \mathbf{r}\! \cdot \boldsymbol{\mu}(\mathbf{r}) \right] d^3\mathbf{r},
\end{equation}
which are respectively the trace, the antisymmetric part, and the symmetric traceless part of the ME multipole tensor $\mathcal{M}_{ij}$ \cite{me_multipoles} introduced in Sec. \ref{sec:multpol}, Eq. \eqref{eq:Memultpol}. ME multipoles have a one-to-one link to the linear ME tensor, with a non-vanishing $ij$ entry of the ME multipole tensor $\mathcal{M}$ implying an $ij$ ME response. 

As discussed in Sec. \ref{sec:multpol}$, \mathcal{M}_{ij}$ can be decomposed into a sum of two terms. One is an origin-dependent, multivalued contribution, referred to as the ``local-moment term'', $\mathcal{M}^{\text{lm}}_{ij}$, and captures the inversion-breaking asymmetries arising from the arrangement of the atomic magnetic dipoles. We used this first term in Part II as a bulk indicator of surface magnetism for magnetically uncompensated surfaces, and showed that it has the virtue of distinguishing between roughness-robust and roughness-sensitive surface magnetizations. The other component of $\mathcal{M}_{ij}$ is the local atomic-site contributions, which describe the inversion-breaking asymmetries in the local magnetization density around the individual ions. In the following we consider this atomic-site component, which we recently showed to be useful in predicting the \textit{local} linear ME effect \cite{Verbeek2023}. Here, by local linear ME effect we denote the change $\boldsymbol{\delta} \mathbf{m}$ in the local \textit{atomic} magnetic moment, induced by an external electric field: 
\begin{equation}
    \label{eq10}
    \delta m_i = \sum_j \alpha^{\text{loc}}_{ji} E_j.
\end{equation}
A non-zero local ME multipole implies a non-vanishing corresponding local linear ME response $\alpha^{\text{loc}}_{ij}$. Whether this results in a net linear ME response $\alpha_{ij}$ depends on the arrangement of the local ME multipoles in the unit cell: specifically, ferroically-ordered ME multipoles entail a net linear ME response, whereas antiferroically ordered ME multipoles imply a linear anti-ME response and, in turn, a vanishing net $\alpha$. Here we will see that the existence of an arrangement of these atomic-site multipoles in the surface plane determines the surface magnetization in many cases when the magnetic dipoles compensate.

We will also find that atomic-site multipoles of higher order than those associated with the linear ME effect can be important in explaining surface magnetism. In particular, we will make the connection to the second-order and third-order ME responses, where the induced magnetization is bilinear and trilinear, respectively, in the applied electric field:
\begin{align}
\label{eq11}
  \delta M_i &= \sum_{jk} \beta_{ijk} E_j E_k, \\
  \delta M_i &= \sum_{jkl} \gamma_{ijkl} E_j E_k E_l.
\end{align}
Similarly to the linear ME response, the second-order response, described by the $\beta_{ijk}$ tensor, has been recently recast in terms of higher-order ME multipoles of the magnetization density, specifically magnetic octupoles \cite{Urru2022}, see Eq \eqref{eq:octupole}. In Appendix \ref{app_mag_oct} we report the irreducible spherical components of $\mathcal{M}_{ijk}$, which will be used later in the discussion. Likewise, the third-order ME response is associated with inversion-breaking magnetic hexadecapoles, although a detailed analysis of the connection, to our knowledge, is still missing in the literature.  

We remark that, in order to have a ferroic order of any ME or higher-order multipole, the breaking of time-reversal symmetry is a \textit{necessary} condition. Given the analogy between surface magnetism and the ME responses discussed earlier, and given the discussion in Part I, this means that a ferroic order of ME or magnetic multipoles provides both a convenient basis for explaining surface magnetism and distinguishes the roughness-robust case, since time-reversal symmetric AFMs can only have roughness-sensitive categories of surface magnetization. We will explain in Sec. \ref{subsec:induced_sense} that roughness-sensitive induced surface magnetization is captured by a ferroic order of atomic-site magnetic multipoles in the surface plane, which switch sign for an equivalent surface connected by a fractional translation along the surface normal, analogously to the roughness-sensitive uncompensated surface magnetization discussed in Sec. \ref{subsec:uncomp_rough_sens}.

\subsection{\label{subsec:induced_robus}Induced, roughness-robust surface magnetization: ($\mathbf{\bar{1}20}$), ($\mathbf{100}$) Cr$_{\mathbf{2}}$O$_{\mathbf{3}}$ and ($\mathbf{110}$) FeF$_{\mathbf{2}}$}

We move now to cases of magnetically compensated planes in the bulk AFM ground state for which roughness-robust magnetization is induced at the surface. Again, let us consider the minimal symmetry requirements for this scenario in terms of $\mathcal{I}$ and $\Theta$ symmetries. As discussed in Secs. \ref{subsec:uncomp_robus} and \ref{subsec:uncomp_rough_sens}, roughness-robust surface magnetization cannot exist for any Miller plane when $\left(\Theta|0\right)$ or $\left(\Theta|\mathbf{t}\right)$, where $\mathbf{t}$ is any translation, is a symmetry of the bulk MSG. Therefore, we know right away that we must have broken $\Theta$ in the bulk.\\
\indent We next consider inversion symmetry $\mathcal{I}$. In Secs. \ref{sec:multpol} and \ref{subsec:uncomp_robus}, we showed that broken $\mathcal{I}$ in the bulk MSG is necessary to obtain the noncentrosymmetric form of the local-moment multipolization lattice in Eq. \eqref{eq:Ibroken_mult} necessary for uncompensated, roughness-robust surface magnetization. However, for surfaces which are  magnetically compensated, the local-moment contribution (Eq. \eqref{eq:M_lm}) for the corresponding multipolization component is \emph{always} zero. Therefore, for compensated surfaces, there is no contribution to the surface magnetization from the lm multipolization lattice. The lack of a local-moment multipolization lattice contribution for compensated surfaces with induced magnetization implies that $\mathcal{I}$ may be either present or absent in the bulk MSG. We show this explicitly by leveraging the group-theory analysis, combined with first-principles DFT calculations, for three example AFM surfaces, two with broken inversion in the bulk and one with a centrosymmetric bulk MSG. In addition, we show that in the roughness-robust case, induced surface magnetization is associated with atomic-site multipoles of the magnetization. Specifically, these are the inversion-broken atomic-site ME multipoles and hexadecapoles for $(\bar{1}20)$ and $(100)$ $\mathrm{Cr_2O_3}$,  and inversion-symmetric atomic-site magnetic octupoles for the $(110)$ surface of centrosymmetric $\mathrm{FeF_2}$.\\

\subsubsection{Induced $(\bar{1}20)$ surface magnetization in non-centrosymmetric $\mathrm{Cr_2O_3}$} \label{-120_chromia} 
\begin{figure}[t]
\includegraphics[width=0.48\textwidth]{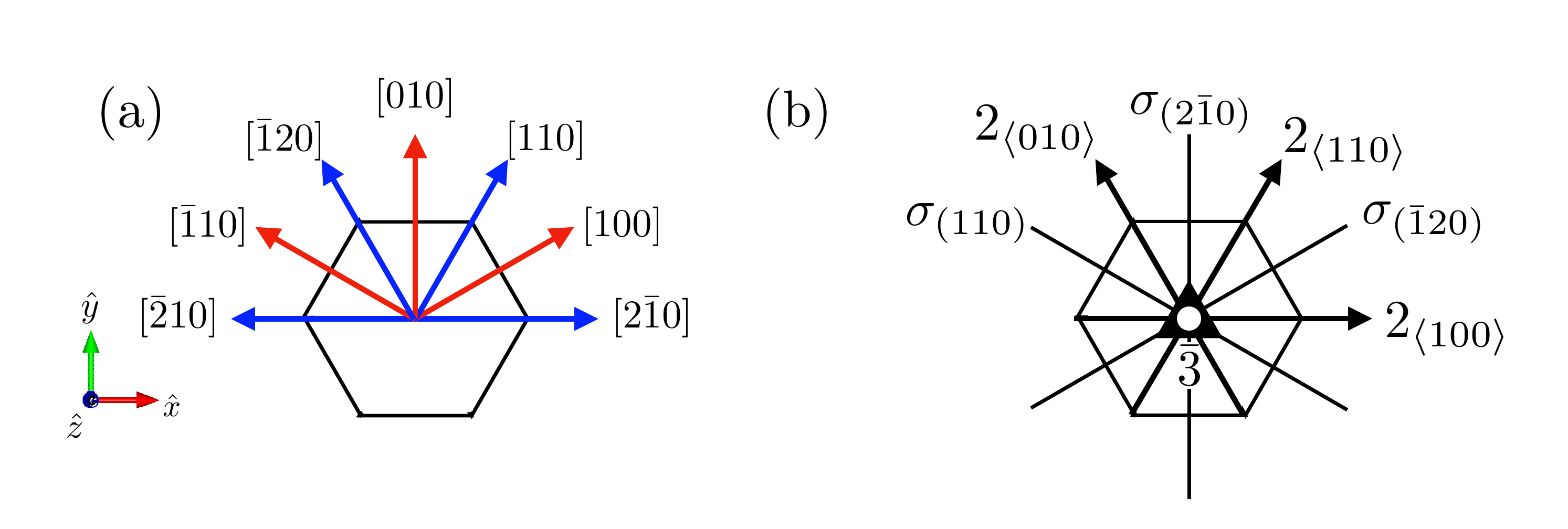}
\caption{(a) High-symmetry directions in the basal plane of a hexagonal cell. (b) Symmetry operations of the $\bar{3}m$ point group.}
\label{fig3}
\end{figure}
\begin{figure}
\includegraphics{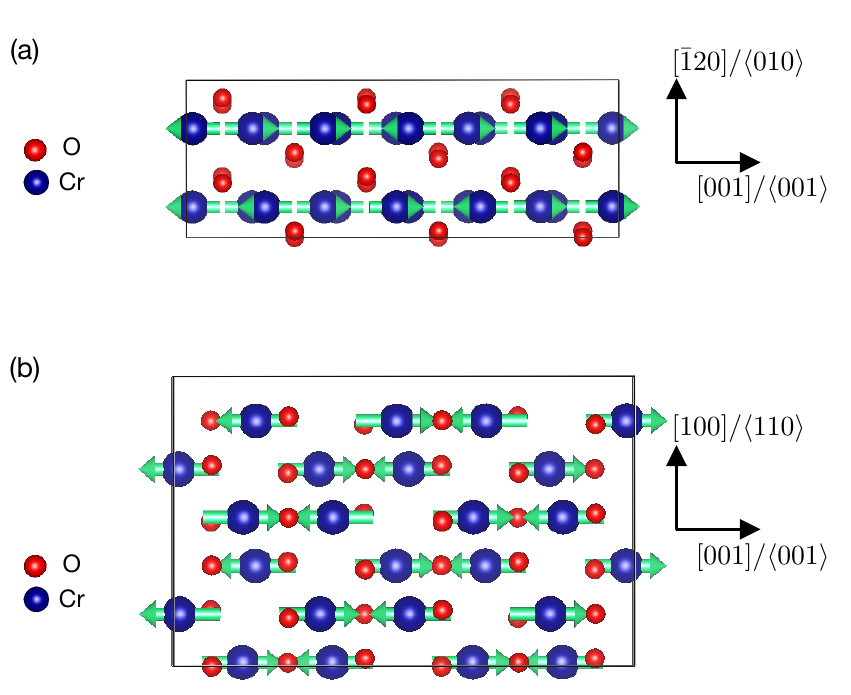}
\caption{\label{fig:Cr2O3_sidewalls} Bulk unit cells which can be tiled semi-infinitely to define the nonpolar atomic terminations of the (a) $(\bar{1}20)$ and (b) $(100)$ surfaces of $\mathrm{Cr_2O_3}$. The magnetic dipoles moments of the $\mathrm{Cr}$ ions, shown in turquoise, are ordered according to the bulk AFM ground state with the N\'{e}el vector parallel to $\langle001\rangle$.}
\end{figure}

We first look again at ME AFM $\mathrm{Cr_2O_3}$, with broken $\mathcal{I}$ and $\Theta$ in the bulk MSG $\mathrm{R\bar{3}'c'}$. Bulk Cr$_2$O$_3$ can be described by a primitive rhombohedral cell, (see Fig. \ref{fig7}(a), where we show also the magnetic dipolar order of Cr atoms), or likewise by a conventional hexagonal cell, whose top view is shown in Fig. \ref{fig7}(b). In contrast to the uncompensated (001) surface we analyzed in Section \ref{subsec:uncomp_robus}, this time we examine a surface perpendicular to the $(001)$ plane. In Fig. \ref{fig3}(a) we indicate the high-symmetry directions in the hexagonal basal plane. Note that for surfaces perpendicular to $(001)$ in hexagonal $\mathrm{Cr_2O_3}$, we label crystallographic directions using reciprocal space vectors $\left[ hkl \right]$ to make the connection between the Miller planes and their corresponding surface normals more clear.\\
\indent We focus first on the $(\bar{1}20)$ surface. The bulk, orthorhombic unit cell which defines the nonpolar $(\bar{1}20)$ surface when tiled semi-infinitely along $[\bar{1}20]$ is shown in Fig. \ref{fig:Cr2O3_sidewalls}(a). The magnetic lattice consists of two layers with $\mathrm{12}$ $\mathrm{Cr}$ each stacked along the $[\bar{1}20]$ direction. Note that within each layer, half of the $\mathrm{Cr}$ magnetic moments are oriented along $[001]$ in the ground state, and half are pointed along $[00\bar{1}]$. Therefore, any flat surface created by terminating a semi-infinite slab with vacuum along the $[\bar{1}20]$ direction with the bulk AFM order will be magnetically compensated.\\
\indent The surface normal, $\mathbf{\hat{n}}$, is parallel to $[\bar{1}20]$ and lies along the $\left<010\right>$ real-space lattice vector, which is a two-fold rotational axis in the bulk MSG. If we go through all space group operations in  $\mathrm{R\bar{3}'c'}$ (see Fig. \ref{fig3}(b)) and select those which leave $\mathbf{\hat{n}}\parallel \langle 010 \rangle$ invariant modulo translations $\perp\mathbf{\hat{n}}$, we find (besides the identity) only the two-fold rotation about the surface normal, $\left(2_{\langle 010 \rangle}|0,0,\frac{1}{2}\right)$. Adopting a rotated basis where $x\parallel\left<010\right>$ (or equivalently, $x\parallel[\bar{1}20]$), and $z\parallel[001]$, the magnetization transforms in the following way:
\begin{equation}
2_{\langle 010 \rangle}: (m_x,m_y,m_z)\rightarrow(m_x,-m_y,-m_z),
\label{eq:2_010}
\end{equation}
implying that a finite surface magnetization is allowed along $x$. Moreover, there are no symmetries in $\mathrm{R\bar{3}'c'}$ which flip the sign of $m_x$ and translate $\mathbf{\hat{n}}$ perpendicular to the surface, so the surface magnetization will be roughness-robust. Note that the only direction which is FM-compatible for the $(\bar{1}20)$ surface, $x\parallel\langle 010 \rangle$, is not parallel to the bulk $[001]$ N\'{e}el vector polarization, but rather along the surface normal. Therefore surface magnetization is allowed by symmetry to develop only via canting along $[\bar{1}20]$.

Next, we rederive from a different standpoint the results obtained in Part I using the group-theory-based prescription to identify surface magnetism. In particular, motivated by the equivalence between surface magnetism and the bulk ME response discussed at the beginning of this section, we address the magnetization induced in the bulk by an electric field parallel to the surface normal. Before doing so, we remind the reader that bulk Cr$_2$O$_3$ allows for a net linear ME response identified by the ME tensor 
\begin{equation}
\alpha = \begin{pmatrix} \alpha_{xx} & 0 & 0 \\ 0 & \alpha_{xx} & 0 \\ 0 & 0 & \alpha_{zz} \end{pmatrix},
\label{eq1}
\end{equation}
which can be interpreted in terms of ME multipoles as net bulk ME monopoles and $z^2$ quadrupoles. Since we are interested in the local atomic response to an electric field, here we consider the local linear ME response for the Cr atoms, which reads
\begin{equation}
    \alpha^{\text{loc}} = \begin{pmatrix} \alpha^{\text{loc}}_{xx} & \overline{\alpha}^{\text{loc}}_{xy} & 0 \\ - \overline{\alpha}^{\text{loc}}_{xy} & \alpha^{\text{loc}}_{xx} & 0 \\ 0 & 0 & \alpha^{\text{loc}}_{zz} \end{pmatrix}.
    \label{eq2}
\end{equation}
Compared to the bulk response (Eq. \eqref{eq1}), the local ME tensor shows additional off-diagonal $xy$ entries, since the site symmetry of the Cr Wyckoff position ($\bar{3}'$), is lower than the point group symmetry ($\bar{3}'m'$). From the atomic-site ME multipoles standpoint, this additional $xy$ component is due to a local toroidal moment along $z$ \cite{Thole_thesis}, arranged antiferroically on the Cr atoms. We use overlined symbols in Eq. \eqref{eq2} to indicate antiferroically ordered local ME multipoles (and the components of the ME response associated with them). Now we consider the local response to an electric field $\mathbf{E} \parallel [\bar{1}20]$. The [$\bar{1}20$] direction makes a 120$^{\circ}$ angle with the $x$ axis in the hexagonal setting (Fig. \ref{fig3}(a)), thus we consider $\mathbf{E} = (-1/2, \sqrt{3}/2, 0) E$. According to Eq. \eqref{eq10} and the shape of the local ME tensor, Eq. \eqref{eq2}, the magnetic moment induced locally at the Cr sites is:
\begin{align}
    \delta m_x &= -\left( \frac{1}{2} \alpha^{\text{loc}}_{xx} + \frac{\sqrt{3}}{2} \overline{\alpha}^{\text{loc}}_{xy} \right) E, \nonumber \\
    \delta m_y &= \left( -\frac{1}{2} \overline{\alpha}^{\text{loc}}_{xy} + \frac{\sqrt{3}}{2} \alpha^{\text{loc}}_{xx} \right) E, \label{eq3} \\
    \delta m_z &= 0. \nonumber
\end{align}
When computing the net induced magnetization we need to take into account that the contributions from $\overline{\alpha}^{\text{loc}}_{xy}$ vanish due to their antiferroic arrangement, thus the net $\boldsymbol{\delta} \mathbf{m}$ per Cr atom is 
\begin{align}
    \delta m_x &= - \frac{1}{2} \alpha^{\text{loc}}_{xx} E, \nonumber \\ 
    \delta m_y &= \frac{\sqrt{3}}{2} \alpha^{\text{loc}}_{xx} E,  \label{eq4} \\
    \delta m_z &= 0, \nonumber
\end{align}
which yields $\boldsymbol{\delta} \mathbf{m} = \alpha^{\text{loc}}_{xx} \mathbf{E} \parallel [\bar{1}20]$, i.e. $\boldsymbol{\delta} \mathbf{m}$ is perpendicular to the surface, see Fig. \ref{fig6}(a). This result is formally equivalent to the surface magnetization predicted by group theory arguments earlier in this Section. 

We remark that the results reported in Eqs. \eqref{eq3}-\eqref{eq4} can be conveniently derived by rotating the reference framework to align the [$\bar{1}20$] axis along the $x$ axis. In this way, the ME response along [$\bar{1}20$] to an electric field parallel to [$\bar{1}20$] corresponds to the $xx$ entry of the rotated tensor. The rotation transformation for $\alpha^{\text{loc}}$ reads 
\begin{equation}
    \alpha'^{\text{loc}}_{i'j'} = \sum_{ij} \mathcal{R}_{i'i} \mathcal{R}_{j'j} \alpha^{\text{loc}}_{ij},
    \label{eq5}
\end{equation}
where the prime ($'$) superscript refers to the rotated framework and 
\begin{equation}
    \mathcal{R} = \begin{pmatrix} \dfrac{1}{2} & \dfrac{\sqrt{3}}{2} & 0 \\ - \dfrac{\sqrt{3}}{2} & \dfrac{1}{2} & 0 \\ 0 & 0 & 1 \end{pmatrix}
\end{equation}
is the matrix representing a clockwise rotation of angle $\theta = 120^{\circ}$. Eq. \eqref{eq5} implies $\alpha'^{\text{loc}}_{xx} = \alpha^{\text{loc}}_{xx}$, thus the magnetic moment induced along $x' \parallel [\bar{1}20$] is $\delta m_{x'} = \alpha^{\text{loc}}_{xx} E$, identical to that reported in Eqs. \eqref{eq3}-\eqref{eq4}.

\subsubsection{Induced $(100)$ surface magnetization in non-centrosymmetric $\mathrm{Cr_2O_3}$} \label{100_chromia} 

The ($100$) surface of chromia, similarly to the ($\bar{1}20$) surface, is obtained by cutting bulk Cr$_2$O$_3$ along a plane perpendicular to the basal hexagonal plane. Its normal $\hat{\mathbf{n}}$ is related to the normal of ($\bar{1}20$) by a clockwise $90^{\circ}$ rotation about the $z$ axis, see Fig. \ref{fig3}(a). Following the group-theory-based guidelines to identify whether this surface allows for roughness-robust surface magnetism, we first identify which symmetry operations among those of the magnetic space group $\mathrm{R\bar{3}'c'}$ of chromia (Fig. \ref{fig3}(b)) leave $\hat{\mathbf{n}}$ invariant up to a fractional translation perpendicular to $\hat{\mathbf{n}}$ itself. By inspection one finds that the operations fulfilling this requirement are the identity $E$ and the mirror plane perpendicular to $[\bar{1}20]$ combined with time-reversal, $\sigma'_{(\bar{1}20)}$. After defining a rotated reference framework ($x$, $y$, $z$), with $x \parallel \hat{\mathbf{n}}$, $y \parallel [\bar{1}20]$, and $z \parallel z$, the action of $\sigma'_{(\bar{1}20)}$ on a generic magnetic moment $\mathbf{m} = (m_{x}, m_{y}, m_{z})$ is: 
\begin{equation}
    \sigma'_{(\bar{1}20)}: (m_{x}, m_{y}, m_{z}) \rightarrow (m_{x}, -m_{y}, m_{z}), 
\end{equation}
which leaves $m_{x}$ and $m_{z}$ invariant. Thus, ($100$) Cr$_2$O$_3$ is  compatible with ferromagnetism both along the surface normal and along the [$001$] direction, within the surface plane. This is distinct from the ($\bar{1}20$) surface analyzed earlier, which was compatible with ferromagnetism only along the surface normal. 

Now, we interpret the surface magnetism in terms of the bulk linear ME response to an electric field along [$100$]. Following the argument discussed above for the ($\bar{1}20$) surface, we take the local linear ME tensor $\alpha^{\text{loc}}$ and we apply a clockwise rotation of angle $\theta = 30^{\circ}$ in order to align the [$100$] direction to $x$. We find $\alpha'^{\text{loc}} = \alpha^{\text{loc}}$, which implies: 
\begin{align}
        \delta m_x &= \alpha^{\text{loc}}_{xx} E, \nonumber \\
        \delta m_y &= -\overline{\alpha}^{\text{loc}}_{xy} E, \\
        \delta m_z & = 0. \nonumber
\end{align} 
After summing over the Cr atoms on the topmost surface atomic layer, $\delta m_y$ vanishes since $\overline{\alpha}^{\text{loc}}_{xy}$ is antiferroically ordered, leaving only a non-vanishing net $\delta m_x$. Thus, using the bulk linear ME response we are able to explain only the component of the surface magnetization perpendicular to the surface, and we are not able to predict the component parallel to the surface, along [$001$]. This feature comes from the local ME response being isotropic in plane, which yields a contribution of surface magnetization perpendicular to the surface for any $(hk0)$ surface, perpendicular to the basal plane. In order to explain components of the magnetization parallel to the surface we need to go beyond the linear ME response and account for higher-order, anisotropic responses, which we discuss next. 

\subsubsection{Higher-order ME-induced surface magnetism: $\mathrm{Cr_2O_3}$ $(100)$ and $({\bar{1}20})$ surfaces} \label{sec:3rd_order_ME}

As discussed in previous Sections, based on symmetry arguments we predict that the ($\bar{1}20$) and ($100$) surfaces of Cr$_2$O$_3$ are not equivalent concerning the allowed components of net surface magnetization: specifically, the ($\bar{1}20$) surface is compatible with ferromagnetism only along [$\bar{1}20$], perpendicular to the surface, whereas the ($100$) surface allows both for ferromagnetism along the [$100$] direction, perpendicular to the surface, and along the [$001$] direction, parallel to the N\'eel vector. As examined earlier in Secs. \ref{-120_chromia} and \ref{100_chromia}, the isotropic linear ME response explains the component of surface magnetization perpendicular to the surface, but does not explain the component parallel to $z \parallel [001]$, since a $zx$ linear ME response is forbidden by symmetry in the bulk. In order to understand, in terms of ME responses, why such a component of the surface magnetization parallel to [$001$] exists and why it is allowed only in the family of ($100$) and equivalent surfaces, we need to analyze higher-order ME responses, beyond the linear one. For clarity, in the following we adopt the right-handed coordinate system with $x \parallel \langle 1 0 0 \rangle$, $y \parallel \langle 1 2 0 \rangle$, and $z \parallel \langle 0 0 1 \rangle$, shown in Fig. \ref{fig3}(a). As such, $x$ lies on the two-fold rotation axis $2_{\langle 100 \rangle}$ of the $\bar{3}m$ point group, whereas $y$ corresponds to the trace of the mirror plane $\sigma_{(2\bar{1}0)}$, and $z$ is the $\bar{3}$ roto-inversion axis (see Fig. \ref{fig3}(b)). 

We start by analyzing the second-order ME response which, as mentioned earlier in Section \ref{ME_response}, is conveniently recast in terms of magnetic octupoles. In Cr$_2$O$_3$, the octupoles allowed by symmetry are $\mathcal{O}_3$, $\mathcal{Q}^{(\tau)}_{z^2}$, which follow a ($+ + - -$) order (here the signs refer to the octupole component on Cr$_1$, Cr$_2$, Cr$_3$, and Cr$_4$ atoms as labelled in Fig. \ref{fig7}(a)), and $\mathcal{O}_{-3}$, $\mathcal{O}_0$, and $t^{(\tau)}_z$, which follow a ($+ - + -$) order. Note that here we follow the same naming convention for magnetic octupoles as in Ref. \onlinecite{Urru2022}. Both families of octupoles order antiferroically, hence despite contributing to the local second-order ME response, they do not provide any net response. The implication for surface magnetism, is that an induced local atomic magnetic moment, bilinear in the effective electric field generated by the surface termination, is expected for the Cr atoms at the surface. However, for $(100)$ and $(\bar{1}20)$ surfaces these induced atomic magnetic moments will arrange antiferroically, since all four Cr atoms appear at the surface plane, and thus will not yield any net surface magnetization. 

Next, we consider the third-order ME response. To simplify the discussion, we do not describe in detail the induced local atomic response, but limit ourselves to the net third-order ME response, for which the induced net magnetization $\boldsymbol{\delta} \mathbf{M}$ is trilinear in the external electric field $\mathbf{E}$: 
\begin{equation}
    \delta M_i = \sum_{j k l} \gamma_{ijkl} E_j E_k E_l,
\end{equation}
with $\gamma_{ijkl}$ the rank-4 third-order ME response tensor. Since $\gamma_{ijkl}$ breaks inversion symmetry, such a response is symmetry-allowed in non-centrosymmetric Cr$_2$O$_3$. As we want to explain the surface magnetization along the $z$ ($\parallel [001]$) direction in the ($100$) and symmetry-equivalent surfaces, we focus on the net magnetization induced along $z$ by a general electric field in the basal plane, thus we consider $\gamma_{zjkl}$, with $j = \{x, y\}$, $k = \{x, y\}$, $l = \{x, y\}$. We use the $\texttt{MTENSOR}$ utility of the Bilbao Crystallographic Server \cite{Bilbao_1,Bilbao_2,Bilbao_3}, to write $\gamma_{zjkl}$ in terms of its independent parameters as 
\begin{align}
\label{eq13}
    \gamma_{zxkl} &= \begin{pmatrix} 0 & c_{zxxy} & c_{zxxz} \\ c_{zxxy} & 0 & 0 \\ c_{zxxz} & 0 & 0  \end{pmatrix} \nonumber, \\
    \gamma_{zykl} &= \begin{pmatrix} c_{zxxy} & 0 & 0 \\ 0 & -c_{zxxy} & c_{zxxz} \\ 0 & c_{zxxz} & 0  \end{pmatrix}.
\end{align}
First we analyze the $(\bar{1}20)$ surface. We note that $x$ ($\parallel [2\bar{1}0]$) is symmetry-equivalent to the $[\bar{1}20]$ direction (see Fig. \ref{fig3}(a)), since the two directions are connected by a $120^{\circ}$ rotation, thus the net magnetization along $z$ induced by an electric field along $x$ is equivalent to the surface magnetization along $z$ for the $(\bar{1}20)$ surface. Since $\gamma_{zxxx} = 0$, our tensor analysis predicts no surface $M_z$ for this surface, in agreement with what we obtained earlier in Sec. \ref{-120_chromia} from group theory arguments. 

\begin{figure}[t!]
\includegraphics[width=0.38\textwidth]{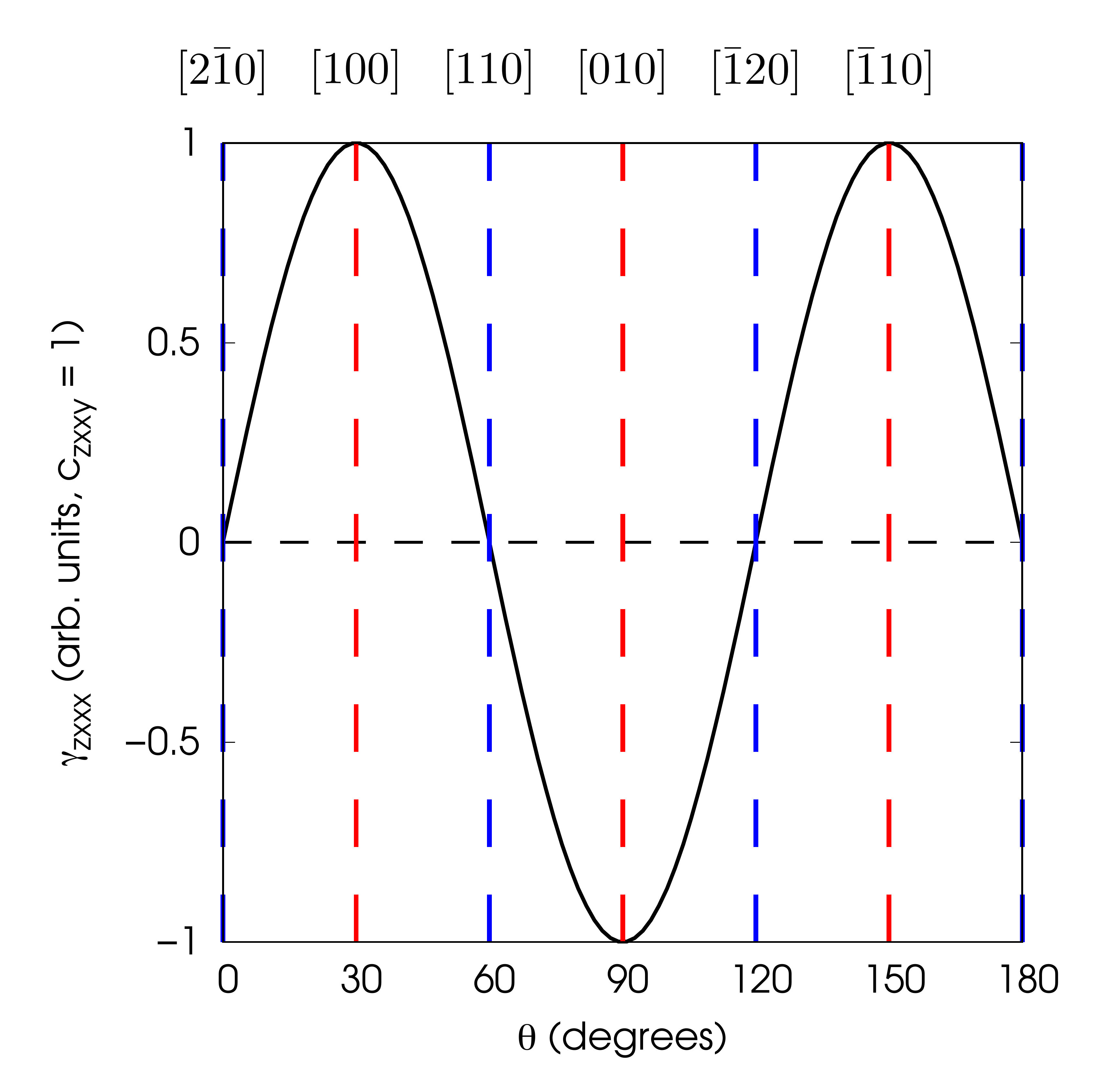}
\caption{Third-order ME response $\gamma_{zxxx}$ to an electric field pointing along any direction in the basal plane of the conventional hexagonal cell. $\theta$ identifies the angle between the direction of the electric field and the $x$ axis.}
\label{fig4}
\end{figure}
Next, we consider any ($hk0$) surface, with normal vector $\hat{\mathbf{n}}$ lying in the hexagonal basal plane (see Fig. \ref{fig3}(a), where we identify the main [$hk0$] high-symmetry directions). For convenience, we apply a rotation to the tensor $\gamma$, such that it aligns the electric field $\mathbf{E} \parallel [hk0]$, perpendicular to the surface, with the $x$ axis, similarly to our earlier procedure for the linear ME contribution. To keep the discussion general, we apply a clockwise rotation $\mathcal{R}$ about the [$001$] axis, for a generic angle $\theta$, 
\begin{equation}
    \mathcal{R} = \begin{pmatrix} \cos \theta & \sin \theta & 0 \\ - \sin \theta & \cos \theta & 0 \\ 0 & 0 & 1 \end{pmatrix}.
\end{equation}
The tensor in the rotated framework is computed as follows:
\begin{equation}
\label{eq14}
    \gamma'_{i'j'k'l'} = \sum_{i j k l} \mathcal{R}_{i'i} \mathcal{R}_{j'j} \mathcal{R}_{k'k} \mathcal{R}_{l'l} \gamma_{ijkl}.  
\end{equation}
As we are interested in the $zxxx$ response in the rotated framework, we set $i'=z$, $j'=k'=l'=x$. From Eqs. \eqref{eq13}-\eqref{eq14} we obtain: 
\begin{equation}
    \begin{split}
            \gamma'_{zxxx} &= \left( 3 \sin \theta \cos^2 \theta - \sin^3 \theta \right) c_{zxxy} \\
            &= - \sin \theta \left( 4 \sin^2 \theta - 3 \right) c_{zxxy}.
    \end{split}
\end{equation}
We note that, in contrast to the linear ME response, the third-order response is anisotropic in plane. We show $\gamma'_{zxxx}$ in Fig. \ref{fig4}, where we arbitrarily set $c_{zxxy}$ to 1. We note that $\gamma'_{zxxx}$ as a function of $\theta$ is periodic with period $120^{\circ}$, arising from the bulk three-fold rotational symmetry. In addition, importantly $\gamma'_{zxxx} \ne 0$ for the ($100$), ($010$), and ($\bar{1}10$) surfaces, which correspond to $\theta = 30^{\circ}, 90^{\circ}, 150^{\circ}$, respectively, whereas $\gamma'_{zxxx} = 0$ for the ($2\bar{1}0$), ($110$), and ($\bar{1}20$) surfaces, corresponding to $\theta = 0^{\circ}, 60^{\circ}, 120^{\circ}$, respectively. Thus, based on the analogy between bulk ME responses and surface magnetism, the anisotropic third-order ME response explains the behavior of the surface magnetization along $z$ that we predicted earlier using group theory for the ($100$) and ($\bar{1}20$) surfaces.  

\subsubsection{\textit{Ab initio} calculations on $(\bar{120})$ and $(100)$ $\mathrm{Cr_2O_3}$}

\begin{figure}[t]
\includegraphics{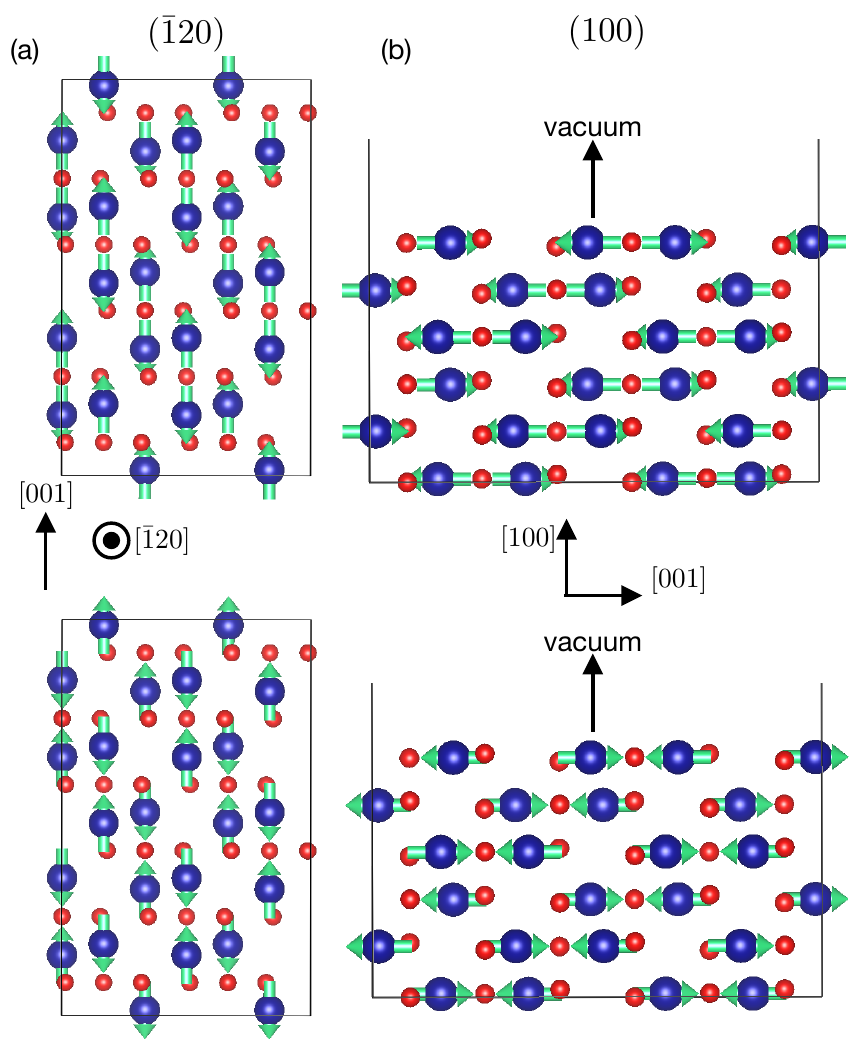}
\caption{(a) Top view of the $(\bar{1}20)$ Cr$_2$O$_3$ surface for the out-pointing (top panel) and in-pointing (bottom panel) domain. (b) Side view of the $(100)$ Cr$_2$O$_3$ surface for the out-pointing (top panel) and in-pointing (bottom panel) domain.}
\label{fig5}
\end{figure}

So far, we have used symmetry arguments to show that surface magnetization can be induced even when the magnetic dipoles in the plane are magnetically compensated. Now, we corroborate our predictions using \textit{ab initio} calculations. This subsection has two purposes. First, to demonstrate with quantitative DFT calculations that a finite magnetization, obtained via canting of the magnetic moments, on the $(\bar{1}20)$ and $(100)$ surfaces of $\mathrm{Cr_2O_3}$ lowers the total energy compared to the surface with zero magnetization. Second, to strengthen the link between surface magnetism and bulk ME effects by testing with first-principles calculations whether the sign change of the ME response upon bulk ME domain reversal, a core ingredient of ME domain selection using ME annealing \cite{Belashchenko2010,Wang2019}, has its counterpart in surface magnetism. If the magnetization induced on a surface can be indeed interpreted as a consequence of a ME effect from an effective electric field due to the surface termination, we expect the surface magnetization to reverse in the opposite bulk magnetic domain. To test this assumption, we consider both ($\bar{1}20$) and (100) surfaces obtained from both bulk magnetic domains, as shown in Fig. \ref{fig5}. We refer to these domains as the \textit{in-pointing} and the \textit{out-pointing} domains, depending on whether the magnetic moments of each pair of nearest-neighbor Cr atoms point towards each other (``in") or away from each other (``out").
\begin{figure}
\includegraphics{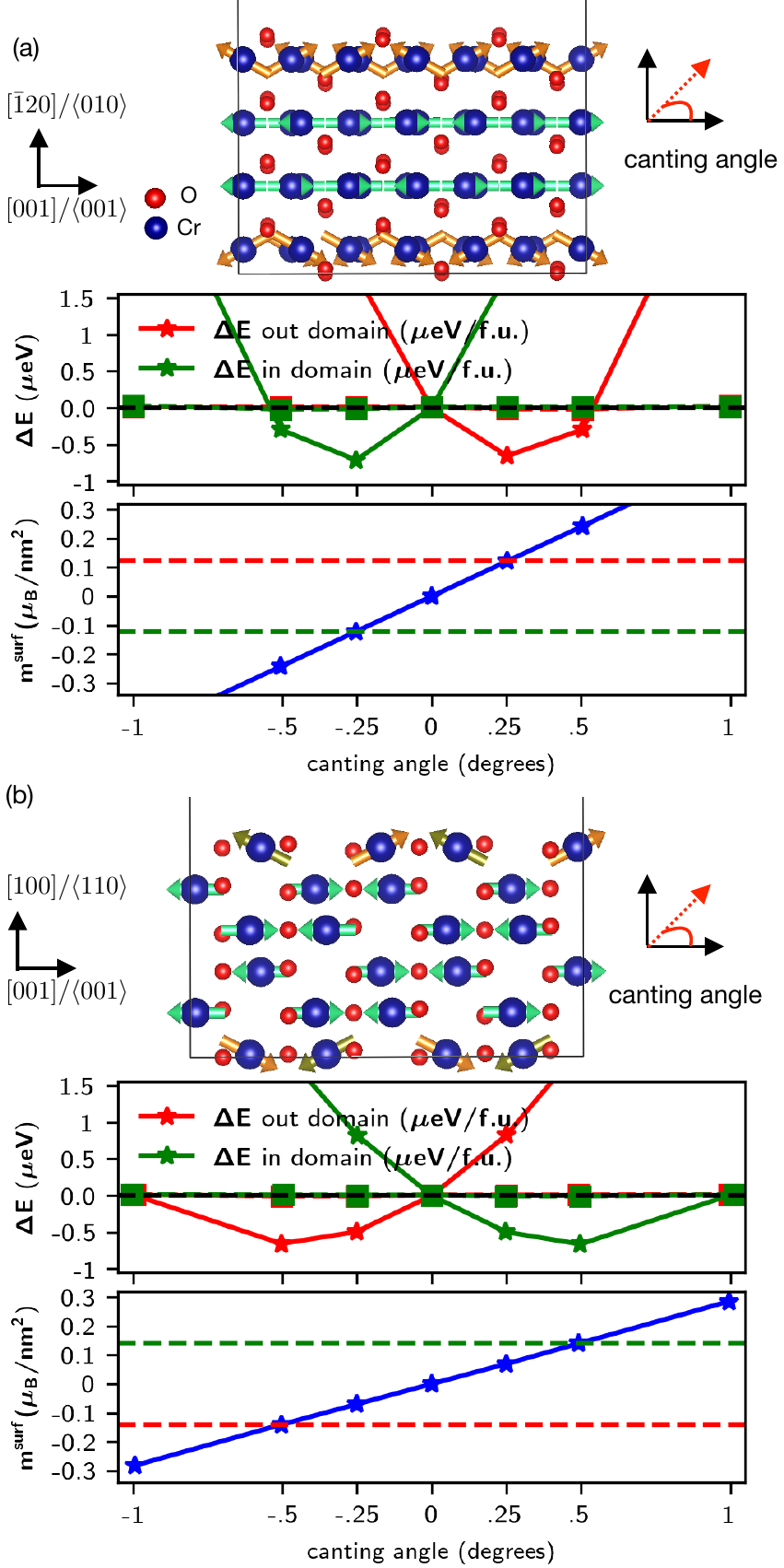}
\caption{Energy, $\Delta E$, as a function of the canting angle for the in-pointing (green line and stars) and out-pointing (red line and stars) domain, and net out-of-plane surface magnetization as a function of the canting angle for (a) $(\bar{1}20)$ and (b) $(100)$ Cr$_2$O$_3$ surfaces. Dashed lines in the bottom panel of both (a) and (b) identify the surface magnetization corresponding to the canting angle that minimizes the energy. Note that for the $(100)$ surface, an in-plane component of surface magnetization along $[001]$ is also allowed to develop; we depict the difference in magnitudes of the $[001]$ components of sublattice magnetization by the different colored arrows for the surface moments in the cartoon of the slab.}
\label{fig6}
\end{figure}

\begin{figure}[t]
\includegraphics{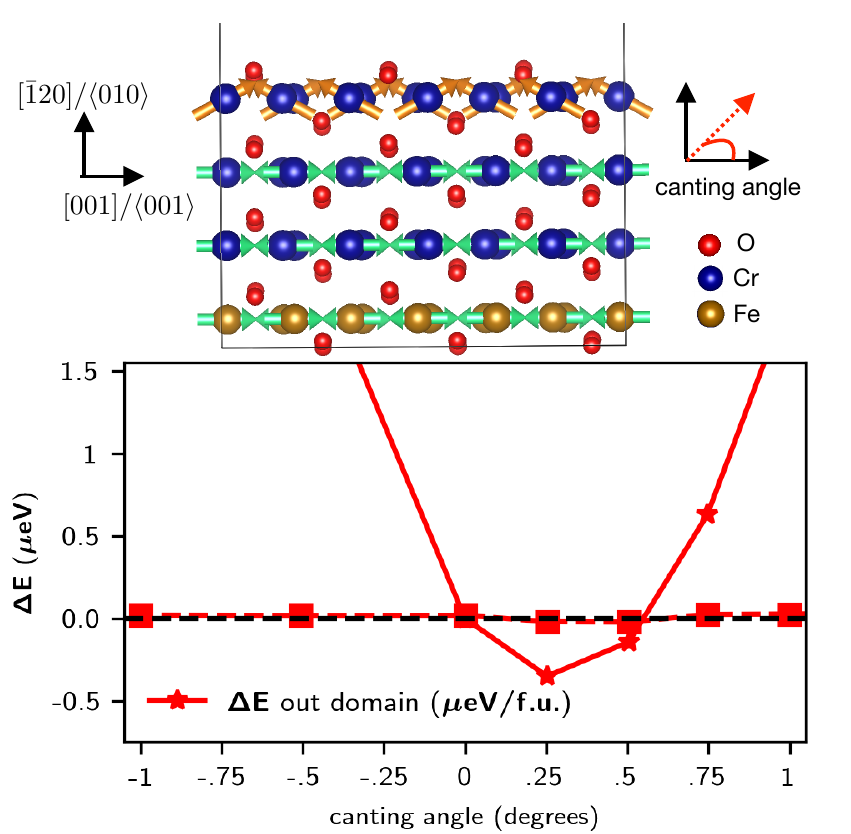}
\caption{\label{fig:Femono}Top: slab of $(\bar{1}20)$ chromia with a monolayer of $\mathrm{Fe}$ replacing the bottom $\mathrm{Cr}$, such that the MPG of the total slab is lowered and matches that of a semi-infinite slab (MPG $\mathbf{2}$). Bottom: Total DFT energy per formula unit for the $\mathrm{Cr}$-terminated surface as a function of canting angle. The penalty energy $E_p$ in the constrained magnetic calculations is plotted with square markers.}
\end{figure}

In our DFT calculations, whose technical details are described in Appendix \ref{m120_Cr2O3deets}, we fix the $\mathrm{Cr}$ moments in the center two (four) layers of our $\mathrm{4}$-layer ($\mathrm{6}$-layer) $(\bar{1}20)$-oriented ($(100)$-oriented) slab along the bulk $\left<001\right>$ direction, while inducing a surface magnetization for the outermost layers on the top and bottom of the slab by canting the moments along the $[\bar{1}20]$ ($[100]$) surface normal, with moments on the top and bottom surfaces canting either both towards vacuum or both towards bulk, as shown in the top panels of Figs. \ref{fig6}(a) and \ref{fig6}(b). The canting angle is defined with respect to the $\langle 001 \rangle$ axis. We define negative angles as a canting towards the bulk, and positive angles as a canting towards vacuum. The change in total energy per formula unit with respect to the energy with $0^{\circ}$ canting (no out-of-plane surface magnetization), and the net out-of-plane surface magnetization, as a function of canting angle, for both domains, are shown in the bottom panels of Figs. \ref{fig6}(a) and \ref{fig6}(b), for the $(\bar{1}20)$ and $(100)$ surfaces, respectively. Beginning with the $(\bar{1}20)$ case, Fig. \ref{fig6}(a), first we notice that, rather than an energy minimum for zero canting, the slab energy instead reaches a minimum when the surface moments are canted around $0.25^{\circ}$ with respect to $[001]$. While the energy difference is tiny ($<1$ $\mu\mathrm{eV}$), the penalty energy for constraining the moments (defined in Appendix \ref{m120_Cr2O3deets}), plotted with red and green squares for the two domains, is orders of magnitude smaller. This, combined with our stringent convergence criterion (we stop the electronic minimization when energy differences between successive iterations differ by $10^{-8}$ $\mathrm{eV}$ or less), makes us confident in the physical significance of the energy lowering. For $0.25^{\circ}$ canting we obtain a $(\bar{1}20)$ surface magnetization with magnitude $\sim 0.12$ $\mu_{\text{B}}/\mathrm{nm^2}$, calculated by taking the projections of all $12$ $\mathrm{Cr}$ surface moments onto the $[\bar{1}20]$ direction and dividing by the $(\bar{1}20)$ surface area. Second, we remark that the energy curves corresponding to the in-pointing and the out-pointing domains (red and green stars) are the mirror images of one another with respect to zero canting angle, hence their minima correspond to oppositely oriented surface magnetizations. This implies that the surface magnetization reverses if we switch from one domain to the other, as we expect from the interpretation in terms of ME responses.  \\
\indent The results for the $(100)$ surface, shown in Fig. \ref{fig6}(b), are qualitatively similar to those for $(\bar{1}20)$, but quantitatively slightly different as the canting perpendicular to the surface is predicted to be approximately $0.5^{\circ}$. Although the induced  moment per $\mathrm{Cr}$ correspondingly doubles for the $(100)$ surface compared to that for ($\bar{1}20)$, the density of surface $\mathrm{Cr}$ per unit area decreases, such that the out-of-plane surface magnetizations in units of $\mu_{\text{B}}/\mathrm{nm^2}$ for the $(\bar{1}20)$ and $(100)$ surfaces are almost identical, as can be seen in Fig. \ref{fig6}. We note that this is consistent with the isotropic linear ME response for $\mathrm{Cr_2O_3}$ in the basal plane, which corresponds to the out-of-plane component of induced surface magnetization. Again, the surface magnetizations for the two opposite domains on the $(100)$ surfaces are oppositely aligned with respect to each other, as expected based on the connection to the ME response.\\
\indent We now comment on the additional surface magnetization along $[001]$ which we expect for the $(100)$ surface based on our symmetry analysis in Secs. \ref{100_chromia} and \ref{sec:3rd_order_ME}. It is worth noting that in our DFT calculations we impose a spin canting perpendicular to the surface, and we constrain only the direction of the atomic magnetic dipoles to be in the plane spanned by the surface normal and the $[001]$ directions, but we do not constrain their magnitude. In this way we in principle allow the magnetic moments to arrange in a way that develops a net component along $[001]$ when summed over the Cr atoms. As discussed in Sec. \ref{-120_chromia}, For the case of the $(\bar{1}20)$ surface, finite magnetization along $[001]$ is forbidden by the symmetry of the MPG. Thus, the sublattice projections along $[001]$ remain equivalent, as indicated by the same-color moments for the slab cartoon in Fig. \ref{fig6}(a). The only surface magnetization induced is that along the $[\bar{1}20]$ surface normal, as plotted in the bottom panel of Fig. \ref{fig6}(a). On the other hand, based on the $(100)$ surface MPG, we expect an additional component of surface magnetization parallel to the N\'eel vector along $[001]$. Indeed, this is confirmed by our DFT calculations, in which the Cr magnetic moments oppositely aligned along $[001]$ on the top and bottom layers of the surface become inequivalent and show a difference in magnitude of $\approx 0.03 \mu_{\text{B}}$ (depicted pictorially by the different-colored surface sublattices) in Fig. \ref{fig6}(b). The sign of the surface magnetization along $[001]$ also switches with opposite domains, as expected from analogy to the third-order ME response. The subsequent in-plane induced surface magnetization is about half the size of the out-of-plane surface magnetization corresponding to the energetic minima in Fig. \ref{fig6}(b).\\
\indent Finally, we point out here a subtle conflict between the DFT calculations corresponding to Figs. \ref{fig6}(a) and \ref{fig6}(b) and the group-theory method used to determine the existence of surface magnetization. The surface MPG is determined by assuming a semi-infinite slab with a single, vacuum-terminated surface with normal $\mathbf{\hat{n}}$. In contrast, it is apparent from Fig. \ref{fig6}(a)-(b) that the nonpolar, symmetric $(\bar{1}20)$ and $(100)$ slabs have two surfaces, and hence actually belong to a higher-symmetry point group than the MPGs which we found above for the semi-infinite case. In fact, for example the point group of the symmetric $(\bar{1}20)$ slab, $2/m'$, is not even FM-compatible. However, we argue that using such a slab to explore surface magnetization is not problematic as long as the number of bulk-like layers between the two surfaces is sufficiently large: then the surfaces behave independently as though they were on two oppositely oriented, semi-infinite slabs. From an energetic perspective, only the local symmetry, i.e. the symmetry of one vacuum-terminated surface and the nearby bulk with which it interacts, is relevant in determining whether an induced magnetization is favorable for this surface. And this local surface symmetry is identical to the surface MPG determined by the group-theory formalism.\\
\indent To confirm that the local symmetry, rather than the symmetry of the entire DFT slab, is the relevant MPG to consider, we force the $(\bar{1}20)$ slab with two surfaces to have the same MPG as the idealized semi-infinite slab with a single $[\bar{1}20]$-oriented surface normal. We do this by substituting the bottom $\mathrm{Cr}$ layer with a monolayer of $\mathrm{Fe}$ as shown in the top of Fig. \ref{fig:Femono}, thereby breaking the extra mirror plane and inversion symmetries created by having two equivalent surfaces in \ref{fig6}(a). Because $\mathrm{Fe}$ naturally adopts a $+3$ valence state, this ``semi-infinite" slab is electrostatically stable and there are no convergence issues. We fix the $\mathrm{Fe}$ moments as well as the $\mathrm{Cr}$ moments in the center two layers along $\left<001\right>$ in the ``out" domain, and we again plot the change in total energy as a function of canting angle for the top layer of $\mathrm{Cr}$. While there are small quantitative differences between Figs. \ref{fig:Femono} and \ref{fig6}(a), the results for the $(\bar{1}20)$ surface with an $\mathrm{Fe}$ monolayer in Fig. \ref{fig:Femono} exhibit the same energy lowering with a minimum at a $0.25^{\circ}$ canting angle towards the vacuum. This demonstrates that indeed, it is the local surface symmetry for the DFT slab, reflecting the semi-infinite slab MPG, which dictates the energetics of the induced surface magnetization. Therefore, for all subsequent DFT calculations, we use the nonpolar symmetric slabs, which have higher global symmetry than the local surface MPG, to reduced computational cost.\\

\subsubsection{Summary of induced roughness-robust surface magnetism in $\mathrm{Cr_2O_3}$\label{sec:surfM_cr2o3_relevance}} Before ending this section, we comment briefly on the practical relevance of our findings for $(\bar{1}20)$ $\mathrm{Cr_2O_3}$. Admittedly, the value of induced surface magnetization, $0.12$ $\mu_{\text{B}}/\mathrm{nm^2}$ at $0$ $\mathrm{K}$ is small compared to even the reduced, room temperature value $\sim2$ $\mu_{\text{B}}/\mathrm{nm^2}$ for the uncompensated surface magnetization on $(001)$ chromia. We stress however that our calculations are not a rigorous indicator of the quantitative, experimentally induced magnetization that one would expect. The number of canting layers, as well as the energetically minimizing canting angle, almost certainly vary with slab thickness. For thicker slabs, it is likely that a surface could be stabilized further by canting the moments for several of the outermost layers, rather than just a single layer, thus increasing the total magnetization compared to the values we obtain here. Additionally, for exchange bias applications where the total surface magnetization summed over all unit cells at the FM-AFM interface is more relevant than the surface magnetization per unit cell, the small induced magnetization on $(\bar{1}20)$ $\mathrm{Cr_2O_3}$ could still have substantial effects. Encouragingly, recent experiments using magnetic circular dichroism have actually detected a nonzero surface magnetization on the nominally compensated surfaces of $\mathrm{Cr_2O_3}$ perpendicular to $(001)$\cite{Du2023}, suggesting that our theoretical prediction of induced surface magnetization is at least in this case large enough to be experimentally detected.\\
\indent Moreover, while the magnitude is small compared to the uncompensated case of Sec. \ref{subsec:uncomp_robus}, we still expect this surface magnetization to be much larger than the bulk magnetization induced by the transverse ME effect ($\mathbf{E}\perp[001]$) for realistic electric field strengths. Previous first-principles calculations of the lattice-mediated contribution to the transverse ME response indicate that for a typical electric field strength of $1$ $\mathrm{V}/\mathrm{nm}$, the induced magnetization in $\mathrm{Cr_2O_3}$ is about $0.002$ $\mu_{\text{B}}$ per $\mathrm{Cr}$ along $\mathbf{E}$\cite{Iniguez2008,Verbeek2023}, corresponding to a $\sim 0.04^{\circ}$ canting angle. In contrast, the surface magnetization toward the surface normal which we find for the $(\bar{1}20)$ and $(100)$ surfaces of chromia for canting angles of $0.25^{\circ}$ and $0.5^{\circ}$ translates to an induced $0.011$ $\mu_{\text{B}}$ and $0.22$ $\mu_{\text{B}}$ respectively per $\mathrm{Cr}$ ion. Thus, the surface magnetization per $\mathrm{Cr}$ induced by a vacuum-terminated surface is substantially larger than that induced due to the bulk ME effect in $\mathrm{Cr_2O_3}$ for a reasonably sized electric field. This is not surprising when considering that the effective electric field of a vacuum-terminated boundary should correspond roughly to the scale of the material crystal field, which can be orders of magnitude larger than an experimentally achievable electric field\cite{Belashchenko2010}.\\

\subsubsection{Induced $(110)$ surface magnetization in centrosymmetric $\mathrm{FeF_2}$}
We next consider induced, roughness-robust surface magnetization in a centrosymmetric AFM. We choose $\mathrm{FeF_2}$, a rutile-structure insulating AFM with a N\'{e}el temperature of $\sim80$ $\mathrm{K}$\cite{Lopez-Moreno2012}, for a few reasons. First, induced magnetization on the $(110)$ $\mathrm{FeF_2}$ surface was recently detected experimentally by analyzing the giant magnetoresistance (GMR) at the interface of an $\mathrm{FeF_2}$-$\mathrm{Cu}$-$\mathrm{Co}$ spin valve\cite{Lapa2020}. Secondly, the $(110)$ surface of $\mathrm{FeF_2}$ as well as isostructural $(110)$ $\mathrm{MnF_2}$ exhibit large exchange bias experimentally, despite the fact that these surfaces are magnetically compensated when considering the bulk AFM ground order\cite{Nogues1999, Munoz2015}. Many possible explanations have been proposed, including structural matching between the AFM and FM surfaces, or non-collinear coupling at the interface\cite{Nogues1999}. However, given the recent theoretical and experimental evidence that these surfaces have a symmetry-based equilibrium magnetization, a connection between induced surface magnetization and the exchange bias for $(110)$ $\mathrm{FeF_2}$ seems likely.\\
\begin{figure}[t]
\includegraphics[width=0.36\textwidth]{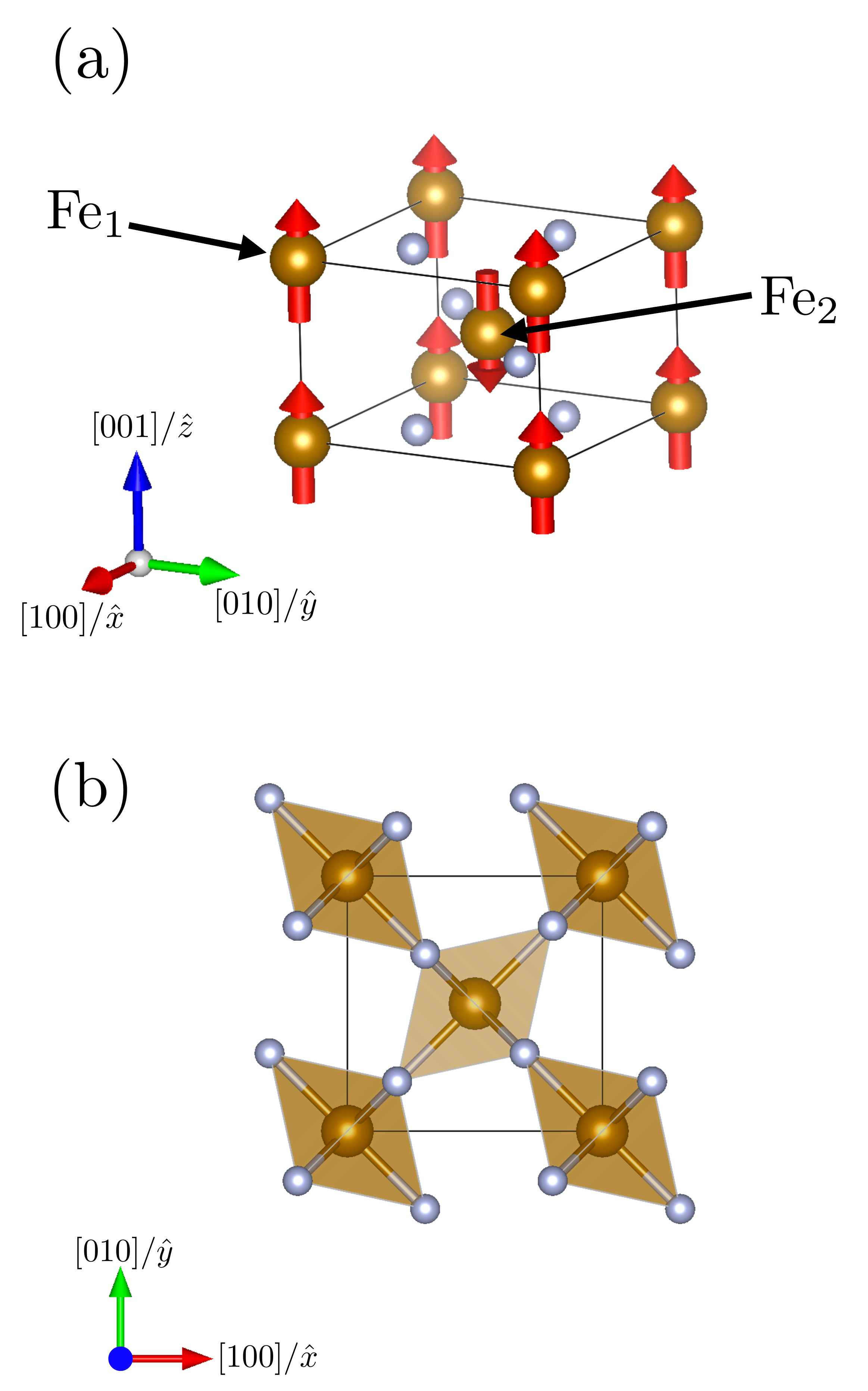}
\caption{(a) Unit cell and magnetic ordering of FeF$_2$. Fe and F atoms are identified by gold and grey spheres, respectively. (b) Top view of the unit cell of FeF$_2$, with FeF$_6$ octahedra highlighted.}
\label{fig2}
\end{figure}
\begin{figure}
\includegraphics{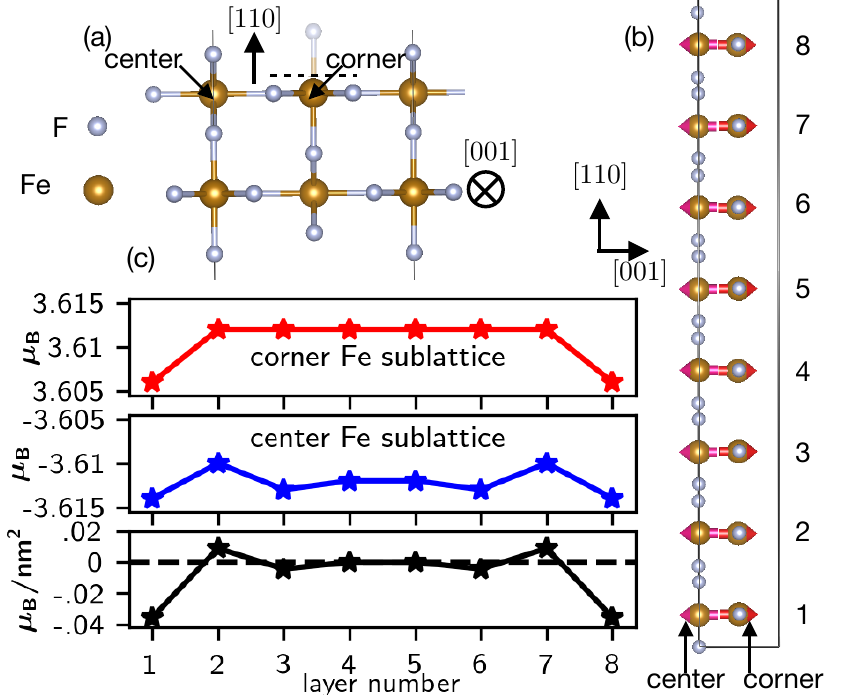}
\caption{\label{fig:FeF2_110} Induced magnetization for $(110)$ $\mathrm{FeF_2}$. (a) Unit cell defining nonpolar $(110)$ surface. The faded-out $\mathrm{F}$ ion above the dotted line for the ``corner", or $\mathrm{Fe}_1$ sublattice, is cut off in the nonpolar termination. (b) $[110]$-oriented slab with eight $\mathrm{Fe}$ layers ($16$ $\mathrm{Fe}$ ions) used in our DFT calculations. (c) Top: site-projected magnetization as a function of layer number for the ``corner" $\mathrm{Fe}$ sublattice. Middle: Same for the ``center" $\mathrm{Fe}$ sublattice. Bottom: Surface magnetization in $\mu_{\text{B}}/\mathrm{nm^2}$ for each layer.} 
\end{figure}
\indent FeF$_2$, isostructural to the recently studied unconventional antiferromagnet MnF$_2$ \cite{Bhowal2022}, crystallizes in the centrosymmetric tetragonal rutile structure described by the $P4_2/mnm$ space group ($4/mmm$ point group). Its unit cell, shown in Fig. \ref{fig2}(a), contains two formula units. The two Fe atoms sit at (0, 0, 0) and ($1/2$, $1/2$, $1/2$) and are octahedrally coordinated by the surrounding F atoms.  The FeF$_6$ octahedra are elongated along one axis (see Fig. \ref{fig2}(b)) and, importantly, for the two Fe atoms in the unit cell they are rotated by $90^{\circ}$ about the [001] direction with respect to each other. Hence, despite belonging to the same Wyckoff orbit (2a) and being thus symmetry-equivalent, the two Fe atoms are not connected by any translation. This has an important consequence when the magnetic ordering is taken into account, since the inequivalent fluorine environment of the Fe atoms makes the AFM order of FeF$_2$, described by the magnetic space group $P4'_2/mnm'$, break time-reversal symmetry.\\
\indent The bulk unit cell defining the nonpolar $(110)$ surface consists of two $\mathrm{Fe}$ layers, see Fig. \ref{fig:FeF2_110}(a). Fig. \ref{fig:FeF2_110}(b) shows the eight-layer slab used in our DFT calculations. Each layer stacked along the $[110]$ direction contains of one $\mathrm{Fe}$ ion per unit cell from the corner of the bulk unit cell and one from the center sublattice, such that the $(110)$ surface is magnetically compensated with oppositely pointed in-plane moments. We now find the symmetries of the bulk MSG which leave $\mathbf{\hat{n}}\parallel[110]$ invariant modulo translations parallel to the surface. Such symmetry analysis for the existence of $(110)$ surface magnetization was already discussed in Ref. \citenum{Lapa2020}, but we repeat it here briefly. In addition to the trivial identity operation, these are time-reversal combined with a 2-fold rotation $\left(2'_{[110]}|0\right)$ about the $[110]$ surface normal, as well as two mirror planes $\left(\sigma_{(001)}|0\right)$ and $\left(\sigma'_{(1\bar{1}0)}|0\right)$. This corresponds to the FM-compatible surface MPG $2'mm'$. We now check how magnetization transforms under these operations, taking a rotated Cartesian basis with $x\parallel[110]$, $y\parallel[1\bar{1}0]$ and $z\parallel[001]$:  
\begin{gather}
2'_{[110]}:\left(m_x,m_y,m_z\right)\rightarrow\left(-m_x,m_y,m_z\right);\label{eq:2_110}\\
\sigma'_{(1\bar{1}0)}:\left(m_x,m_y,m_z\right)\rightarrow\left(m_x,-m_y,m_z\right);\label{eq:M_1m10}\\
\sigma_{(001)}:\left(m_x,m_y,m_z\right)\rightarrow\left(-m_x,-m_y,m_z\right)\label{eq:M_001}.
\end{gather}
Eqs. \eqref{eq:2_110}-\eqref{eq:M_001} indicate that the only direction along which a magnetization can develop is $z\parallel[001]$. Thus, in contrast to $(\bar{1}20)$ $\mathrm{Cr_2O_3}$, for which surface magnetization developed via a relativistic canting of the surface moments perpendicular to the bulk N\'{e}el vector direction, for $(110)$ $\mathrm{FeF_2}$ the surface magnetization develops \emph{parallel} to the bulk N\'{e}el vector.\\
\indent In order for a finite magnetization to develop parallel to the N\'{e}el vector, the moment magnitudes of the two sublattices in the surface layer must necessarily become inequivalent. This is allowed by symmetry, since none of the operations \eqref{eq:2_110}-\eqref{eq:M_001} in the surface MPG connect the corner sublattice moment to the center sublattice moment for \emph{any} $(110)$ layer. Moreover, as pointed out in Ref. \citenum{Lapa2020}, we can understand intuitively why the symmetry-allowed imbalance in sublattice magnitude might be particularly enhanced for the vacuum-terminated surface layers (layers $1$ and $8$ in Fig. \ref{fig:FeF2_110}(b)) by considering the $\mathrm{F}$ coordination of the surface $\mathrm{Fe}$ ions. This is more easily seen by rotating the $(110)$ slab such that the $[001]$ N\'{e}el vector polarization points into the page, as done in Fig. \ref{fig:FeF2_110}(a). From this, it is apparent that for the electrostatically stable $(110)$ surface, the upper $\mathrm{F}$ atom for the ``corner" $\mathrm{Fe}$ sublattice is cut off (indicated by its greater transparency in the figure), whereas the ``center" $\mathrm{Fe}$ sublattice retains its full bulk coordination of $\mathrm{F}$ ions. Thus, the chemical environments around the $\mathrm{Fe}$ ions on the surface layers are strongly dissimilar, implying that the inequivalence in the magnitude of their moments will be large for the surface layers.\\
\indent Next, we show how the surface magnetism for the (110) surface of FeF$_2$ is interpreted in terms of ME effects. Since FeF$_2$ is centrosymmetric, with the Fe atoms at inversion centers, a linear ME response, both \textit{local} and \textit{net}, is forbidden by symmetry. Instead, the lowest-order ME response is the second-order ME effect, in which the induced magnetization is bilinear in the applied electric field, $\mathbf{M} \propto \beta \mathbf{E} \otimes \mathbf{E}$, see Eq. \eqref{eq11}. In FeF$_2$, this manifests both as a local, atomic response, and as a bulk net effect due to the breaking of the time-reversal symmetry. Following Ref. \onlinecite{Urru2022}, the local second-order ME response can be conveniently interpreted in terms of magnetic octupoles. FeF$_2$, similarly to MnF$_2$, allows for ferroically ordered $\mathcal{O}_{-2}$ octupoles and antiferroically ordered $\mathcal{O}_0$ octupoles. Additionally, FeF$_2$ hosts ferro-type $x^2-y^2$ quadrupoles of the toroidization density, $\mathcal{Q}^{(\tau)}_{x^2-y^2}$, and antiferro-type $z$ components of the moment of the toroidization density, $t^{(\tau)}_z$ (note: here we follow the naming convention of Ref. \onlinecite{Urru2022}). Overall, the full magnetic octupole tensor (see Appendix \ref{app_mag_oct}) reads 
\begin{widetext}
\begin{align}
  \mathcal{M}_{xjk} &= \begin{pmatrix} 0 & 0 & -\dfrac{1}{10} \mathcal{O}_0 + \dfrac{1}{3} t_z^{(\tau)} \\ 0 & 0 & \mathcal{O}_{-2} - \dfrac{1}{6} \mathcal{Q}_{x^2-y^2}^{(\tau)} \\ -\dfrac{1}{10} \mathcal{O}_0 + \dfrac{1}{3} t_z^{(\tau)} & \mathcal{O}_{-2} - \dfrac{1}{6} \mathcal{Q}_{x^2-y^2}^{(\tau)} & 0 \end{pmatrix}, \nonumber \\
  \mathcal{M}_{yjk} &= \begin{pmatrix} 0 & 0 & \mathcal{O}_{-2} - \dfrac{1}{6} \mathcal{Q}_{x^2-y^2}^{(\tau)} \\ 0 & 0 & - \dfrac{1}{10} \mathcal{O}_{0} + \dfrac{1}{3} t_z^{(\tau)} \\ \mathcal{O}_{-2} - \dfrac{1}{6} \mathcal{Q}_{x^2-y^2}^{(\tau)} & - \dfrac{1}{10} \mathcal{O}_0 + \dfrac{1}{3} t_z^{(\tau)} & 0 \end{pmatrix}, \\
  \mathcal{M}_{zjk} &= \begin{pmatrix} - \dfrac{1}{10} \mathcal{O}_{0} - \dfrac{2}{3} t_z^{(\tau)}& \mathcal{O}_{-2} + \dfrac{1}{3} \mathcal{Q}_{x^2-y^2}^{(\tau)} & 0 \\ \mathcal{O}_{-2} + \dfrac{1}{3} \mathcal{Q}_{x^2-y^2}^{(\tau)} & - \dfrac{1}{10} \mathcal{O}_{0} - \dfrac{2}{3} t_z^{(\tau)} & 0 \\ 0 & 0 & \dfrac{1}{5} \mathcal{O}_0 \end{pmatrix}. \nonumber
\end{align}
\end{widetext}
We note that, besides being symmetric upon exchange of $j$ and $k$ by construction, $\mathcal{M}_{xjk}$ and $\mathcal{M}_{yjk}$ are equivalent on exchange of $x$ and $y$, due to the tetragonal symmetry (with four-fold rotation axis parallel to $z$) of both the crystal and the magnetic structure. For the same reason, $\mathcal{M}_{zjk}$ is not equivalent to $\mathcal{M}_{xjk}$ and $\mathcal{M}_{yjk}$. 

As a next step, we consider the local second-order ME response. The Fe magnetic moments $\boldsymbol{\delta} \mathbf{m}$ induced by an external electric field $\mathbf{E}$ are given by
\begin{equation}
  \label{eq6}
  \delta m_i = \sum_{jk} \beta^{\text{loc}}_{ijk} E_j E_k.
\end{equation}
The tensor $\beta^{\text{loc}}_{ijk}$ describes the local second-order ME response and has the same symmetry properties as the magnetic octupole tensor \cite{Urru2022}, thus it can be written in terms of independent parameters in the following way:
\begin{align}
  \beta^{\text{loc}}_{xjk} &= \begin{pmatrix} 0 & 0 & \overline{b_{xxz}} \\ 0 & 0 & b_{xyz} \\ \overline{b_{xxz}} & b_{xyz} & 0 \end{pmatrix}, \nonumber \\
  \beta^{\text{loc}}_{yjk} &= \begin{pmatrix} 0 & 0 & b_{xyz} \\ 0 & 0 & \overline{b_{xxz}} \\ b_{xyz} & \overline{b_{xxz}} & 0 \end{pmatrix}, \label{eq12} \\
  \beta^{\text{loc}}_{zjk} &= \begin{pmatrix} \overline{b_{zxx}} & b_{zxy} & 0 \\ b_{zxy} & \overline{b_{zxx}} & 0 \\ 0 & 0 & \overline{b_{zzz}}, \end{pmatrix},  \nonumber
\end{align} 
where the overlined (non-overlined) symbols identify the antiferroic (ferroic) responses, i.e. those that have opposite (same) sign on the two Fe atoms. Following again the analogy between surface magnetism and ME effects, in order to study the magnetism of the (110) surface we analyze the bulk ME response to an electric field along the [110] direction. Previously, we showed that this can be done by either explicitly setting an electric field along the desired direction, or by applying a rotation to the tensor, such that it brings the electric field from the [$110$] direction to the $x$ direction. We adopt the latter approach, and we apply a clockwise rotation $\mathcal{R}$ of angle $\theta = 45^{\circ}$ about the $z$ axis, 
\begin{equation}
\mathcal{R} = \begin{pmatrix} \dfrac{\sqrt{2}}{2} & \dfrac{\sqrt{2}}{2} & 0 \\ -\dfrac{\sqrt{2}}{2} & \dfrac{\sqrt{2}}{2} & 0 \\ 0 & 0 & 1  \end{pmatrix},
\end{equation}
to the rank-3 tensor $\beta^{\text{loc}}$, which accordingly transforms as follows: 
\begin{equation}
\label{eq7}
\beta'^{\text{loc}}_{i'j'k'} = \sum_{i j k} \mathcal{R}_{i'i} \mathcal{R}_{j'j} \mathcal{R}_{k'k} \beta^{\text{loc}}_{ijk}.
\end{equation}
In the rotated framework, we are interested in the atomic magnetic moments induced by an electric field along $x$. From Eqs. \eqref{eq6}-\eqref{eq7}, we obtain: 
\begin{align}
  \delta m_x &= \beta'^{\text{loc}}_{xxx} E^2 = 0, \nonumber \\
  \label{eq8}
  \delta m_y &= \beta'^{\text{loc}}_{yxx} E^2 = 0, \\
  \delta m_z &= \beta'^{\text{loc}}_{zxx} E^2 = (\overline{b_{zxx}} + b_{zxy}) E^2, \nonumber 
\end{align}
which implies that the only second-order ME response allowed in this case is off-diagonal and that it does not induce any spin canting, but rather a change in size of the zero-field magnetic moment along the N\'eel direction $z \parallel [001]$. Interestingly, $\delta m_z$ contains both a contribution from the antiferroically ordered $\mathcal{O}_0$ and $t^{(\tau)}_{z}$ and the ferroically ordered $\mathcal{O}_{-2}$ and $\mathcal{Q}^{(\tau)}_{x^2-y^2}$. As a consequence, the local magnetic moments induced on the two Fe atoms,
\begin{equation}
\begin{split}
\label{eq9}
  \delta m_z (\text{Fe}_1) &= (\lvert \overline{b_{zxx}} \lvert + b_{zxy}) E^2, \\
  \delta m_z (\text{Fe}_2) &= (- \lvert \overline{b_{zxx}} \lvert + b_{zxy}) E^2,
\end{split}
\end{equation}
are different in size. Thus, the second-order $zxx$ ME response in this case is a \textit{ferri}-ME response. We remark that the different response of the two Fe atoms stems from their geometrically inequivalent fluorine environment, which is also crucial for the time-reversal symmetry breaking and, consequently, of the non-relativistic spin-splitting of the electronic energy bands, similarly to MnF$_2$ \cite{Bhowal2022}. Since the (110) surface magnetization can be interpreted in terms of a second-order bulk ME response to an electric field perpendicular to the surface, our results suggest a net magnetic moment parallel to the (110) surface, which is due to inequivalent anti-parallel magnetic moments, with different magnitude on the two Fe atoms, in agreement with the group-theory-based analysis discussed earlier.  

\subsubsection{\textit{Ab initio} calculations for $(110)$ $\mathrm{FeF_2}$}
\indent In this subsection, we confirm our group theory and second-order ME tensor analyses with first-principles DFT calculations, the details of which are described in Appendix \ref{110_FeF2deets}. First, we confirm that a finite surface magnetization does indeed develop along the $[001]$ in-plane direction for the $(110)$ surface. We plot our calculated, site-projected $\mathrm{Fe}$ magnetization (in $\mu_{\text{B}}$) as a function of layer number for corner and center $\mathrm{Fe}$ sublattices on the top and center subplots respectively in Fig. \ref{fig:FeF2_110}(c). The corresponding ``layer magnetization" in $\mu_{\text{B}}/\mathrm{nm^2}$, calculated by summing the two sublattice moments in each layer and dividing by the cross-sectional $(110)$ surface area, is plotted in the bottom of Fig. \ref{fig:FeF2_110}(c). For the central layers $3$-$6$ the magnitudes of the oppositely pointed moments are nearly constant and identical in size ($3.12$-$3.13$ $\mu_{\text{B}}$), and correspondingly the bulk layers are almost perfectly compensated as seen in the bottom plot. For the surface layers $1$ and $8$ however, we see an appreciable $0.008$ $\mu_{\text{B}}$ difference in sublattice magnitudes, corresponding to a negative surface magnetization of $-0.035$ $\mu_{\text{B}}/\mathrm{nm^2}$ along $[001]$. For our choice of AFM domain (Fig. \ref{fig2}(a)), this corresponds to a larger moment magnitude on the center $\mathrm{Fe}$ sublattice. Clearly, the precise numerical values for all three subplots in Fig. \ref{fig:FeF2_110}(c) are dependent on the specifics of the DFT parameters, such as the Hubbard $\mathrm{U}$ and Hund $\mathrm{J}$ as well as the radius of the integration sphere used to calculate the site-projected magnetization (all of which we report in Appendix \ref{110_FeF2deets}). Nevertheless, our results provide an unambiguous, quantitative confirmation of the symmetry analysis.\\
\begin{figure}[t]
\includegraphics[width=0.48\textwidth]{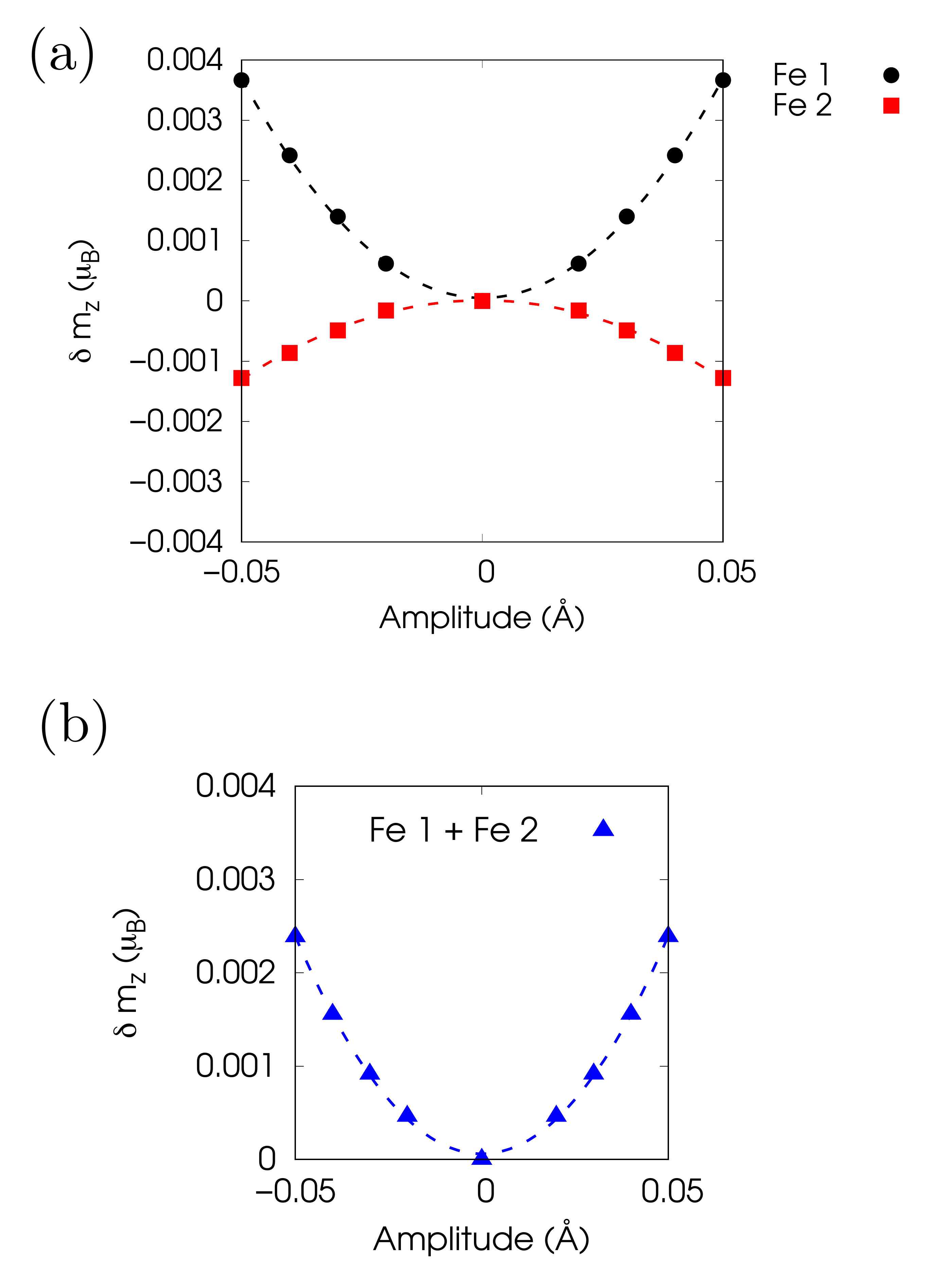}
\caption{(a) Change in the magnetic moment along $z$ on the two Fe atoms in FeF$_2$, induced by an infrared-active phonon mode polarized along [110]. (b) Sum of the Fe magnetic moments along $z$ induced by an infrared-active phonon mode polarized along [110].}
\label{fig1}
\end{figure}
\indent Second, we confirm our predictions for the local and net ME response  with explicit \textit{ab initio} calculations. These are carried out by freezing in infrared-active phonon modes polarized along the [110] direction, corresponding to the atomic displacements induced by an electric field pointing along [110]. According to our calculations, FeF$_2$ shows seven infrared-active phonon modes. Six of these transform as the $E_u$ irreducible representation of the $4/mmm$ point group: they are polarized in the $(x,y)$ plane, identified by the $[100]$ and $[010]$ directions, see Fig. \ref{fig2}, and are grouped into three families of degenerate modes, with frequencies 138 cm$^{-1}$, 267 cm$^{-1}$, and 454 cm$^{-1}$. The remaining infrared-active mode, with frequency 274 cm$^{-1}$, transforms as the $A_{2u}$ irreducible representation of $4/mmm$, thus it is polarized along the $z$ direction. Since we are interested in the response to an electric field along the $[110]$ direction, here we consider the contribution of the $E_u$ modes to the ME response. As a proof of concept, we focus on the contribution of the highest-frequency $E_u$ mode. We construct an appropriate linear combination of the two degenerate modes such that the resulting mode is polarized along $[110]$. Subsequently, we freeze this mode into the structure and study the induced magnetic moments on the Fe atoms as a function of the mode's amplitude. Fig. \ref{fig1} summarizes our results. Panels (a)-(b) show the induced magnetic moment along $z$ for the two Fe atoms. The response is quadratic in the mode amplitude and has a different magnitude for the two Fe atoms, thus rendering them inequivalent, as predicted by Eq. \eqref{eq9}. The difference in responses on the Fe sites results in a net induced magnetic moment when summing over the two ions, as shown in Fig. \ref{fig1}(c), in agreement with our tensor and symmetry analyses above. 

\subsubsection{Summary of induced roughness-robust surface magnetism in $\mathrm{FeF_2}$}
\indent Before concluding our discussion of roughness-robust, induced surface magnetization, we point out one additional exciting implication of surface magnetization for AFMs such as $\mathrm{FeF_2}$. A recently identified class of AFMs exhibiting nonrelativistic spin-split band structures along particular directions in reciprocal space\cite{Yuan2020,Yuan2021,Smejkal2022,Bhowal2022} has generated great excitement due to their potential use in applications such as spin-charge conversion. The phenomenon requires broken $\Theta\mathcal{I}$ symmetry in the bulk MSG, and thus does not occur in linear ME AFM $\mathrm{Cr_2O_3}$\footnote{However, there are others MSGs with all of $\Theta$, $\mathcal{I}$ and $\Theta\mathcal{I}$ broken such that both nonrelativistic spin-splitting and the linear ME effect are allowed}, but is allowed in centrosymmetric $\mathrm{FeF_2}$ and its isotructural counterpart $\mathrm{MnF_2}$\cite{Yuan2020,Yuan2021}.\\
\indent Crucially for us, the directions in the Brillouin zone along which nonrelativistic spin-splitting is allowed correspond to k points with a little group of $\mathbf{k}$ (the subgroup of the MPG leaving $\mathbf{k}$ invariant modulo a reciprocal vector) that does not reverse the spin direction\cite{Yuan2020}. Note that this is similar to the definition of a FM-compatible surface MPG, with the surface normal $\mathbf{\hat{n}}$ corresponding to the wave vector $\mathbf{k}$. Of course, there is not a one-to-one correspondence, firstly because unlike $\mathbf{\hat{n}}$, $\mathbf{k}$ switches sign with time-reversal, and secondly because real and reciprocal space directions are not parallel for non-orthogonal lattices. For cubic and tetragonal AFMs such as $\mathrm{FeF_2}$ however, the paths in reciprocal space with nonrelativistic spin-splitting coincide exactly with surfaces allowing induced surface magnetization. Specifically, in $\mathrm{FeF_2}$ and $\mathrm{MnF_2}$, the high-symmetry paths with nonrelativistic spin splitting are $\Gamma$-$\mathrm{M}$ and $\mathrm{Z}$-$\mathrm{A}$, corresponding to $\mathbf{k}=(u,u,0)$ and $\mathbf{k}=(u,u,1/2)$, as well as parallel paths at other constant $k_z$ values. These directions are parallel to the real space $(110)$ surface normal for which we and previous authors have demonstrated the existence of roughness-robust induced surface magnetization. Therefore, experimental detection of surface magnetization could be an alternative indicator of non-relativistic spin splitting, and vice versa.

\subsection{\label{subsec:induced_sense}Induced, roughness-sensitive surface magnetization}
We note that it is also possible to have AFM surfaces that have an induced magnetization which averages to zero in the presence of roughness. Given the combination of small surface magnetization in the induced case and roughness sensitivity, this fourth category of surface magnetization is likely not very useful. Moreover, thus far we have not found an explicit material example of this category. We nevertheless discuss the symmetry and multipole requirements for roughness-sensitive, induced surface magnetization and leave the analysis of a concrete example for future work.\\
\indent First, let us note the minimal bulk symmetry requirements in terms of $\mathcal{I}$ and $\Theta$ which allow for induced, roughness-sensitive surface magnetization. Based on analogous arguments to those given in Section \ref{subsec:induced_robus}, this surface magnetization can occur in AFMs with either broken or preserved inversion symmetry $\mathcal{I}$ in the bulk. Moreover, analogously to the case of roughness-sensitive, \emph{uncompensated} surface magnetization, roughness-sensitive induced surface magnetization can exist in AFMs with $\Theta$ preserved in the bulk, as long as there is a symmetry $\left(\Theta|\mathbf{t}\right)$ that combines $\Theta$ with a translation perpendicular to the surface in question.\\
\indent Let us now consider the form of the bulk magnetic multipoles which should correspond to this category of surface magnetization. Because we are dealing with magnetically compensated surfaces, the local-moment ME multipolization tensor must be zero; therefore, as explained in Sec. \ref{ME_response}, an induced surface magnetization should instead correspond to ferroically ordered atomic-site multipoles in the surface plane. The order of the relevant multipoles, that is, whether they are atomic-site ME multipoles, magnetic octupoles, or hexadecapoles, depends on the bulk MSG and the symmetries of the Wyckoff sites.\\
\indent In contrast to the roughness-robust induced surface magnetization discussed for $\mathrm{Cr_2O_3}$ and $\mathrm{FeF_2}$ in Part III however, roughness-sensitive induced surface magnetization should exist for an atomically flat surface only. This will occur if the atomic-site multipoles ordere ferroically within a plane, but if we move to an adjacent, parallel plane, the multipoles in this plane will be ferroically ordered, but with a direction opposite to that for the previous surface.  This is completely analogous to roughness-sensitive, uncompensated surface magnetization discussed in Sec. \ref{sec:multpol} for which the local-moment multipole tensor, and correspondingly the surface magnetization, changes sign for tensors corresponding to surfaces separated by atomic steps.

\section{\label{subsec:trivial}The ``null" case: $(1\bar{1}0)$ $\mathrm{NiO}$}
\begin{figure}
\includegraphics{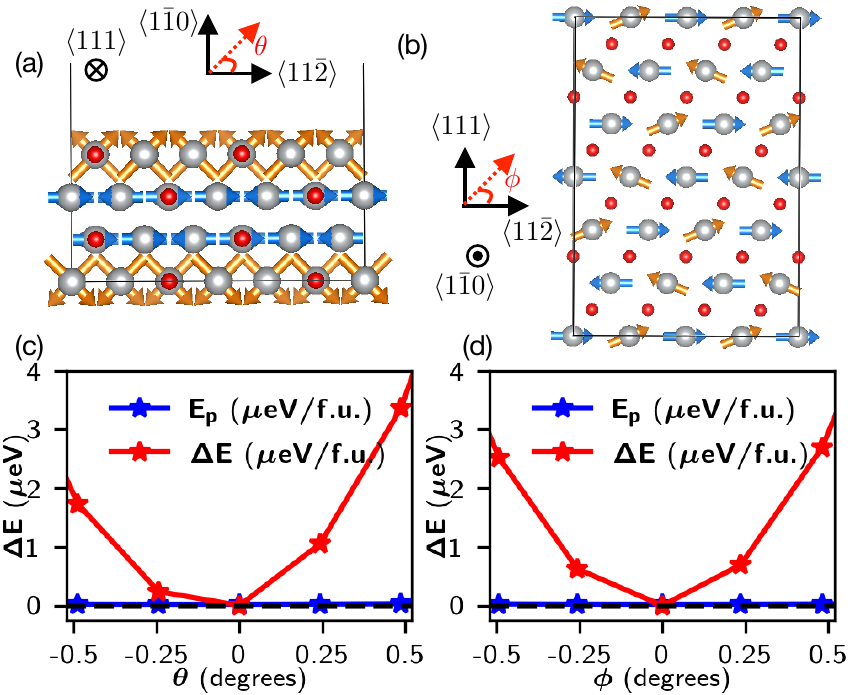}
\caption{\label{fig:1m10NiO_nosurfM} Induced magnetization on the non-FM-compatible $(1\bar{1}0)$ surface of $\mathrm{NiO}$. (a) Nonpolar slab used in our DFT+U calculations for the $(1\bar{1}0)$ surface, demonstrating the out-of-plane rotation $\theta$ for the surface $\mathrm{Ni}$ moments (in orange) which we enforce in our constrained magnetic calculations. (b) View of the slab looking down at the $(1\bar{1}0)$ surface, showing the in-plane rotation $\phi$ for the surface moments. (c) and (d) show the changes compared to the energy with the bulk $\left<11\bar{2}\right>$ polarization as a function of the rotation angles in (a) and (b) respectively. Blue lines in (c) and (d) show the penalty energy, $E_p$, in the constrained magnetic calculations.} 
\end{figure}

For $(100)$ and $(\bar{1}20)$ $\mathrm{Cr_2O_3}$ and $(110)$ $\mathrm{FeF_2}$, we have demonstrated that for surfaces which are magnetically compensated in the bulk AFM ground state, a symmetry-allowed induced magnetization energetically stabilizes the vacuum-terminated slab. In our final example, we show with DFT calculations that when a nominally compensated surface is not FM-compatible by symmetry, then inducing a finite surface magnetization is not energy lowering.\\
\indent We again take rock-salt AFM $\mathrm{NiO}$ as an example (the primitive bulk unit cell was shown in Fig. \ref{fig:NiO_Fe2O3}(a)). As a reminder, the bulk MSG is type IV $C_c2/c$ [15.90] with preserved $\mathcal{I}$ as well as $\Theta$ combined with several linearly independent translations. We already showed in Section \ref{subsec:uncomp_rough_sens} that the ideal, non-reconstructed $(111)$ surface of $\mathrm{NiO}$ has uncompensated but roughness-sensitive in-plane magnetization along $\left<11\bar{2}\right>$. We now consider the $(1\bar{1}0)$ surface, with surface normal $\mathbf{\hat{n}}\parallel\left<1\bar{1}0\right>$ that is perpendicular to both the $(111)$ surface normal and the bulk N\'{e}el vector polarization $\left<11\bar{2}\right>$. Looking at a nonpolar slab terminated with vacuum along $\left<1\bar{1}0\right>$ as shown in Fig. \ref{fig:1m10NiO_nosurfM}(a), it is clear that the $(1\bar{1}0)$ surface is magnetically compensated when the surface moments retain the bulk $\left<11\bar{2}\right>$ N\'{e}el vector direction. A search for the operations of $C_c2/c$ that leave the unit normal of the $(1\bar{1}0)$ surface invariant modulo translations $\perp{\mathbf{\hat{n}}}$ yields the following four operations: $\mathrm{1}$, $\left(\mathrm{2}_{\left<1\bar{1}0\right>}|1, 1, 1\right)$, $\left(\mathrm{2}'_{\left<1\bar{1}0\right>}|0\right)$ and $\left(\Theta|1,1,1\right)$. The corresponding surface MPG is $21'$, which is not FM-compatible because it contains $\Theta$. Indeed, the time-reversal operation $\Theta$ reverses every component of magnetization, so there is no direction along which FM is symmetry-allowed on the $(1\bar{1}0)$ surface. \\
\indent We check this by performing DFT calculations for a slab of NiO with $(1\bar{1}0)$ surfaces (details are in Appendix \ref{1m10_NiOdeets}). We keep the $\mathrm{Ni}$ moments in the center two layers of the $(1\bar{1}0)$ slab fixed along the bulk $\left<11\bar{2}\right>$ N\'{e}el vector direction, and we cant the $\mathrm{Ni}$ moments in the two surface layers about two orthogonal rotation axes. We first rotate about the in-plane $\left<111\right>$ lattice vector, corresponding to an out-of-plane canting angle $\theta$ towards vacuum ($\theta>0$) or towards bulk ($\theta<0$) for both surfaces, as shown pictorially in Fig. \ref{fig:1m10NiO_nosurfM}(a). Separately, we cant around the out-of-plane $\left<1\bar{1}0\right>$ surface normal, corresponding to an in-plane rotation $\phi$ with respect to the bulk $\left<11\bar{2}\right>$ polarization. This is shown for the top surface with a bird's eye view of the surface in Fig. \ref{fig:1m10NiO_nosurfM}(b) (the blue non-rotated moments are from the bulk layers below). Both rotations induce a finite magnetization per unit area on the surface $\mathrm{Ni}$ layers, along $\left<1\bar{1}0\right>$ and $\left<111\right>$ (for $\theta$ and $\phi$ respectively).\\
\indent The changes in total DFT energy, given in $\mu\mathrm{eV}$ per formula unit, are shown in Figs. \ref{fig:1m10NiO_nosurfM}(c) and \ref{fig:1m10NiO_nosurfM}(d) for the out-of-plane canting $\theta$ and in-plane canting $\phi$ respectively. In contrast to Figs. \ref{fig6}(a)-(b), where a finite surface magnetization lowers the slab energy compared to the completely magnetically compensated surface, here the energetic minimum is at $\phi=\theta=0$ when there is zero surface magnetization, as expected.\\
\indent Note that qualitatively, Figs. \ref{fig:1m10NiO_nosurfM}(c) and (d) differ: for the in-plane canting in Fig. \ref{fig:1m10NiO_nosurfM}(d), the energy change is symmetric about $\phi=0$, whereas the energy change for out-of-plane canting is asymmetric about $\theta=0$ in Fig. \ref{fig:1m10NiO_nosurfM}(c). Specifically, the energy increase is more rapid when the canting on both surfaces is towards vacuum ($\theta>0$) than when the canting is toward the bulk. This is to be expected since the vacuum termination on one side of the $(1\bar{1}0)$ layer introduces a \emph{unidirectional} anisotropy to the local environment. Thus, we would expect the energy change for a canting towards vacuum to differ from a canting towards bulk. Nevertheless, it is clear that although the rate of energy increase differs for positive and negative $\theta$, any finite canting either toward vacuum or toward bulk destabilizes the $(1\bar{1}0)$ surface of $\mathrm{NiO}$. We have therefore shown that the group-theory formalism to identify AFM planes with surface magnetization also correctly predicts the cases where surface magnetization is not symmetry-allowed.\\
\indent Finally, we note that the surfaces with no expected equilibrium surface magnetization are characterized by atomic-site ME and higher-order multipoles which are antiferroically ordered \emph{in the surface plane}. Our analysis of the $\mathrm{Ni}$ octupoles for the $(1\bar{1}0)$ surface $\mathrm{Ni}$ ions shown in Fig. \ref{fig:1m10NiO_nosurfM} reveals that the octupolar ordering in the $(1\bar{1}0)$ surface is antiferroic, consistent with the group-theory analysis. 
\section{Summary and Conclusions} \label{conclusions}
\begin{table*}
\caption{Classification of the four categories of AFM surface magnetization in terms of the structure of their corresponding local-moment (lm) or atomic-site (as) multipoles and the order of the allowed bulk ME response. RR and RS refer to roughness-robust and roughness-sensitive respectively. For the atomic-site multipoles characterizing the induced surface magnetization, $r$ refers to the rank of the tensor ($r=2$ for ME multipoles, $r=3$ for magnetic octupoles, etc.) which depends on the bulk AFM MSG.}\label{tab:surfM_multpolorder_MEreponse} 
	\begin{tabular}{| c | c | c | c | c | c |}
 \hline
&  \multicolumn{2}{|c|}{\textbf{uncompensated}}
&  \multicolumn{2}{|c|}{\textbf{induced}}
&  \multicolumn{1}{|c|}{\textbf{no surface magnetization}}\\
\hline
& \textbf{RR} & \textbf{RS} & \textbf{RR} & \textbf{RS} & \textbf{-}\\
\hline
multipole & $\mathcal{I}$-broken $\widetilde{\mathcal{M}}_{ij}^{\text{lm}}$ & $\mathcal{I}$-preserved $\widetilde{\mathcal{M}}_{ij}^{\text{lm}}$ & ferroic $\mathcal{M}_r^{\text{as}}$ & ferroic $\mathcal{M}_r^{\text{as}}$ in-plane, & antiferroic $\mathcal{M}_r^{\text{as}}$\\
description & lattice & lattice &  in unit cell  &  antiferroic along $\mathbf{\hat{n}}$ & in-plane\\
\hline
bulk ME & linear & none & $\mathbf{M}\sim\mathbf{E}^{r-1}$  & none & none\\
responses for $\mathbf{E}\parallel\mathbf{\hat{n}}$ & $\mathbf{M}\sim\mathbf{E}$ & & & &\\
\hline
\end{tabular}
\end{table*}
In this work, we have provided a comprehensive classification of surface magnetization in AFMs. This decades-old topic has experienced a recent revival due to the appeal of controllable surface magnetization in ME AFMs, although as pointed out long before our work\cite{Andreev1996}, AFM surface magnetization as a general property is ubiquitous, as evidenced in the flowchart in Fig. \ref{fig:categories}.\\
\indent In Part I (Sec. \ref{sec:partI}), we have introduced a classification system of AFM surface magnetization in terms of its sensitivity to roughness and whether it exists considering the bulk AFM order with no modifications, or if it is induced by symmetry lowering at the surface. We also presented a modified group-theory formalism, building on the seminal works by Andreev and Belashchenko\cite{Andreev1996,Belashchenko2010}, which allows identification of planes with surface magnetization based on the bulk AFM symmetry, and distinguishes between roughness-robust and roughness sensitive cases.\\
\indent In Part II (Sec. \ref{sec:partII}), we focused on uncompensated surface magnetization and showed that it can be quantitatively described by local-moment ME multipoles. Moreover, we demonstrated with symmetry arguments and concrete examples of surfaces for $\mathrm{Cr_2O_3}$, $\mathrm{Fe_2O_3}$ and $\mathrm{NiO}$ that the character with respect to roughness is determined by whether the bulk local-moment multipolization lattice is centrosymmetric or noncentrosymmetric.\\
\indent Finally, in Part III (Sec. \ref{sec:partIII}) , we show that the remaining, induced cases of surface magnetization can be described by atomic-site ME multipoles, magnetic octupoles, and even higher order atomic-site terms in the magnetization density, depending on the particular bulk and surface symmetries. We discussed how the in-plane isotropic linear ME effect explains the component of surface magnetization perpendicular to the surface for $(\bar{1}20)$ and $(100)$ chromia, and how the anisotropic third-order ME response explains the additional component of surface magnetization parallel to the surface in $(100)$ chromia. In a similar way, we showed how the surface magnetization in $(110)$ FeF$_2$ follows from the inequivalence of the induced magnetic moments at the Fe sites and how this is a consequence of the local second-order ME effect, arising from local magnetic octupoles.\\  
\indent In Part III we also showed that induced, roughness-sensitive surface magnetization is associated with atomic-site multipoles that are ferroic in the surface plane, but which change sign upon a translation parallel to $\mathbf{\hat{n}}$. Finally, in Sec. \ref{subsec:trivial} we gave the ``null" example of a surface for which no surface magnetization is expected based on group theory, e.g. $(1\bar{1}0)$ $\mathrm{NiO}$, and explained this in terms of multipoles which are ordered antiferroically in the surface plane of the bulk unit cell tiling the surface. In Tab. \ref{tab:surfM_multpolorder_MEreponse}, as a succinct conclusion of Parts II and III we summarize the classification of the four discussed categories of AFM surface magnetization in terms of their bulk magnetic multipoles and ME responses.\\
\indent We conclude this manuscript with some outlook. While surface magnetization in the roughness-sensitive cases is undoubtedly more challenging to realize and control than roughness-robust surface magnetization, we believe that further investigation of roughness-sensitive surface magnetization is highly worthwhile given the huge number of previously overlooked possibilities it opens up for devices. To this end, the comprehensive group-theory formalism in Part I (Sec. \ref{sec:partI}) which we have built based on previous works\cite{Belashchenko2010,Andreev1996} serves as a straightforward recipe for distinguishing between roughness-robust and roughness-sensitive cases and for interpreting differences and similarities in their properties.\\
\indent Regarding induced, roughness-robust surface magnetization, our DFT calculations indicate that such surface magnetization occurs in agreement with non-material-specific symmetry arguments for both centrosymmetric $\mathrm{FeF_2}$ and non-centrosymmetric $\mathrm{Cr_2O_3}$. To our knowledge, this is the first \emph{ab initio} confirmation of such surface  magnetization. While induced surface magnetization has yet to generate the same attention as the uncompensated cases of $(001)$ $\mathrm{Cr_2O_3}$, it has been detected experimentally in at least three AFMs ($(110)$ $\mathrm{FeF_2}$\cite{Lapa2020}, $(110)$ $\mathrm{Fe_2TeO_6}$\cite{Wang2014} and very recently in surfaces perpendicular to $(001)$ $\mathrm{Cr_2O_3}$\cite{Du2023}, in agreement with our DFT calculations in Part III). Induced surface magnetization also has likely connections to the mechanism of exchange bias as well as nonrelativistic spin-splitting, and could be used as a zero-field alternative for some applications of the bulk ME effect, for example, AFM domain selection. Thus, roughness-robust induced surface magnetization appears to hold great practical promise. Its intimate symmetry connection to the bulk ME response at linear and higher orders, which we discuss at length in Part III (Sec. \ref{sec:partIII}), also makes roughness-robust induced magnetization fascinating from a fundamental perspective.\\
\indent Overall, we hope that this work motivates further studies, both experimental and theoretical, on the identification and the properties of various types of AFM surface magnetization. We also hope it spurs a comprehensive search for AFM candidates, both linear ME and otherwise, with experimentally useful surface magnetization. 
\appendix
\section{\label{m120_Cr2O3deets}DFT calculation details for $(\bar{1}20)$ and $(100)$ $\mathrm{Cr_2O_3}$}
For our density functional calculations of total energies of $(\bar{1}20)$ and $(100)$-oriented slabs of $\mathrm{Cr_2O_3}$ with respect to moment canting angle, we employ the Vienna \textit{ab initio} simulation package (VASP)\cite{Kresse1996} with the local density approximation (LDA) within the projector augmented wave method (PAW)\cite{Blochl1994}. We use the standard VASP PAW pseudopotentials with the following valence electron configurations: $\mathrm{Cr}$ $(3p^64s^23d^5)$ and $\mathrm{O}$ $(2s^2 2p^4)$. To approximately account for the localized nature of $3d$ electrons in $\mathrm{Cr}$, we add a Hubbard U correction\cite{Anisimov1997} using the rotationally invariant method by Dudarev et al.\cite{Dudarev1998}. We set $\mathrm{U}=4$ $\mathrm{eV}$ for consistency with our previous DFT+U work on $\mathrm{Cr_2O_3}$\cite{Weber2022}.\\
\indent To model the $(\bar{1}20)$ surface, we use an orthorhombic slab with in-plane lattice vectors $(-7.37,4.25,0.0)$ and $(0,0,13.53)$ in units of $\mathrm{\AA}$ using a Cartesian coordinate system with $z$ along the hexagonal $\left<001\right>$ axis. We double the minimal bulk cell defining the nonpolar $(\bar{1}20)$ surface along the $[\bar{1}20]$ direction for a total of four $12$-$\mathrm{Cr}$ layers, and add $15$ $\mathrm{\AA}$ of vacuum along $[\bar{1}20]$. We confirm that the slab is sufficiently thick to have bulklike behavior in the center by checking the layer-projected density of states. The slab is shown in the main text Fig. \ref{fig6}(a). We fix the lattice parameters to the relaxed values of bulk $\mathrm{Cr_2O_3}$ obtained in Ref. \citenum{Weber2022} and allow all internal coordinates to relax until forces on all atoms are less than $0.01$ $\mathrm{eV}/\mathrm{\AA}$. We expand the pseudo wavefunctions in a plane wave basis set with kinetic energy cut-off of 800 eV, and sample the Brillouin Zone (BZ) with a $\Gamma$-centered \textbf{k}-point Monkhorst-Pack mesh \cite{Monkhorst_Pack_mesh} with $6 \times 4 \times 1$ points. For the $(100)$ surface, we use an orthorhombic slab with in-plane lattice vectors $(2.46,4.26,0.0)$ and $(0,0,13.53)$. We use a slab with the size of the minimal bulk unit cell, with $6$ $\mathrm{Cr}$ layers ($24$ $\mathrm{Cr}$ atoms total) and $15$ $\mathrm{\AA}$ vacuum along $[100]$ (shown in Fig. \ref{fig6}(b)). We expand the pseudo wavefunctions in a plane-wave basis set with the same kinetic energy cut-off as for the $(\bar{1}20)$ surface and we sample the BZ with a $\Gamma$-centered k-point mesh of $14\times5\times1$ points for the total energy calculations.\\
\indent Because the energy differences we are exploring are very small (order $\mu\mathrm{eV}$), we use constrained magnetic calculations\cite{Ma2015} to ensure that all the $\mathrm{Cr}$ moments remain along the directions we initialize them in. The detailed implementation of constrained DFT for noncollinear magnetism is described in Ref. \citenum{Ma2015}. For our calculations, we select a Lagrange multiplier of $\lambda=10$, which was sufficient to both fix all moments along the desired directions, and keep the penalty energy (an energy contribution in the constrained magnetism formulation of DFT which forces magnetic moments to lie along the desired spin axis) well below the total energy differences.\\
\indent For the calculations of the $(\bar{1}20)$ surface in which we substitute one of the surface $\mathrm{Cr}$ layers for a monolayer of $\mathrm{Fe}$ in order to enforce the surface MPG of a semi-infinite slab, we use the same Hubbard $\mathrm{U}$ of $4$ $\mathrm{eV}$ for $\mathrm{Fe}$ and $\mathrm{Cr}$ $3d$ states. We use the same convergence criteria as for the symmetric $(\bar{1}20)$ slab, and use the same Lagrange multiplier $\lambda=10$ for the constrained magnetic calculations.
\section{\label{110_FeF2deets}DFT calculation details for $(110)$ $\mathrm{FeF_2}$}
For density-functional calculations for the $(110)$ $\mathrm{FeF_2}$ surface we again use VASP. We employ the generalized gradient approximation using the Perdew-Burke-Ernzerhof functional\cite{Perdew1996}, and we select a Hubbard $\mathrm{U}$ of $6$ $\mathrm{eV}$ and a Hund's exchange $\mathrm{J}$ of $0.95$ $\mathrm{eV}$. These parameters are based on previous DFT+U calculations of $\mathrm{FeF_2}$ and its $(110)$ surface\cite{Lopez-Moreno2012,Munoz2015}. Because surface magnetization in the case of $\mathrm{FeF_2}$ develops collinearly along the bulk N\'{e}el vector in contrast to $(\bar{1}20)$ $\mathrm{Cr_2O_3}$, we do not include spin-orbit coupling in our calculations and instead use collinear spin-polarized DFT. In calculating the site-projected magnetization for the $\mathrm{Fe}$ and $\mathrm{F}$ ions, we use the default integration spheres of $\mathrm{1.058} \mathrm{\AA}$ and $\mathrm{0.794}\mathrm{\AA}$, respectively. We relax the bulk $\mathrm{FeF_2}$ structure using the plane-wave and force criteria described for $\mathrm{Cr_2O_3}$ in the previous section, and a $6\times6\times9$ Gamma-centered k-point mesh for the bulk tetragonal cell. We then create a nonpolar $(110)$ slab with eight $\mathrm{Fe}$ layers and $15$ $\mathrm{\AA}$ of vacuum along $[110]$. Note that we increase the slab thickness of $(110)$ $\mathrm{FeF_2}$ compared to $(\bar{1}20)$ $\mathrm{Cr_2O_3}$ due to previous reports that $\sim8$ layers is required to obtain bulk-like behavior in the center of the slab\cite{Lopez-Moreno2012,Munoz2015}, which was not the case for $(\bar{1}20)$ $\mathrm{Cr_2O_3}$. Finally, we relax all internal coordinates of the slab while keeping the lattice vectors fixed to our bulk relaxed values.

For the calculation of the lattice-mediated ME response, phonon frequencies and eigendisplacements are computed using density functional perturbation theory \cite{DFPT_review_Baroni}, as implemented in the \texttt{Quantum ESPRESSO} (QE) \cite{QE_1,QE_2} and \texttt{thermo\_pw} \cite{thermo_pw} packages. Self-consistent ground state and linear response calculations are performed within the spin-polarized generalized gradient approximation scheme, and the exchange-correlation energy is treated using the Perdew-Burke-Ernzerhof parametrization \cite{perdew-burke-ernzerhof}. Ion cores are described using scalar-relativistic ultrasoft PPs \cite{US_Vanderbilt}, with 4$s$ and 3$d$ valence electrons for Fe (PP \texttt{Fe.pbe-n-rrkjus\_psl.1.0.0.UPF} from pslibrary 1.0.0 \cite{pslibrary,pslibrary_2}) and with 2$s$ and 2$p$ valence electrons for F (PP \texttt{F.pbe-n-rrkjus\_psl.1.0.0.UPF}). Correlation effects are dealt with by introducing a Hubbard U correction \cite{U_cococcioni}, with U $= 6$ eV and J $= 0.95$ eV, selected to be consistent with VASP calculations on the $(110)$ surface. Pseudo wavefunctions and charge density are expanded in a plane-wave basis set with kinetic energy cut-offs of 80 Ry and 400 Ry, respectively. BZ integrations are performed using a $\Gamma$-centered uniform Monkhorst-Pack \cite{Monkhorst_Pack_mesh} mesh of \textbf{k}-points with $4 \times 4 \times 6$ points.

Frozen phonon calculations of the induced Fe magnetic moments are performed at the distorted structure obtained by freezing in the phonon ionic displacements. The induced local magnetic moments are computed using the all-electron linearized augmented-plane-wave (LAPW) code \textsc{ELK} \cite{ELK}, to have a more accurate description of the atomic magnetic moments, within the local spin-density approximation, with spin-orbit coupling included. The APW wavefunctions and potential inside the muffin-tin spheres are expanded in a spherical harmonic basis set, with cut-off $l_{\text{max}} = 12$. Similarly to the calculations performed in QE, a Hubbard U correction with the same values of U and J is applied. The BZ is sampled with the same mesh of \text{k}-points used in QE.
\section{\label{1m10_NiOdeets}DFT calculation details for $(1\bar{1}0)$ $\mathrm{NiO}$}
For density-functional calculations of $(1\bar{1}0)$ $\mathrm{NiO}$ we again use VASP and choose the generalized gradient approximation with the Perdew-Burke-Ernzerhof functional\cite{Perdew1996}. We take $\mathrm{U}=4.6$ $\mathrm{eV}$ within the Dudarev implementation\cite{Dudarev1998} of DFT+U\cite{Anisimov1997}. This is based on the parameters used for DFT+U calculations of $\mathrm{NiO}$ in Ref. \citenum{Yuan2021}. We fully relax the bulk structure, and then relax the internal coordinates of a four-layer slab with a nonpolar $(1\bar{1}0)$ surface and $15$ $\mathrm{\AA}$ of vacuum along $\left<1\bar{1}0\right>$. We note that the slab is quite thin $\sim6$ $\mathrm{\AA}$; due to the large number of atoms in the $(\bar{1}10)$ surface plane ($96$ for the slab shown in Figs. \ref{fig:1m10NiO_nosurfM}(a) and (b)), a thicker slab would be computationally prohibitive. By checking the layer-projected density of states, we confirm that the center of the slab is sufficiently bulk-like. Analogously to the case of $(\bar{1}20)$ $\mathrm{Cr_2O_3}$, we perform constrained magnetic calculations with a Lagrange multiplier $\lambda=10$. 

\section{Magnetic octupole tensor}
\label{app_mag_oct}

Here we report the spherical irreducible components of the magnetic octupole tensor $\mathcal{M}_{ijk}$ introduced in Section \ref{ME_response}. For the full derivation see Ref.\ \onlinecite{Urru2022}. $\mathcal{M}$ can be decomposed into a totally symmetric tensor $\mathcal{S}$ and a residue tensor $\mathcal{R}$: 
\begin{equation}
    \mathcal{M}_{ijk} = \mathcal{S}_{ijk} + \mathcal{R}_{ijk}.
\end{equation}
The traceless part of the totally symmetric tensor $\mathcal{S}$ is expressed in terms of seven independent parameters ($\mathcal{O}_{-3}$, $\mathcal{O}_{-2}$, \dots, $\mathcal{O}_{2}$, $\mathcal{O}_{3}$) as follows: 
\begin{widetext}
\begin{equation}
\label{eq35}
    \begin{split}
        \mathcal{S}_{1jk} &= \frac{1}{20} \begin{bmatrix} 5 \mathcal{O}_3 - 3 \mathcal{O}_1 & 5 \mathcal{O}_{-3} - \mathcal{O}_{-1} & 2 \left( 5 \mathcal{O}_2 - \mathcal{O}_0 \right) \\ 
                                            5 \mathcal{O}_{-3} - \mathcal{O}_{-1} & - \left( 5 \mathcal{O}_3 + \mathcal{O}_1 \right) & 20 \mathcal{O}_{-2} \\ 
                                            2 \left( 5 \mathcal{O}_2 - \mathcal{O}_0 \right) & 20 \mathcal{O}_{-2} & 4 \mathcal{O}_1 \end{bmatrix}, \\
        \mathcal{S}_{2jk} &= \frac{1}{20} \begin{bmatrix} 5 \mathcal{O}_{-3} - \mathcal{O}_{-1} & -\left( 5 \mathcal{O}_{3} + \mathcal{O}_{1} \right) & 20 \mathcal{O}_{-2} \\ 
                                            - \left( 5 \mathcal{O}_{3} + \mathcal{O}_{1} \right) & - \left( 5 \mathcal{O}_{-3} + 3 \mathcal{O}_{-1} \right) & - 2 \left( 5 \mathcal{O}_2 + \mathcal{O}_0 \right) \\ 
                                            20 \mathcal{O}_{-2} & - 2 \left( 5 \mathcal{O}_2 + \mathcal{O}_0 \right) & 4 \mathcal{O}_1 \end{bmatrix}, \\
        \mathcal{S}_{3jk} &= \frac{1}{20} \begin{bmatrix} 2 \left( 5 \mathcal{O}_2 - \mathcal{O}_0 \right) & 20 \mathcal{O}_{-2} & 4 \mathcal{O}_1 \\ 20 \mathcal{O}_{-2} & -2 \left( 5 \mathcal{O}_2 + \mathcal{O}_0 \right) & 4 \mathcal{O}_{-1} \\ 4 \mathcal{O}_1 & 4 \mathcal{O}_{-1} & 4 \mathcal{O}_0 \end{bmatrix}.
    \end{split}
\end{equation}
\end{widetext}
We refer to these parameters as \textit{proper} magnetic octupoles, because they transform into each other upon rotations in the same way as the atomic $f$ orbitals. 

The residual tensor $\mathcal{R}$ can be further decomposed into two tensors, written by using five and three independent parameters respectively. The five-dimensional residue tensor, $\mathcal{R}^{(\mathbf{5})}$, is written in terms of the quadrupole moments $\mathcal{Q}^{(\tau)}$ of the toroidal moment density $\boldsymbol{\tau} (\mathbf{r}) = \mathbf{r} \times \boldsymbol{\mu} (\mathbf{r})$ as: 
\begin{widetext}
\begin{equation}
    \begin{split}
        R^{(\mathbf{5})}_{1jk} &= \frac{1}{3} \begin{bmatrix} 0 & - \mathcal{Q}^{(\tau)}_{xz} & \mathcal{Q}^{(\tau)}_{xy} \\ - \mathcal{Q}^{(\tau)}_{xz} & -2 \mathcal{Q}^{(\tau)}_{yz} & - \frac{1}{2} \left( \mathcal{Q}^{(\tau)}_{x^2-y^2} + 3 \mathcal{Q}^{(\tau)}_{z^2} \right) \\ \mathcal{Q}^{(\tau)}_{xy} & - \frac{1}{2} \left( \mathcal{Q}^{(\tau)}_{x^2-y^2} + 3 \mathcal{Q}^{(\tau)}_{z^2} \right) & 2 \mathcal{Q}^{(\tau)}_{yz} \end{bmatrix}, \\
        R^{(\mathbf{5})}_{2jk} &= \frac{1}{3} \begin{bmatrix} 2 \mathcal{Q}^{(\tau)}_{xz} & \mathcal{Q}^{(\tau)}_{yz} & - \frac{1}{2} \left( \mathcal{Q}^{(\tau)}_{x^2-y^2} - 3 \mathcal{Q}^{(\tau)}_{z^2} \right) \\ \mathcal{Q}^{(\tau)}_{yz} & 0 & - \mathcal{Q}^{(\tau)}_{xy} \\ - \frac{1}{2} \left( \mathcal{Q}^{(\tau)}_{x^2-y^2} - 3 \mathcal{Q}^{(\tau)}_{z^2} \right) & - \mathcal{Q}^{(\tau)}_{xy} & -2 \mathcal{Q}^{(\tau)}_{xz} \end{bmatrix}, \\
        R^{(\mathbf{5})}_{3jk} &= \frac{1}{3} \begin{bmatrix} -2 \mathcal{Q}^{(\tau)}_{xy} & \mathcal{Q}^{(\tau)}_{x^2-y^2} & - \mathcal{Q}^{(\tau)}_{yz} \\ \mathcal{Q}^{(\tau)}_{x^2-y^2} & 2 \mathcal{Q}^{(\tau)}_{xy} & \mathcal{Q}^{(\tau)}_{xz} \\ - \mathcal{Q}^{(\tau)}_{yz} & \mathcal{Q}^{(\tau)}_{xz} & 0 \end{bmatrix},
    \end{split}
\end{equation}
\end{widetext}
and the three-dimensional residue tensor $\mathcal{R}^{(\mathbf{3})}$ is cast into the three components of the moment of the toroidal moment density, $\mathbf{t}^{(\tau)} = \int \mathbf{r} \times \boldsymbol{\tau}(\mathbf{r}) d^3 \mathbf{r}$, in the following way: 
\begin{widetext}
\begin{equation}
    \begin{split}
        R^{(\mathbf{3})}_{1jk} &= \frac{1}{3} \begin{bmatrix} 0 & t^{(\tau)}_{y} & t^{(\tau)}_{z} \\ t^{(\tau)}_{y} & -2 t^{(\tau)}_{x} & 0 \\ t^{(\tau)}_{z} & 0 & - 2 t^{(\tau)}_{x} \end{bmatrix}, \\
        R^{(\mathbf{3})}_{2jk} &= \frac{1}{3} \begin{bmatrix} - 2 t^{(\tau)}_{y} & t^{(\tau)}_{x} & 0 \\ t^{(\tau)}_{x} & 0 & t^{(\tau)}_{z} \\ 0 & t^{(\tau)}_{z} & -2 t^{(\tau)}_{y} \end{bmatrix}, \\
        R^{(\mathbf{3})}_{3jk} &= \frac{1}{3} \begin{bmatrix} -2 t^{(\tau)}_{z} & 0 & t^{(\tau)}_{x} \\ 0 & - 2 t^{(\tau)}_{z} & t^{(\tau)}_{y} \\ t^{(\tau)}_{x} & t^{(\tau)}_{y} & 0 \end{bmatrix}.
    \end{split}
\end{equation}
\end{widetext}

\bibliography{surfmag_zoology.bib,references_Andrea.bib}

\end{document}